\documentclass[%
aps,
prb,
reprint,
superscriptaddress,
amsmath,
amssymb,
]{revtex4-2}

\usepackage{graphicx}
\usepackage{diagbox}
\usepackage{xcolor}
\usepackage{amsmath,amssymb,amsthm,amsfonts}
\usepackage{bbm}
\usepackage{bm}
\usepackage[normalem]{ulem}
\usepackage{amsthm}
\usepackage{mathtools}
\usepackage{physics}
\usepackage{xcolor}
\usepackage{graphicx}
\usepackage{adjustbox}
\usepackage{placeins}
\usepackage[T1]{fontenc}
\usepackage{lipsum}
\usepackage{csquotes}
\usepackage[compat=1.1.0]{tikz-feynhand} 
\usepackage{braket}
\usepackage{booktabs}
\usepackage{array} 
\usepackage{graphicx}
\usepackage{extarrows}
\usepackage{multirow}
\usepackage{float}

\usepackage[colorlinks,linkcolor=blue,anchorcolor=blue,citecolor=blue]{hyperref}

\usepackage{stackengine}
\newcommand\xrowht[2][0]{\addstackgap[.5\dimexpr#2\relax]{\vphantom{#1}}}

\usepackage{titlesec}
\titleformat{\paragraph}
  {\normalfont\normalsize\bfseries}{\theparagraph}{1em}{}

\titlespacing{\paragraph}{0pt}{5pt}{5pt} 
\begin{document}

\title{Bicircular Light Induced Multi-State Geometric Current}

\author{Zhichao Guo}
\thanks{These authors contributed equally to this work.}
\affiliation{Center for Quantum Matter, School of Physics, Zhejiang University, Zhejiang 310058, China}

\author{Zhuocheng Lu}
\thanks{These authors contributed equally to this work.}
\affiliation{Center for Quantum Matter, School of Physics, Zhejiang University, Zhejiang 310058, China}

\author{Hua Wang}
\email{daodaohw@zju.edu.cn}
\affiliation{Center for Quantum Matter, School of Physics, Zhejiang University, Zhejiang 310058, China}

\author{Kai Chang}
\email{kchang@zju.edu.cn}
\affiliation{Center for Quantum Matter, School of Physics, Zhejiang University, Zhejiang 310058, China}

\begin{abstract}
We investigate the photocurrent induced by bicircular light (BCL) in materials, with a focus on its multi-state geometric nature. BCL, a combination of left- and right-circularly polarized light, can generate both injection and shift currents, originating from the geometric properties of gauge-invariant shift vectors, quantum geometric tensors, and triple-phase products. Crucially, the real parts of the quantum geometric tensors and triple-phase products remain nonzero in centrosymmetric systems, facilitating photocurrent generation in contrast to the traditional shift current bulk photovoltaic effect. Using a diagrammatic approach, we systematically analyze the BCL-induced photocurrents and demonstrate the multi-state geometric nature within a one-dimensional three-site Rice-Mele model. Our findings provide a quantum geometric understanding of BCL-induced photocurrents, underscoring the importance of considering multi-band contributions in real materials.
\end{abstract}

\maketitle

\section{Introduction} \label{sec:Introduction_geometry}
    The bulk photovoltaic effect (BPVE) has attracted significant interest due to its broad applications in photodetectors and solar cells~\cite{BPVE_Wolfgang_1981, BPVE_Sipe_2000, BPVE_Fridkin_2001, BPVE_Andrew_2012, BPVE_Tan_2016, BPVE_Morimoto_2016}. Unlike traditional photovoltaics, which rely on heterojunction~\cite{CPL_Nagaosa_2017, CPL_Orenstein_2021, CPL_Watanabe_2021, CPL_Dai_2023}, the BPVE can occur in homogeneous, noncentrosymmetric materials illuminated by monochromatic light. This unique characteristic opens up possibilities for high-efficiency solar cells and polarization-sensitive photodetectors\cite{Application_BPVE_Grinberg_2013, Application_BPVE_Nie_2015}. The centrosymmetric materials, including Dirac and Weyl semimetals (e.g., Na\textsubscript{3}Bi and TaAs) \cite{Na3Bi_2014, HongDing_TaAs_2015}, and two-dimensional topological materials like 1T$^\prime$ transition-metal dichalcogenides (TMDCs) \cite{TMDs_Qian_2014, TMDs_Wang_2019, BCL_Habara_PRB_2023}, also exhibit strong optical responses. However, these materials often present challenges for BPVE studies under monochromatic light due to the constraints imposed by their inversion symmetry.
    
    To overcome this limitation, bicircular light (BCL) has emerged as a promising method for generating BPVE in centrosymmetric materials \cite{BCL_Sipe_1996, BCL_Ikeda_Morimoto_2021, BCL_Habara_PRB_2023, BCL_Zhan_Blcak_Phosphorus_2024}. BCL, consisting of two counter-rotating circularly polarized beams with frequencies $n_1\omega$ and $n_2\omega$, possesses unique dynamical symmetries that can break centrosymmetry, enabling nonlinear optical effects not achievable with monochromatic light. This field can be represented by the vector potential:
        \begin{equation}
            \boldsymbol{A}(t) = \boldsymbol{A}_R e^{in_1 \omega t} + \boldsymbol{A}_L e^{i(n_2 \omega t + \theta)} + \text{c.c.},
        \end{equation}
    where $\boldsymbol{A}_{L/R}$ are the amplitudes of the left- and right-circularly polarized light (LCP and RCP), respectively, and $\theta$ denotes the phase difference. The characteristics of BCL, including its symmetry-breaking effects, depend strongly on the frequency ratio $n_1/n_2$. Among various configurations, $(n_1, n_2) = (1,2)$ has been widely studied, where it induces a third order photocurrent in centrosymmetric systems as the leading nonlinear response.\cite{BCL_Ikeda_Morimoto_2023, BCL_Ikeda_Morimoto_2024, BCL_Kanega_Graphene_2024}.
    
    Given the unique properties of BCL, recent theoretical studies have explored the impact of BCL on various photocurrent mechanisms in centrosymmetric materials. In particular, BCL has been shown to induce both injection and shift currents \cite{BCL_Ikeda_Morimoto_2024}. While early works suggested that injection current, scaling with the relaxation time, dominates BCL-induced photocurrents \cite{BCL_tau1_Sipe_2000, BCL_tau1_Sipe_2012, BCL_tau1_Sipe_2014}, later studies found that shift current, independent of relaxation time, also plays a significant role. Notably, Ikeda et al.~\cite{BCL_Ikeda_Morimoto_2024} predicted the $\theta$-dependence of both types of currents in Dirac electron systems, a finding that was subsequently verified by Kanega et al.~\cite{BCL_Kanega_Graphene_2024} via quantum master equation simulations in a tight-binding model. Despite these theoretical advancements, a rigorous framework for first-principles calculations of BCL-induced photocurrents remains lacking. One critical aspect that has not been sufficiently explored is the role of multi-band contributions, which involve interactions beyond just two bands. 
    
    Furthermore, incorporating multi-band contributions is expected to reveal the composite and multi-state quantum geometry in BPVE. In this regard, BPVE serves as a valuable platform for studying geometric properties beyond single-state descriptions. Quantum geometry plays a crucial role in understanding nonlinear optical phenomena, especially photocurrent generation. While the connection between conventional BPVE (driven by monochromatic light) and the quantum geometric tensor and Levi-Civita connection is established~\cite{Geometry_Ahn_PRX_2020, Geometry_Ahn_NP_2022,Geometry_Avdoshkin_2023,Geometry2NLO_Wang_2024}, the geometric interpretation of BCL-induced BPVE, particularly the geometric features of multi-band contributions, remains an open question. 
    \usetikzlibrary{,fit}
\newcommand\addvmargin[1]{
  \node[fit=(current bounding box),inner ysep=#1,inner xsep=0]{};
}
\begin{table*}[t]
	\centering
	\small
	\begin{tabular}{m{5cm}<{\centering} m{4cm}<{\centering} m{4.5cm}<{\centering}}
		\toprule
		\toprule
		Components & Diagram & Expression \\
		\midrule
        Classical Photon propagator &
        \begin{tikzpicture}[baseline=0] 
            \begin{feynhand}
                \vertex (a) at (0, 0) {};
                \vertex (b) at (2, 0) {};
                \propag[pho] (a) to [edge label = {$\omega$, $\mu$}]  (b);
            \end{feynhand}
        \end{tikzpicture}
        & 1 
        \\     [0.8cm]
        Electron propagator &
        \begin{tikzpicture}[baseline=0]
            \begin{feynhand}
                \vertex (a) at (0, 0) {};
                \vertex (b) at (2, 0) {};
                \propag[fer] (a) to [edge label = {$a$}]  (b);
            \end{feynhand}
        \end{tikzpicture}
        & $ G_a(\omega) = \frac{1}{\hbar\omega - \varepsilon_{a} + i\gamma}$ 
        \\  [0.8cm]   
        Input vertex &
        \begin{tikzpicture}[baseline=0]
            \begin{feynhand}
                \vertex[dot] (a) at (0,0) {};
                \vertex (b) at (-0.71, 1.22) {$\omega_1$, $\alpha_1$};
                \vertex (c) at (-1.22, 0.71) {$\omega_2$, $\alpha_2$};
                \vertex (d) at (-1.0, -1.0) {$\omega_n$, $\alpha_n$};
                \propag[pho] (a) to (b);
                \propag[pho] (a) to (c);
                \propag[pho] (a) to (d);
                \vertex (e) at (-0.5, -0.2);
                \vertex (f) at (-0.6, 0.2);
                \propag[gho] (e) to [out = 120, in = -90] (f);
                \vertex (g) at (1, -1) {$a$};
                \vertex (h) at (1, 1) {$b$};
                \propag[fer] (g) to (a) ;
                \propag[fer] (a) to (h) ;
            \end{feynhand}
        \end{tikzpicture}
        & $ \frac{1}{n!}\prod\limits^{n}_{j=1} \left(\frac{ie}{\hbar \omega_j}\right)h^{\alpha_1...\alpha_n}_{ba}$
        \\     [1.5cm]
        Output vertex &
        \begin{tikzpicture}[baseline=0]
            \begin{feynhand}
                \vertex[crossdot] (a) at (0,0) {};
                \vertex (b) at (+0.71, 1.22) {$\omega$, $\mu$};
                \vertex (c) at (+1.22, 0.71) {$\omega_1$, $\alpha_1$};
                \vertex (d) at (+1.0, -1.0) {$\omega_n$, $\alpha_n$};
                \propag[pho] (a) to (b);
                \propag[pho] (a) to (c);
                \propag[pho] (a) to (d);
                \vertex (e) at (+0.5, -0.2);
                \vertex (f) at (+0.6, 0.2);
                \propag[gho] (e) to [out = 60, in = -90] (f);
                \vertex (g) at (-1, 1) {$a$};
                \vertex (h) at (-1, -1) {$b$};
                \propag[fer] (g) to (a) ;
                \propag[fer] (a) to (h) ;
            \end{feynhand}
        \end{tikzpicture}
        & $ \frac{1}{n!}\frac{e}{\hbar}\prod\limits^{n}_{j=1} \left(\frac{ie}{\hbar \omega_j}\right)h^{\mu\alpha_1...\alpha_n}_{ba}$
        \\ [1.5cm]
		\bottomrule
		\bottomrule
	\end{tabular}
	\setlength{\abovecaptionskip}{0.4cm}
	\caption{Diagram components and Feynman rules for nonlinear optical responses. The input vertex includes only the contributions from input fields $\{\omega_j, \alpha_j\}$, while the output vertex may contain both the output current $\{\omega, \mu\}$ and the contributions from input fields  $\{\omega_j, \alpha_j\}$.}
	\label{feynman rule}
\end{table*}
    
    In this paper, we present a comprehensive analysis of BCL-induced photocurrents and related composite and multi-state quantum geometries. The paper is organized as follows. Section \ref{sec:Theoretical Background} presents a concise overview of the field theory formalism and the associated Feynman rules. Section \ref{BCL Induced Current}  applies the diagrammatic approach to calculate the conductivities corresponding to the BCL-induced current, examining two resonant terms, injection and shift currents, successively, with a focus on their distinct relaxation time dependencies. We then categorize these conductivities based on their composite geometric quantities. In addition to the pairwise-band terms (involving two bands), we identify multi-band terms featuring both single and double singularities. These are analyzed in detail, considering their geometric nature and symmetry properties. To further elucidate the physical mechanisms underlying BCL-induced photocurrents, Section \ref{sec:Model Example} employs a three-site Rice-Mele model, with particular emphasis on multi-band contributions and the $\theta$-dependent photocurrent. Finally, Section \ref{sec:discussion} summarizes the key findings. The Supplementary Material provides an in-depth analysis of the frequency integrals, a detailed exposition of the conductivity calculations and classifications, and a symmetry analysis.
 
\section{Theoretical Background} \label{sec:Theoretical Background}
    In this section, we briefly review the field theory formalism and construction of Feynman rules \cite{Diagrammatic_Parker_Moore_2019, Diagrammatic_Chan_2024, Diagrammatic_Bradlyn_2024} for calculating BCL conductivity.
    \par
    For a crystalline material, the single-particle Hamiltonian in second quantization is expressed as:
    \begin{equation}
        \hat{H}_0 = \sum\limits_{a} \int [d\boldsymbol{k}] \varepsilon_{a}(\boldsymbol{k})\hat{c}^{\dagger}_{a}(\boldsymbol{k})\hat{c}_{a}(\boldsymbol{k}),
    \end{equation}
    $[d\boldsymbol{k}]$ is defined as $d^{d}\boldsymbol{k}/(2\pi)^d$ with $d$ being the dimension of system. $\hat{c}^{\dagger}_{a}$ and $\hat{c}_{a}$ denote the fermionic creation and annihilation operator of eigenstate $\ket{a}$ of $\hat{H}_0$. $\varepsilon_{a}$ is the band energy.
    \par
    When a classical external electric field is introduced as a perturbation, the full Hamiltonian $\hat{H}$ in the velocity gauge comprises the unperturbed component $\hat{H_0}$ and a perturbation term $\hat{V}$. The perturbation term $\hat{V}$ can be written as 
    \begin{equation}
		\begin{aligned}
            \hat{V}
			=& \sum\limits_{n=1}^{\infty}\frac{1}{n!} \prod\limits_{j=1}^{n}\int d\omega_{j} e^{-i\omega_j t} \left(\frac{ie}{\hbar \omega_{j}}\right)E_{\alpha_j}^{(\omega_{j})} \\
            &\times \sum\limits_{a, b}h^{\alpha_1...\alpha_n}_{ab}\hat{c}^{\dagger}_{a}(\boldsymbol{k})\hat{c}_{b}(\boldsymbol{k}).
		\end{aligned}
    \end{equation}
    Here, $e = -|e|$ is the electron charge and the $E_{\alpha_j}^{(\omega_{j})}$ denotes the $\alpha_j$-component of the electric field in Cartesian coordinates. Here we introduce the notation $\hat{h}^{\alpha_1...\alpha_n} = \left[\hat{D}^{\alpha_n}, ...[\hat{D}^{\alpha_1}, \hat{H}_0]\right]$ as $n$-th order covariant derivative. The covariant derivative is defined as $[\hat{D}^{\alpha}, \mathcal{\hat{O}}]_{ab} = \partial_{\alpha} \mathcal{O}_{ab} - i[\hat{r}^{\alpha}, \mathcal{\hat{O}}]_{ab}$, where $\partial_{\alpha} = \partial / \partial k_{\alpha}$ denotes the partial derivative with respect to $k_{\alpha}$, and $r^{\alpha}_{ab} = \bra{a}i\partial_{\alpha}\ket{b}$ corresponds to the Berry connection. Covariant derivatives have deep connections with some physical quantities. For example, the first-order covariant derivative is associated with the unperturbed velocity matrix $h^{\alpha}_{ab} = \braket{a|\partial_{\alpha} \hat{H}_0|b} = \hbar \braket{a|\hat{v}^{\alpha}|b}$. And the second-order covariant derivative $h^{\alpha\beta}_{ab} = \braket{a|\partial_{\alpha}\partial_{\beta}\hat{H}_0|b}$ represents the mass term~\cite{Non_local_potential_Cook_2017}, the off-diagonal components of which encode non-local potential effects, crucial in realistic material calculations \cite{Non_local_potential_Duan_2017}.
    \par
    To determine the optical conductivity, we begin by expressing the photocurrent as an ensemble average:
    \begin{equation}
	    \begin{aligned}
            \braket{J^{\mu}(t)} &= \frac{1}{Z} \text{Tr}\left[\hat{J}^{\mu}(t)e^{-\beta \hat{H}(t)}\right],
	    \end{aligned}
    \end{equation}
    \begin{equation}
	\begin{aligned}
            Z = \int \mathcal{D}c^{\dagger}\mathcal{D}c \text{ } \text{  exp}\left(-S_{E}\right),
	\end{aligned}
    \end{equation}
    \begin{equation}
	\begin{aligned}
            S_{E} =\int^{\beta}_{0} d\tau \left[\sum\limits_{a} \int [d\boldsymbol{k}] c^{*}_a(\boldsymbol{k}, \tau) \partial_{\tau} c_a(\boldsymbol{k}, \tau) + H_0 + V\right],
	\end{aligned}
    \end{equation}
    where $Z$ is the partition function in the imaginary time $\tau$ and $S_E$ is the Euclidean action. The creation and annihilation operators within are replaced by Grassmann variables, which also obey anticommutation relations. The current operator $\hat{J}^{\mu}(t)$ can be expressed using covariant derivatives of the full Hamiltonian
    \begin{equation}
        \begin{aligned}
            \hat{J}^{\mu}(t) =& \frac{e}{\hbar}[\hat{D}^{\mu}, \hat{H}(t)] \\
            =&\frac{e}{\hbar}\sum\limits_{n=0}^{\infty}\frac{1}{n!} \prod\limits_{j=1}^{n}\int d\omega_{j} e^{-i\omega_j t} \left(\frac{ie}{\hbar \omega_{j}}\right)E_{\alpha_j}^{(\omega_{j})} \\
            &\times \sum\limits_{a, b} h^{\mu\alpha_1 ...\alpha_n}_{ab}\hat{c}^{\dagger}_a(\boldsymbol{k}, t) \hat{c}_b(\boldsymbol{k}, t).
        \end{aligned}
    \end{equation}
    \par
    The corresponding conductivity is obtained via functional differentiation:
	\begin{equation}
		\begin{aligned}
			\sigma^{\mu\alpha_1...\alpha_n}(\omega; \omega_j) =& 
				\int \frac{dt}{2\pi}\prod\limits_{j=1}^{n}\int dt_{j} e^{i\omega_j t_j}  \\
                &\times\frac{\delta}{\delta E_{\alpha_{j}}^{(t_j)}} \braket{J^{\mu}(t)}|_{E=0}.
		\end{aligned}
	\end{equation}
    \par
    The $N$th-order nonlinear conductivity is computed diagrammatically using the following set of Feynman rules. The key components of the diagrammatic expansion are summarized in Table~\ref{feynman rule}, subject to the following conditions:
    (a) The conductivity diagrams contain two types of vertices: input vertices and output vertices, with the output vertex corresponding to the final current. Frequency conservation is enforced at each vertex.
    (b) Each diagram represents a fully connected Fermi loop with $N+1$ photon propagators, at least one of which must be attached to the output vertex.
    (c) To account for dissipative effects, such as scattering and relaxation processes, each Green's function is regularized by introducing an imaginary broadening term $i\gamma$, where $\gamma$ is inversely proportional to the relaxation time and is assumed to be small and positive.

\section{BCL Induced Currents} \label{BCL Induced Current}
    Using the path integral formalism outlined above, we now turn to the analysis of the BCL-induced photocurrent. We specifically examine the case where the input frequencies for left- and right-circularly polarized (LCP and RCP) light are $\Omega$ and $2\Omega$, respectively, which induces a third-order direct current response. The expression for the BCL-induced photocurrent is given by:
    \begin{equation}
        \label{photocurrent}
        J^{\mu}_{\text{BCL}}(\Omega) = \text{Re}\left[\sum\limits_{(\alpha\beta\gamma)}\sigma^{\mu\alpha\beta\gamma}_{\text{BCL}}E^{(\Omega)}_{\alpha}E^{(\Omega)}_{\beta}E^{(-2\Omega)}_{\gamma}\right],
    \end{equation}
    where $\boldsymbol{E}^{(\Omega)} = - \Omega A_0 (i, 1)$ and $\boldsymbol{E}^{(-2\Omega)} = 2e^{-i\theta} \Omega A_0 (i, 1)$ are the electric fields corresponding to LCP and RCP, and $A_0$ is the amplitude of the vector field. We defined $\sum_{(\alpha, \beta, \gamma)}$ as the summation over all combinations of the input Cartesian indices. The symmetric conductivity $\sigma^{\mu\alpha\beta\gamma}_{\text{BCL}} \equiv \sum_{\{\alpha \beta\gamma\}}\sigma^{\mu \alpha \beta \gamma}(0; \Omega, \Omega, -2\Omega)$ is defined as the sum over all permutations of the input Cartesian indices, e.g. $\sigma^{xxyy}_{\text{BCL}} = \sigma^{xxyy} + \sigma^{xyxy} + \sigma^{xyyx}$ \cite{BCL_Ikeda_Morimoto_2021}.
    \par
    Our primary focus here is on the resonant component, which generally dominates the response when the light frequency exceeds the band gap. We categorize the resonant contribution into two distinct types: injection current, which scales as $\mathcal{O}(\gamma^{-1})$, and shift current, which scales as $\mathcal{O}(\gamma^0)$. The properties of these contributions will be discussed separately.
        \begin{figure}[ht]
    \centering
    \includegraphics[scale=0.6]{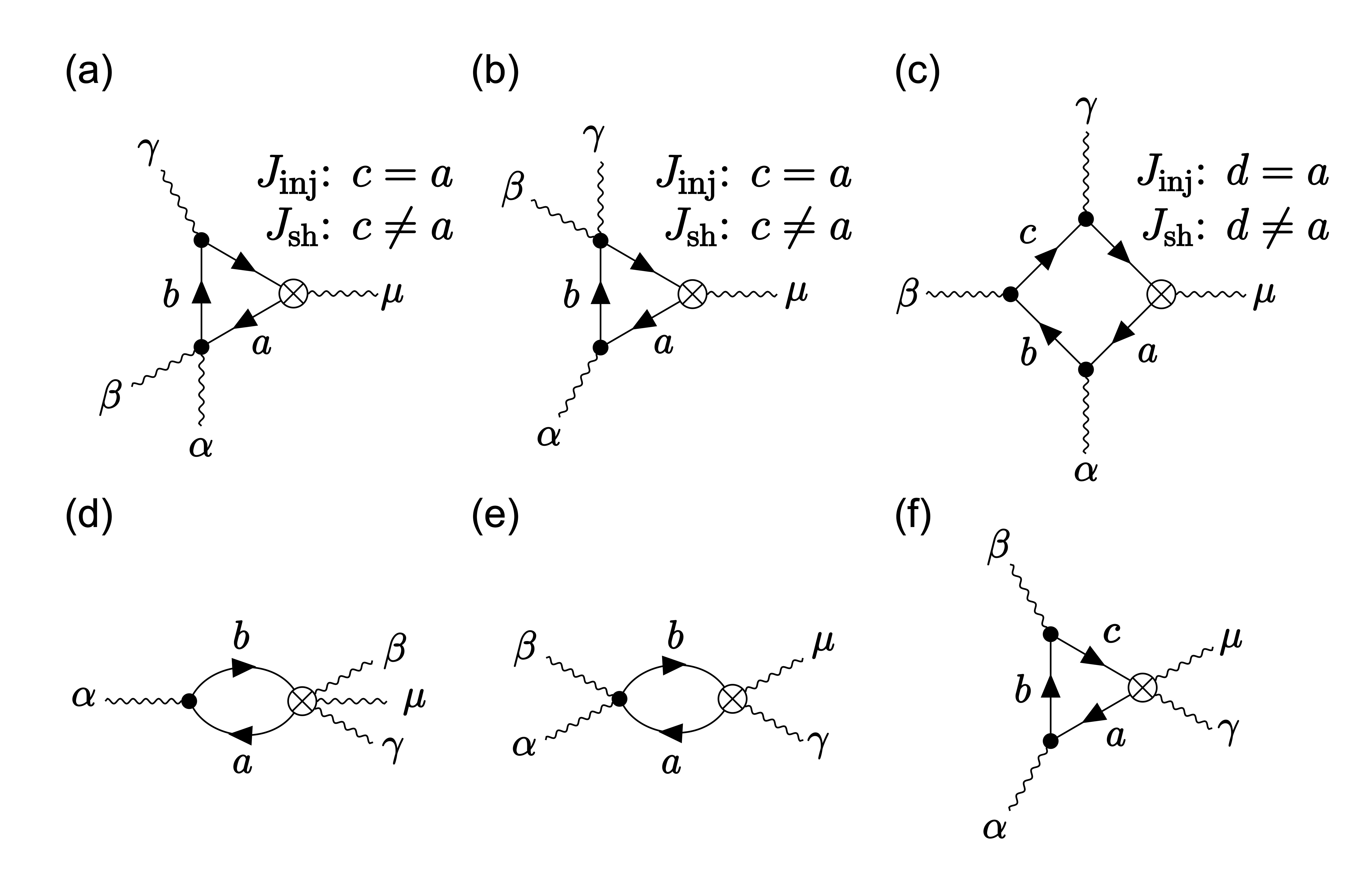}
    \caption{Feynman diagram illustrating the third order resonant photocurrent induced by BCL driving. Diagrams (a-c) contribute to both injection current ($J_{\text{inj}}$) and shift current ($J_{\text{sh}}$). Diagrams (d-e) contribute solely to shift current. The indices $\{\alpha, \beta, \gamma\}$ and $\{\mu\}$ represent the Cartesian coordinates of the three incident light beams and the output current, respectively, and $\{a, b, c, d\}$ indicate the band indices of each Green's function. Within diagrams (a-c), contribution of injection current (shift current) can be obtained by requiring the band indices to satisfy $c=a$/$d=a$ ($c \neq a$/$d \neq a$), respectively.}
    \label{feyn diagram for BCL current}
\end{figure}
        \begin{table*}[ht]
\centering
\small
\begin{tabular}{|m{4cm}<{\centering} | m{9cm}<{\centering} |m{2cm}<{\centering} |m{2cm}<{\centering}|}
\hline\xrowht[()]{18pt}
\textbf{Composite Geometry} & \textbf{Expression} & \textbf{Band Num.} & \textbf{Singularity} \\
\hline\xrowht[()]{24pt}
 $\Delta \times R \times Q$& 
$
\begin{aligned}
\frac{C}{3\gamma} \sum f_{ab} \Delta^{\mu}_{ab}\varepsilon^2_{ab}Q^{\alpha\gamma}_{ab} \left(  2R^{\beta,\gamma}_{ab}\delta^{\Omega}_{ab} + 
R^{\beta,\alpha}_{ba} \delta^{2\Omega}_{ab}\right)                   
\end{aligned}
$& \multirow{3}*{$\text{Pairwise}$} & ~ \\
\cline{1-2}\xrowht[()]{24pt}
 $\Delta \times \Delta \times Q$&
$
\begin{aligned}
\frac{iC}{3\gamma} \sum  f_{ab}\varepsilon_{ab}Q^{\alpha\beta}_{ab}\Delta^{\mu}_{ab}\Delta^{\gamma}_{ab}\left(5\delta^{\Omega}_{ab}-2\delta^{2\Omega}_{ab}\right)
\end{aligned}
$& ~ & \multirow{3}*{$\text{1S}$} \\
\cline{3-3}\cline{1-2}\xrowht[()]{72pt}
 $\Delta \times T$
& 
$
\begin{aligned}
&\frac{C}{3\gamma} \sum  \varepsilon_{ba}\varepsilon_{cb}\varepsilon_{ac} \\
&\left\{f_{ab} \Delta^{\mu}_{ab}\left(\frac{1}{\varepsilon_{cb}} - \frac{1}{\varepsilon_{ac}}\right)\left[2T^{\alpha\beta\gamma}_{abc} \delta^{\Omega}_{ab} - \left(T^{\alpha\beta\gamma}_{abc}\right)^*\delta^{2\Omega}_{ab}\right]\right. \\
&+ \left. 2f_{ab}\Delta^{\mu}_{ab} \left[T^{\alpha\beta\gamma}_{abc}\left(\tilde{d}^{2\Omega}_{ac} + \tilde{d}^{-\Omega}_{ac}\right)\delta^{\Omega}_{ab}-\left(T^{\alpha\beta\gamma}_{abc}\right)^*\tilde{d}^{\Omega}_{ac}\delta^{2\Omega}_{ab}\right] \right\}
\end{aligned}
$&\multirow{3}*{$\text{Multi}$} & ~ \\
\cline{1-2}\cline{4-4}\xrowht[()]{24pt}
$\Delta \times T $& 
$
\begin{aligned}
\frac{2i\pi C}{3\gamma} \sum \varepsilon_{ba}\varepsilon_{cb}\varepsilon_{ac}T^{\alpha\beta\gamma}_{abc}\left(f_{a}\Delta^{\mu}_{cb} + f_{b}\Delta^{\mu}_{ac} + f_{c}\Delta^{\mu}_{ba}\right) \delta^{\Omega}_{ab} \delta^{\Omega}_{bc}
\end{aligned}
$&~ & $\text{2S}$ \\
\hline
\end{tabular}
\caption{Classification of BCL-induced injection current conductivities based on composite geometry. Geometric and physical quantities are defined as follows: $\varepsilon_{ab} = \varepsilon_a - \varepsilon_b$ represents the energy difference; $\Delta^{\mu}_{ab} = \partial_{\mu}\varepsilon_{ab}$ denotes the group velocity difference; $R^{\mu, \alpha}_{ab} = i\partial_{\mu}\ln{r^{\alpha}_{ab}} + r^{\mu}_{aa} - r^{\mu}_{bb}$ denotes the shift vector; $Q^{\alpha\beta}_{ab} = r^{\alpha}_{ba}r^{\beta}_{ab}$ is the quantum geometric tensor; and $T^{\alpha\beta\gamma}_{abc} = r^{\alpha}_{ba}r^{\beta}_{cb}r^{\gamma}_{ac}$ represents the triple phase product. The composite geometry of each conductivity term is identified by products of these quantities, e.g., $\Delta \times R \times Q$ denotes a conductivity term involving group velocity differences, the shift vector, and the quantum geometric tensor. For conciseness, we introduce the following shorthand notations: $C = \frac{\pi e^4}{3! \hbar^4 \Omega^3}$ and $\sum = \sum\limits_{\{\alpha\beta\gamma\}} \sum\limits_{a \neq b (\neq c)} \int [d\boldsymbol{k}]$. Categories are further classified by the number of bands involved (`Pairwise' or `Multi') and the number of singularities (`$1\text{S}$' or `$2\text{S}$').} 
\label{injection currenty contribution}
\end{table*}

    \subsection{Resonant BCL-Induced Injection Current} \label{Resonant Effect I: Injection Current}
        In this subsection, we discuss the injection current conductivity. First, we identify the Feynman diagrams contributing to the third-order resonant response, which is proportional to the relaxation time. As outlined in Appendix \ref{appendix: Frequency Integral}, only three diagrams contribute to the injection current, as illustrated in Fig. \ref{feyn diagram for BCL current}.
        \par
        The calculated conductivity is generally proportional to a product of $d^{\omega}_{ab} = \left(\varepsilon_{ab} + \hbar\omega + i\gamma\right)^{-1}$, where $\varepsilon_{ab}$ denotes the energy difference between bands $a$ and $b$ and the $\gamma$ is a small positive smearing factor. To extract the resonant part involving the Dirac delta function, we further decompose $d^{\omega}_{ab}$ into its real and imaginary part. The real part is denoted as $\tilde{d}^{\omega}_{ab} \equiv \text{Re}d^{\omega}_{ab} $, and the imaginary part is proportional to a delta function $\text{Im} d^{\omega}_{ab} \equiv -\pi\delta^{\omega}_{ab}$, which becomes an ordinary Dirac delta function $\delta(\hbar\omega + \varepsilon_{ab})$ when $\gamma \rightarrow 0^+$. 
        Our analysis reveals not only conventional resonant contributions with a single Dirac delta function but also contributions involving products of two Dirac delta functions. While mathematically ill-defined in their standard form, these double-delta contributions are physically meaningful. To address this, we employ a regularization strategy, replacing the ordinary Dirac delta functions with their regularized counterparts. The validity of this approach is demonstrated in Appendix \ref{appendix: Properties of Regularized Dirac delta function}.
        \par
        We classify injection current conductivity involving a single Dirac delta function as "one singularity" ($\sigma^{\mu\alpha\beta\gamma}_{\text{BCL, inj, 1S}}$) and those involving a product of two Dirac delta functions as "double singularities" ($\sigma^{\mu\alpha\beta\gamma}_{\text{BCL, inj, 2S}}$).  The detailed derivation is presented in Appendix \ref{appendix: Derivation of Injection Current}. The corresponding expressions are as follows:
		\begin{equation}
            \label{Injection current expression 1S}
			\begin{aligned}
				\sigma^{\mu\alpha\beta\gamma}_{\text{BCL, inj, 1S}}
				=&\frac{1}{3!}\frac{i\pi e^4}{3\hbar^4\Omega^3\gamma}\sum\limits_{\{\alpha\beta\gamma\}}\int [d\boldsymbol{k}]  \sum\limits_{ab} f_{ab} (h^{\mu}_{aa} - h^{\mu}_{bb}) \\
                &\left\{\left(  2h^{\alpha}_{ba}h^{\beta\gamma}_{ab} \delta^{\Omega}_{ab}+ 
				h^{\alpha\beta}_{ba}h^{\gamma}_{ab} \delta^{2\Omega}_{ab}\right)\right.\\
                &+  2\left.\sum\limits_{c}\left[h^{\alpha}_{ab}h^{\beta}_{bc}h^{\gamma}_{ca} \tilde{d}^{\Omega}_{ac}\delta^{2\Omega}_{ab}\right.\right.\\
                &+\left.\left.h^{\alpha}_{ba}h^{\beta}_{cb}h^{\gamma}_{ac}\left(\tilde{d}^{2\Omega}_{ac} +  \tilde{d}^{-\Omega}_{ac}\right)\delta^{\Omega}_{ab}\right]\right\},
			\end{aligned}
		\end{equation}
        \par
        \begin{equation}
            \label{Injection current expression 2S}
			\begin{aligned}
				\sigma^{\mu\alpha\beta\gamma}_{\text{BCL, inj, 2S}}
				=& -\frac{1}{3!}\frac{2\pi^2 e^4}{3\hbar^4\Omega^3\gamma}\sum\limits_{\{\alpha\beta\gamma\}}\int [d\boldsymbol{k}] \sum\limits_{a \neq b \neq c} h^{\alpha}_{ba}h^{\beta}_{cb}h^{\gamma}_{ac} \\
                &\left[f_{a}\left(h^{\mu}_{cc} - h^{\mu}_{bb}\right) + f_{b}\left(h^{\mu}_{aa} - h^{\mu}_{cc}\right)  \right.\\
                &\left.+ f_{c}\left(h^{\mu}_{bb} - h^{\mu}_{aa}\right)\right]\delta^{\Omega}_{ab}\delta^{\Omega}_{bc},
			\end{aligned}
		\end{equation}
        where $f_{ab} = f_{a} - f_{b}$ is the difference between the Fermi-Dirac distribution functions of bands $a$ and $b$. In the one-singularity contribution, we encounter terms containing the second-order covariant derivative, which corresponds to a mass term  closely related to non-local potentials. Therefore, the first-principles calculations of BCL-induced injection current conductivity must consider both one-singularity and double-singularity contributions in realistic materials. The one-singularity contribution includes both $1\Omega$-resonant (involving $\delta^{\Omega}$) and $2\Omega$-resonant (involving $\delta^{2\Omega}$) components, corresponding to absorption of one photon with frequencies $\Omega$ and $2\Omega$, respectively. In contrast, the double-singularity contribution involves the product of two Dirac delta functions, both at the the $1\Omega$-resonant region, corresponding to a consecutive absorption of two photons with frequency $\Omega$. These distinct features will be further illustrated in the model discussed in Section \ref{sec:Model Example}.
        \par
        To better understand the physical origin of the BCL-induced photocurrent response, we categorize the contributions according to their dependence on fundamental physical and geometric quantities. The injection current conductivity involves four important quantities: the difference in group velocity ($\Delta$), the shift vector ($R$), the quantum geometric tensor ($Q$), and the triple phase product ($T$). The shift vector is a geometric quantity that quantifies the coordinate shift of the charge center between two bands \cite{Macdonald_Physical_illustration_of_shift_vector}. The real and imaginary parts of the quantum geometric tensor correspond to the quantum metric, which measures the distance between quantum states, and the Berry curvature, which describes the phase acquired by a quantum state traversing a closed loop in parameter space, respectively \cite{QGT_Provost_1980, QGT_PATI_1991, QGT_Bernevig_2020}. The triple phase product, $T$, has received less attention in prior studies. The geometric interpretation of $T$ is discussed in Section \ref{symmetry analysis}. Each category is characterized by a product of these quantities, termed "composite geometry". Four categories are summarized in Table \ref{injection currenty contribution}, with labels indicating the composite geometry, the number of bands involved, and the number of singularities. The detailed derivation is provided in Appendix \ref{appendix: Derivation of Injection Current}. Pairwise-band contributions, involving interactions between two bands, constitute the first two categories. These contributions exhibit a single singularity and are characterized by the quantum geometric tensor, $Q$. The final two categories, representing multi-band contributions, involve both single and double singularities and are characterized by the triple phase product, $T$. This three-band geometric quantity, analogous to Bargmann invariants  \cite{2023_Avdoshkin_Bargmann_invariants, 1964_original_Bargmann}, plays a critical role in gauge-invariant semiconductor Bloch equations, which is essential for modeling and interpreting light-matter interactions in solids\cite{parks2023gauge}.

    \begin{table*}[ht]
			\centering
			\small
			\begin{tabular}{|m{4cm}<{\centering} |m{9cm}<{\centering} |m{2cm}<{\centering} |m{2cm}<{\centering}|}
			\hline\xrowht[()]{18pt}
                \textbf{Composite Geometry} & \textbf{Expression} &\textbf{Band Num.} & \textbf{Singularity} \\
                \hline \xrowht[()]{18pt}
                $ \partial R \times Q $ 
                &
                $
                \begin{aligned}
                    \frac{iC}{2}\sum f_{ab}\varepsilon^2_{ab}Q^{\alpha\gamma}_{ab}\times \left(2\partial_{\mu}R^{\beta,\gamma}_{ab} \delta^{\Omega}_{ab} -  \partial_{\mu}R^{\beta,\alpha}_{ba} \delta^{2\Omega}_{ab}\right)
                \end{aligned}
                $& ~ & ~ 
                \\
                \cline{1-2}\xrowht[()]{18pt}
				$ R \times R \times Q $
                &
				$
                \begin{aligned}
				\frac{C}{2}\sum f_{ab}\varepsilon^2_{ab}Q^{\alpha\gamma}_{ab} \left(R^{\mu,\gamma}_{ab} - R^{\mu,\alpha}_{ba}\right)
				\left( R^{\beta,\gamma}_{ab}2\delta^{\Omega}_{ab} + R^{\beta,\alpha}_{ba}\delta^{2\Omega}_{ab} \right)    
                \end{aligned}
                $ &\multirow{5}*{Pairwise} & ~ \\
                \cline{1-2}\xrowht[()]{18pt}
				$ \Delta \times R \times Q$
                &
				$
                \begin{aligned}    
                    &\frac{iC}{2}\sum f_{ab}\varepsilon_{ab}Q^{\alpha\gamma}_{ab} \Delta^{\beta}_{ab}\left(R^{\mu, \gamma}_{ab} - R^{\mu, \alpha}_{ba}\right)\left(5\delta^{\Omega}_{ab} -2\delta^{2\Omega}_{ab}\right)
                \end{aligned}
                $ &~& ~ \\
                \cline{1-2}\xrowht[()]{18pt}
				$ \Delta \times \Delta \times Q$
				&
				$
                \begin{aligned}
                    C \sum f_{ab}\Delta^{\alpha}_{ab}\Delta^{\mu}_{ab}Q^{\beta\gamma}_{ab}\left( 2\delta^{\Omega}_{ab} - \delta^{2\Omega}_{ab}\right)
                \end{aligned}
                $& ~ & ~\\
                \cline{1-2}\xrowht[()]{18pt}
				$ w \times Q$ 
                &
				$
                \begin{aligned}      
                    -\frac{C}{2}\sum f_{ab}w ^{\mu\gamma}_{ab} \varepsilon_{ab}Q^{\alpha\beta}_{ab}\left(5\delta^{\Omega}_{ab} +2 \delta^{2\Omega}_{ab}\right)
                \end{aligned}
                $&~ & \multirow{3}*{1S} \\
                \cline{1-3}\xrowht[()]{96pt}
				$R \times T$
                &
				$
                \begin{aligned}
                    &C \sum \varepsilon_{ba}\varepsilon_{cb}\varepsilon_{ac} \\
                    &\left\{\frac{1}{2}f_{ab}\left(\frac{1}{\varepsilon_{cb}} - \frac{1}{\varepsilon_{ac}}\right)\times \left[2T^{\alpha\beta\gamma}_{abc}\left(R^{\mu, \gamma}_{ac} + R^{\mu, \beta}_{cb} - R^{\mu, \alpha}_{ba}\right)\delta^{\Omega}_{ab}\right.\right. \\
                    &\left.\left.+ \left(T^{\alpha\beta\gamma}_{abc}\right)^* \left(R^{\mu, \beta}_{bc} + R^{\mu, \gamma}_{ca} - R^{\mu, \alpha}_{ab}\right)\delta^{2\Omega}_{ab}\right]\right\} \\
                    &+ \left\{ f_{ab} \times \left[T^{\alpha\beta\gamma}_{abc}\left(R^{\mu, \gamma}_{ac} + R^{\mu, \beta}_{cb} - R^{\mu, \alpha}_{ba}\right)\left(\tilde{d}^{-\Omega}_{ac} + \tilde{d}^{2\Omega}_{ac}\right)\delta^{\Omega}_{ab} \right. \right.\\
                    &\left.\left. + \left(R^{\mu, \gamma}_{ca} + R^{\mu, \beta}_{bc} - R^{\mu, \alpha}_{ab}\right)\left(T^{\alpha\beta\gamma}_{abc}\right)^*\tilde{d}^{\Omega}_{ac}\delta^{2\Omega}_{ab}\right]\right\}
                \end{aligned}
                $&\multirow{13}*{Multi}&~\\
                \cline{1-2}\xrowht[()]{88pt}
				$\Delta \times T$
                &
				$
                \begin{aligned}  
                    & iC \sum \varepsilon_{ba}\varepsilon_{cb}\varepsilon_{ac} \\
                    &\left\{\frac{1}{2}f_{ab}\left(\frac{1}{\varepsilon_{cb}} - \frac{1}{\varepsilon_{ac}}\right)\times\left(\frac{\Delta^{\mu}_{cb} - \Delta^{\mu}_{ac}}{\varepsilon_{cb} - \varepsilon_{ac}} - \frac{\Delta^{\mu}_{ba}}{\varepsilon_{ba}}\right) \right. \\
                    &\times \left. \left[2T^{\alpha\beta\gamma}_{abc} \delta^{\Omega}_{ab} + \left(T^{\alpha\beta\gamma}_{abc}\right)^* \delta^{2\Omega}_{ab}\right]\right\}+ \left\{f_{ab}\left(\frac{\Delta^{\mu}_{ac}}{\varepsilon_{ac}}+ \frac{\Delta^{\mu}_{cb}}{\varepsilon_{cb}} - \frac{\Delta^{\mu}_{ba}}{\varepsilon_{ba}} \right) \right. \\
                    &\times \left. \left[T^{\alpha\beta\gamma}_{abc}\left(\tilde{d}^{-\Omega}_{ac} + \tilde{d}^{2\Omega}_{ac}\right)\delta^{\Omega}_{ab} +\left(T^{\alpha\beta\gamma}_{abc}\right)^*\tilde{d}^{\Omega}_{ac}\delta^{2\Omega}_{ab}\right]\right\}
                \end{aligned}
                $ &~&~\\
				\cline{1-2}\cline{4-4}\xrowht[()]{40pt}
				$R \times T$
                &
				$
                \begin{aligned}
				    &-i\pi C \sum  \varepsilon_{ba}\varepsilon_{cb}\varepsilon_{ac} f_{ab} T^{\alpha\beta\gamma}_{abc}\left[f_{a}\left(R^{\mu,\alpha}_{ba} + R^{\mu,\gamma}_{ac} - R^{\mu,\beta}_{cb}\right)\right.
				    \\
				    &\left.+ f_{b} \left(R^{\mu, \beta}_{cb} + R^{\mu, \alpha}_{ba} - R^{\mu, \gamma}_{ac}\right) + f_{c} \left(R^{\mu, \gamma}_{ac} + R^{\mu, \alpha}_{cb} - R^{\mu, \gamma}_{ba}\right)\right]\delta^{\Omega}_{ab} \delta^{\Omega}_{bc}
                \end{aligned}
                $&~& \multirow{4}*{2S} \\
                \cline{1-2}\xrowht[()]{45pt}
				$\Delta \times T $
                &
				$
                \begin{aligned}
				    &\pi C \sum  f_{ab}\varepsilon_{ba}\varepsilon_{cb}\varepsilon_{ac}T^{\alpha\beta\gamma}_{abc}\left[f_a \left( \frac{\Delta^{\mu}_{ba}}{\varepsilon_{ba}}+ \frac{\Delta^{\mu}_{ac}}{\varepsilon_{ac}} - \frac{\Delta^{\mu}_{cb}}{\varepsilon_{cb}}\right)\right.
				    \\
				    &\left.+ f_{b} \left(\frac{\Delta^{\mu}_{cb}}{\varepsilon_{cb}} + \frac{\Delta^{\mu}_{ba}}{\varepsilon_{ba}}- \frac{\Delta^{\mu}_{ac}}{\varepsilon_{ac}}\right) + f_{c} \left(\frac{\Delta^{\mu}_{ac}}{\varepsilon_{ac}} + \frac{\Delta^{\mu}_{cb}}{\varepsilon_{cb}} - \frac{\Delta^{\mu}_{ba}}{\varepsilon_{ba}}\right)\right]\delta^{\Omega}_{ab} \delta^{\Omega}_{bc}
                \end{aligned}
                $ &~&~\\
				\hline
			\end{tabular}
			\caption{Classification of BCL-induced shift current conductivities based on composite geometry.  Key geometric and physical quantities are defined as follows: $\varepsilon_{ab} = \varepsilon_a - \varepsilon_b$ represents energy differences; $\Delta^{\mu}_{ab} = \partial_{\mu}\varepsilon_{ab}$ denotes group velocity differences; $w^{\mu\alpha}_{ab} = \partial_{\mu}\partial_{\alpha}\varepsilon_{ab}$ is the inverse mass difference. $R^{\mu, \alpha}_{ab} = i\partial_{\mu}\ln{r^{\alpha}_{ab}} + r^{\mu}_{aa} - r^{\mu}_{bb}$ denotes the shift vector $\partial_{\mu}R^{\beta,\gamma}_{ab}$ is the shift vector dipole; $Q^{\alpha\beta}_{ab} = r^{\alpha}_{ba}r^{\beta}_{ab}$ is the quantum geometric tensor; $T^{\alpha\beta\gamma}_{abc} = r^{\alpha}_{ba}r^{\beta}_{cb}r^{\gamma}_{ac}$ represents the triple phase product. Composite geometries are identified by products of these quantities, e.g., $\Delta \times R \times Q$ denotes a conductivity term involving group velocity differences and the shift vector. For brevity, we use $C = \frac{\pi e^4}{3! \hbar^4 \Omega^3}$ and $\sum = \sum\limits_{\{\alpha\beta\gamma\}} \sum\limits_{a \neq b (\neq c)} \int [d\boldsymbol{k}]$. Categories are further classified by the number of bands involved (`Pairwise' or `Multi') and the number of singularities (`$1\text{S}$' or `$2\text{S}$').}
			\label{shift currenty contribution}
		\end{table*}
    \subsection{Resonant BCL-Induced  Shift Current}  \label{Resonant Effect II:  Shift Current}
        In this subsection, we discuss the BCL-induced shift current, which corresponds to resonant contributions that are independent of the relaxation time. Following the analysis in Appendix \ref{appendix: Frequency Integral}, the Feynman diagrams contributing to the shift current are listed in Fig. \ref{feyn diagram for BCL current}.
        Similar to the injection current, the shift current exhibits both "one singularity" ($\sigma^{\mu\alpha\beta\gamma}_{\text{BCL, sh, 1S}}$) and "double singularity" ($\sigma^{\mu\alpha\beta\gamma}_{\text{BCL, sh, 2S}}$) contributions, which refer to terms containing single  Dirac delta function and a product of two Dirac delta functions, respectively. The detailed derivation is presented in Appendix \ref{appendix: Derivation of Shift Current}, and the resulting expressions are given below:
        \begin{equation}
            \label{Shift current 1S}
			\begin{aligned}
				&\sigma^{\mu\alpha\beta\gamma}_{\text{BCL, sh, 1S}} \\
				=& -\frac{1}{3!}\frac{\pi e^4}{\hbar^4\Omega^3}\sum\limits_{\{\alpha\beta\gamma\}}\int [d\boldsymbol{k}]  \sum\limits_{ab} f_{ab} \\
                &\left\{\frac{1}{2}\left[  2\left(h^{\alpha}_{ba}h^{\beta\gamma}_{ab; \mu} - h^{\alpha}_{ba; \mu}h^{\beta\gamma}_{ab}\right)\delta^{\Omega}_{ab}  \right.\right.\\-
				&\left.\left.\left(h^{\alpha}_{ab}h^{\beta\gamma}_{ba; \mu}
                - h^{\alpha}_{ab; \mu}h^{\beta\gamma}_{ba}\right)\delta^{2\Omega}_{ab}\right]\right.\\
                &+  \left.\sum\limits_{c}\left[-h^{\alpha}_{ab}h^{\beta}_{bc}h^{\gamma}_{ca}\left(\frac{h^{\gamma}_{ca; \mu}}{h^{\gamma}_{ca}} + \frac{h^{\beta}_{bc; \mu}}{h^{\beta}_{bc}} - \frac{h^{\alpha}_{ab; \mu}}{h^{\alpha}_{ab}}\right)\tilde{d}^{\Omega}_{ac}\delta^{2\Omega}_{ab}\right.\right.\\
                &+\left.\left.h^{\alpha}_{ba}h^{\beta}_{cb}h^{\gamma}_{ac}\left(\frac{h^{\gamma}_{ac; \mu}}{h^{\gamma}_{ac}} + \frac{h^{\beta}_{cb; \mu}}{h^{\beta}_{cb}} - \frac{h^{\alpha}_{ba; \mu}}{h^{\alpha}_{ba}}\right)
                \right. \right.\\
                &\left.\left.\left(\tilde{d}^{2\Omega}_{ac} +  \tilde{d}^{-\Omega}_{ac}\right)\delta^{\Omega}_{ab}\right]\right\},
			\end{aligned}
		\end{equation}
        \begin{equation}
            \label{Shift current 2S}
			\begin{aligned}
				&\sigma^{\mu\alpha\beta\gamma}_{\text{BCL, sh, 2S}} \\
				=& \frac{1}{3!}\frac{i \pi^2 e^4}{\hbar^4\Omega^3}\sum\limits_{\{\alpha\beta\gamma\}}\int [d\boldsymbol{k}] \sum\limits_{a \neq b \neq c} h^{\alpha}_{ba}h^{\beta}_{cb}h^{\gamma}_{ac} \\
                &\left[f_a\left( \frac{h^{\alpha}_{ba; \mu}}{h^{\alpha}_{ba}}+ \frac{h^{\gamma}_{ac; \mu}}{h^{\gamma}_{ac}} - \frac{h^{\beta}_{cb; \mu}}{h^{\beta}_{cb}}\right) \right. \\
                &+ f_b \left(\frac{h^{\beta}_{cb; \mu}}{h^{\beta}_{cb}} + \frac{h^{\alpha}_{ba; \mu}}{h^{\alpha}_{ba}} - \frac{h^{\gamma}_{ac; \mu}}{h^{\gamma}_{ac}}\right)\\
                &\left.+ f_c \left(\frac{h^{\gamma}_{ac; \mu}}{h^{\gamma}_{ac}} + \frac{h^{\beta}_{cb; \mu}}{h^{\beta}_{cb}} - \frac{h^{\alpha}_{ba; \mu}}{h^{\alpha}_{ba}}\right)\right] \delta^{\Omega}_{ab}\delta^{\Omega}_{bc}.
			\end{aligned}
		\end{equation}
        Here, we adopt the shorthand notation for the generalized derivative $\mathcal{O}_{ab; \mu} = \partial_{\mu}\mathcal{O}_{ab}-\left(r^{\mu}_{aa} - r^{\mu}_{bb}\right)\mathcal{O}_{ab}$ as introduced by Sipe et. al. \cite{Generalized_derivative_Sipe_1995}. The BCL-induced shift current shares several similarities with the injection current. Notably, the one-singularity conductivity terms include the mass term, which is crucial for accurate calculations of real materials. Furthermore, the resonant behavior mirrors that observed in the injection current.
        \par
        We classify the shift current conductivity according to its composite geometry, with detailed derivation provided in Appendix \ref{Identification of Geometry Quantities for Shift Current}. In addition to the geometric quantities identified for injection current, the shift current involves two additional high-order quantities: the shift vector dipole ($\partial R$) and the inverse mass difference ($w$). The shift vector dipole describes the inhomogeneity of $R$ in $k$-space and, to our knowledge, has not been previously reported. The inverse mass difference, $w$, quantifies the interband changes in the electron's effective mass. As summarized in Table \ref{shift currenty contribution}, nine categories are identified, each labeled by its composite geometry, the number of bands involved, and the number of singularities. Despite the increased complexity due to the additional high-order quantities, the shift current conductivity retains the same fundamental features as the injection current conductivity. The first five categories correspond to pairwise-band contributions, each containing a single singularity and characterized by the quantum geometric tensor $Q$. The last four categories correspond to multi-band contributions, involving both one-singularity and double-singularity terms, and characterized by the the triple phase product $T$.
        
    \subsection{Multi-band Contribution: Geometry and Symmetry} \label{symmetry analysis}

        Having established the presence of multi-band contributions in BCL-induced photocurrent, we now comprehensively examine its characteristics from two perspectives: geometric features and symmetry properties.
        \par
\begin{figure}[ht]
            \centering
            \includegraphics[scale=0.4]{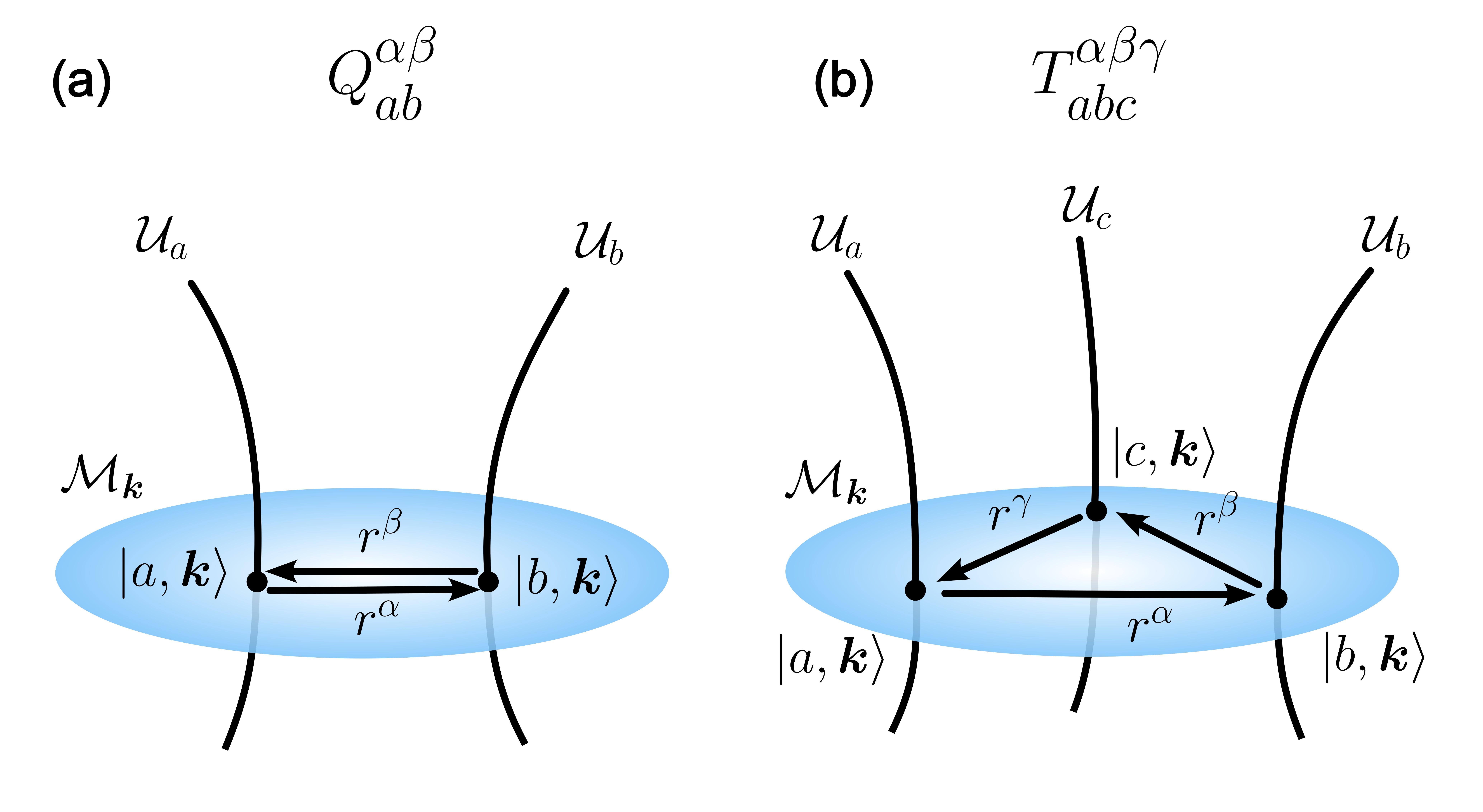}
            \caption{Schematic representation of (a) quantum geometric tensor and (b) triple phase product by generalized Wilson loop. The black rays illustrate the space $\mathcal{U}_a$, representing all physical states characterized by the band index $a$ across all k-points. The blue disk depicts the space $\mathcal{M}_{\boldsymbol{k}}$, representing all physical states for a specific k-point $\boldsymbol{k}$ across all bands. The black dots indicate specific physical states $\ket{a, \boldsymbol{k}}$ at the intersection of $\mathcal{U}_a$ and $\mathcal{M}_{\boldsymbol{k}}$. The black arrows represent the generalized Wilson loop, corresponding to the interband Berry connection $r^{\alpha}_{ba}$.}
            \label{illustration of geometry quantities}
        \end{figure}

        First, let’s consider the geometric in BCL-induced photocurrent. As discussed previously, multi-band contributions are characterized by the triple phase product. This geometric feature was first introduced by Ahn et al. \cite{Geometry_Ahn_PRX_2020} as `virtual transitions', which serve as multi-band terms in the Christoffel symbol and have been recognized as the non-vanishing torsion of a multi-band manifold. We propose that triple phase product can also be considered as a multi-band counterpart of the quantum geometric tensor in a uniform framework using Wilson loops \cite{Wilson_loop_Wang_2022}, as illustrated in Fig. \ref{illustration of geometry quantities}. Quantum geometric tensor can be represented by a two-point Wilson loop. For the geometric triple phase product involving three bands, a three-point Wilson loop can be constructed.
        \par
        Next, we examine the symmetry constraints on the multi-state geometric current in materials possessing both time reversal symmetry ($\mathcal{T}$) and inversion symmetry ($\mathcal{P}$). We begin by analyzing the symmetry properties of the key geometric quantities involved in the BCL-induced photocurrent: the shift vector $R$, the shift vector dipole $\partial R$, the quantum geometric tensor $Q$, and the triple phase product $T$. The real and imaginary parts of these quantities exhibit distinct symmetry properties, determining which terms remain nonvanishing under $\mathcal{T}$ and $\mathcal{P}$ symmetries. The detailed derivation is presented in Appendix \ref{Symmetry Consideration of Geometry Quantities}. Specifically, for $R$ and $\partial R$, only the imaginary part survives under $\mathcal{T}$ and $\mathcal{P}$. For $Q$, the real part, corresponding to the quantum metric, survives, while the imaginary part, associated with the Berry curvature, vanishes. Conversely, for $T$, only the imaginary part remains finite. These results allow us to deduce the symmetry constraints of composite geometric quantities. Finally, we find that all categories listed in Table \ref{injection currenty contribution} and \ref{shift currenty contribution} retain non-zero contributions even in the presence of both $\mathcal{T}$ and $\mathcal{P}$ symmetries.
        The conductivity within each category is restricted to being either purely real or purely imaginary under $\mathcal{T}$ and $\mathcal{P}$. Consequently, the photocurrent can be expressed in the following unified form:  
        \begin{equation}
            \begin{aligned}
                J^{\mu}_{\text{BCL, type}} =& A^3_0 C_1 2\Omega^3\left[\sigma^{\mu xxx}_{\text{BCL, type}} - \sigma^{\mu    xyy}_{\text{BCL, type}}\right]  \\
                &- A^3_0 C_2 2\Omega^3 \left[\sigma^{\mu yyy}_{\text{BCL, type}} - \sigma^{\mu yxx}_{\text{BCL, type}}\right].
	   	    \end{aligned}
        \end{equation}
        Here, $(C_1, C_2) = (\cos\theta, \sin\theta)$ for `type' $\in$ \{`inj, 1S', `sh, 2S'\}, and $(C_1, C_2) = (\sin\theta, \cos\theta)$ for `type'$\in$ \{`inj, 2S', `sh, 1S'\}. This indicates that $J^{\mu}_{\text{BCL, inj, 1S}}$ and $J^{\mu}_{\text{BCL, sh, 2S}}$ exhibit a common $\theta$-dependence, while $J^{\mu}_{\text{BCL, inj, 2S}}$ and $J^{\mu}_{\text{BCL, sh, 1S}}$, shares a distinct, complementary $\theta$-dependence. The $\theta$-dependence of the pairwise-band current is straightforward and depends only on the relaxation time, as it involves a single singularity. In contrast, the $\theta$-dependence of the multi-band contributions is influenced by both the relaxation time and the number of singularities, due to the coexistence of both single and double singularities.

        

\section{Model Example For Multi-band Contribution} \label{sec:Model Example}
\begin{figure}[ht]
		\centering
		\includegraphics[scale=0.45]{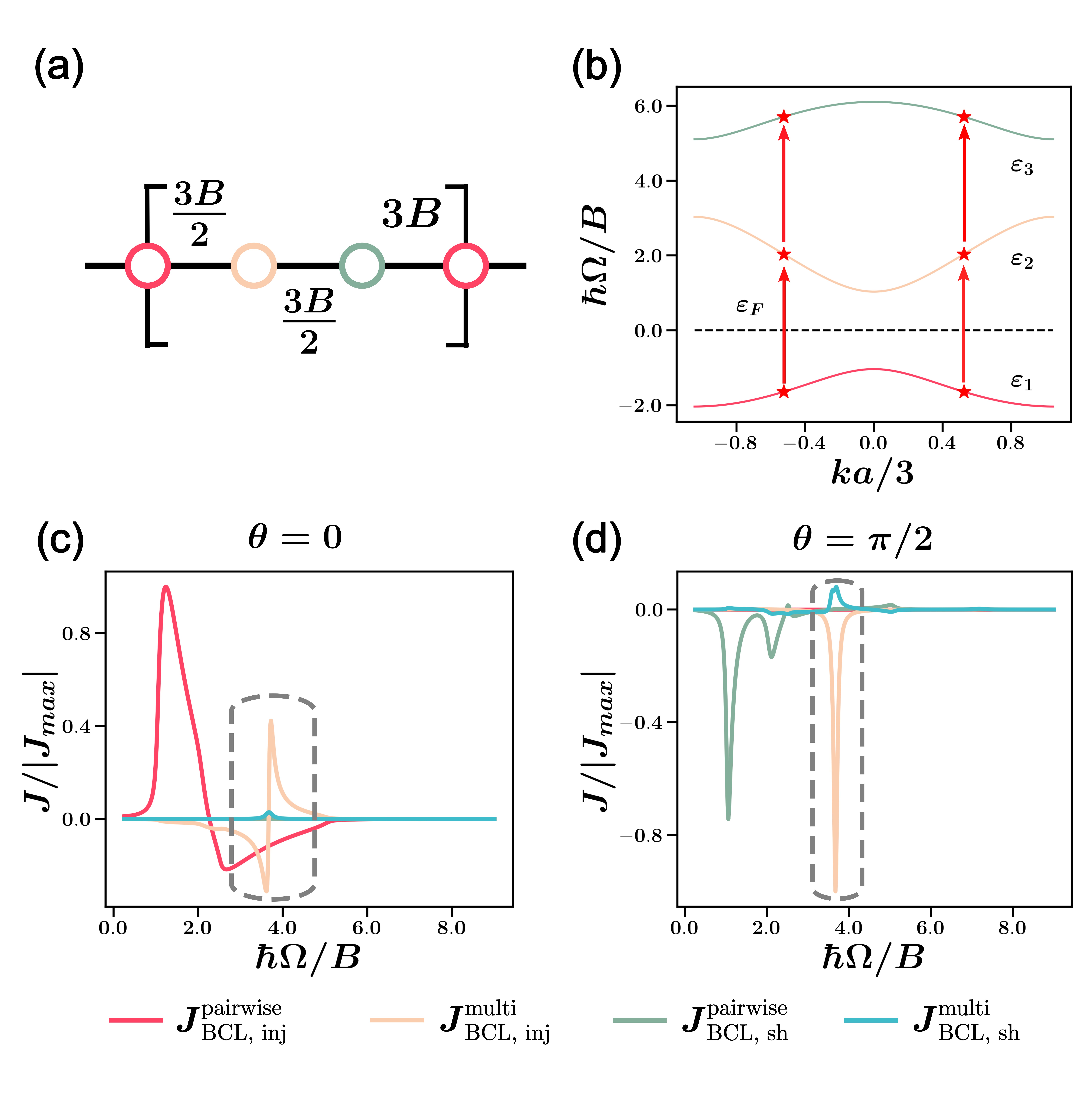}
		\caption{BCL induced photocurrent in the 1D three-site Rice-Mele Model. (a) Schematic representation of the Rice-Mele model. (b) Band dispersion, with the Fermi level positioned between the first band and the second bands. Red stars indicate the k-points where energy levels are equidistant. (c, d) Dependence of induced photocurrent intensity on BCL frequency for phase differences $\theta$ is $0$ and $\pi/2$, respectively. Injection and shift currents are further separated into pairwise-band and multi-band contribution, i.e. $J^{\text{pairwise}}_{\text{BCL, inj}}$, $J^{\text{multi}}_{\text{BCL, inj}}$, $J^{\text{pairwise}}_{\text{BCL, sh}}$ and $J^{\text{multi}}_{\text{BCL, sh}}$. Frequency regions dominated by multi-band contributions are indicated by gray dashed lines.}
		\label{Three-band model result}
    \end{figure}

    To illustrate the role of multi-band contributions to the BCL-induced photocurrent, we analyze a 1D three-site Rice-Mele model, which involves three bands and serves as an effective framework for exploring multi-band effects. The Hamiltonian of this model can be written as
	\begin{equation}
		\begin{aligned}
			H(k) =
			\begin{pmatrix}
				0   & t_1 e^{ika/3} & t_3 e^{-ika/3}\\
				t_1 e^{-ika/3}   & 0 & t_2 e^{ika/3}\\
				t_3 e^{ika/3}   & t_2 e^{-ika/3} & 0\\
			\end{pmatrix}
			,
		\end{aligned}
	\end{equation}
    where the hopping parameters are defined as $t_j = B\cos(2\pi j / 3 - \alpha) + C$, with $C/B = 2, \alpha = 0$ chosen to preserve the centrosymmetry of the Hamiltonian. Specifically, the inversion center is located at the site marked with a yellow ball in Fig. \ref{Three-band model result}(a), and the inversion operator is defined as $\mathcal{P}_{ij} = \delta_{i, 1}\delta_{j, 3} + \delta_{i, 2}\delta_{j, 2} + \delta_{i, 3}\delta_{j, 1}$. The Hamiltonian satisfies $\mathcal{P}^{\dagger}H(k)\mathcal{P} = H(-k)$, ensuring the centrosymmetry. 
    \par
    A unique feature of this model is that $\partial^2_{k}H(k)$ and $H(k)$ shares identical eigenstates, indicating a vanishing off-diagonal mass term is zero, and thus no mass-term correction is required.
    The band dispersion for this model is shown in Fig. \ref{Three-band model result}(b), and the corresponding analytic form can be found in Ref. \cite{Multibandmodel_Moore_2017}. The Fermi level is set such that only the lowest band ($\varepsilon_1$) is occupied. The energy gap between $\varepsilon_1$ and $\varepsilon_2$ ranges within $[2.07B, 5.07B]$ and that from $\varepsilon_1$ to $\varepsilon_3$ is within $[7.14B, 7.34B]$. Based on the identified energy gaps and the Dirac delta functions in the conductivity, the BCL-induced photocurrent is expected to concentrate in the $1\Omega$-resonant ($[2.07B, 5.07B]\cup [7.14B, 7.34B]$) and $2\Omega$-resonant ($[1.04B, 2.54B]\cup [3.57B, 3.67B]$) frequency regions. We then numerically calculate the complete resonant photocurrent using Equation \ref{photocurrent}, with a smearing factor of $\gamma = 0.1B$. The resulting currentexhibits a significant dependence on the BCL phase difference $\theta$, with distinct behaviors observed for different photocurrent types as $\theta$ varies from 0 to $\pi/2$.
    
    \par
    As discussed in Section \ref{symmetry analysis}, we categorize the photocurrent into four components according to the dependence on relaxation time and the number of singularities. For this one-dimensional model preserving both $\mathcal{P}$ and $\mathcal{T}$, the $\theta$-dependence of the photocurrent is
    \begin{equation}
        \begin{aligned}
		J^{x}_{\text{BCL, inj, 1S}} \sim & A^3_0 \cos(\theta) 2\Omega^3 \sigma^{xxxx} \sim J^{x}_{\text{BCL, sh, 2S}} \\
            J^{x}_{\text{BCL, inj, 2S}} \sim & A^3_0 \sin(\theta) 2\Omega^3 \sigma^{xxxx} \sim J^{x}_{\text{BCL, sh, 1S}}.
	\end{aligned}
    \end{equation}
    Both $J^{x}_{\text{BCL, inj, 1S}}$ and $J^{x}_{\text{BCL, sh, 2S}}$ peak at $\theta=0$ and decrease to zero as $\theta$ approaches to $\pi/2$, while $J^{x}_{\text{BCL, sh, 1S}}$ and $J^{x}_{\text{BCL, inj, 2S}}$ display the opposite trend, peaking at $\theta=\pi/2$ and vanishing at $\theta=0$. The photocurrent at $\theta=0$ and $\theta=\pi/2$ are shown in Fig. \ref{Three-band model result}(c, d). These results demonstrate that multi-band contributions lead to the coexistence of injection and shift currents in specific frequency domains.
    \par
    Finally, we analyze the influence of multi-band contributions on the total photocurrent. These contributions arising from the products of $\tilde{d}$ terms and $\delta$ terms, as well as products of two $\delta$ terms, are theoretically confined to the $\Omega$-resonant region. They are particularly prominent around $\hbar\Omega = 3.67B$, where they significantly alter the current intensity, generating distinct peaks, as shown in Fig. \ref{Three-band model result}(c, d). The frequency domain associated with these multi-band contributions corresponds to k-points with uniformly spaced energy levels, marked by red stars in Fig. \ref{Three-band model result}(b). Physically, this suggests that the multi-band contributions arise from the sequential absorption of two photons of identical frequency, enabling transitions from band $\varepsilon_1$ to band $\varepsilon_3$ via the intermediate band $\varepsilon_2$, as sketched by red arrows in Fig. \ref{Three-band model result}(b).
\section{Discussion} \label{sec:discussion}
    In this work, we employed a diagrammatic approach to systematically investigate the BCL-induced photocurrent, categorizing the resulting conductivity based on composite geometries and high-order quantum geometric quantities, such as the gauge-invariant shift vector dipole and triple phase product. This classification framework facilitates a deeper physical understanding of the BCL-induced photocurrent. Using a three-site Rice-Mele model, we demonstrated that multi-band conductivities, often neglected in two-band models, play a significant role in modulating the photocurrent intensity at specific k-points where energy levels are uniformly spaced. This finding highlights the importance of incorporating multi-band contributions when calculating photocurrents in real materials. Additionally, we validated the expected $\theta$-dependence of the BCL-induced current, revealing distinct behaviors characterized by single and double singularities. These insights are crucial for designing innovative optoelectronic devices and enabling precise control of photocurrent responses via BCL, particularly in centrosymmetric materials. The ability to generate photocurrents in centrosymmetric materials opens up new possibilities for solar energy harvesting and other photovoltaic applications. Future research directions include exploring the effects of BCL on topological materials, investigating the role of excitonic effects and electron-phonon couplings, and developing new materials with enhanced BCL-induced photocurrent responses.
\section*{Acknowledgements} \label{sec:acknowledgements}
    We thank Jiaming Hu, Junwen Zhao at Westlake University for their helpful discussions. H.W. acknowledges the support from the NSFC under Grants Nos. 12474240 and 12304049. K. C. acknowledges the support from the Strategic Priority Research Program of the Chinese Academy of Sciences (Grants Nos. XDB28000000 and XDB0460000), the NSFC under Grants Nos. 92265203 and 12488101, and the Innovation Program for Quantum Science and Technology under Grant No. 2024ZD0300104.

\bibliography{ref}

\clearpage
\onecolumngrid
\setcounter{section}{0}
\setcounter{figure}{0}
\setcounter{equation}{0}
\setcounter{table}{0}
\section*{Appendix}

\section{Frequency Integral and Resonant Contribution} \label{appendix: Frequency Integral}
    Here, we present a method for extracting (a) the injection current conductivity, which is resonant and scales with the relaxation time; and (b) the shift current conductivity, which is resonant and independent of the relaxation time. According to the diagrammatic method, the expression for conductivity containing \textit{n} vertices is characterized by frequency integral of \textit{n} Green's functions. All information about Dirac delta function and relaxation time $\tau \propto \gamma^{-1}$ is encoded in the frequency integrals. Thus we can analyze frequency integral to determine which diagram satisfies our requirement. 
    The frequency integral can be calculated using contour integral techniques. The BCL-induced photocurrent involves the evaluation of four such frequency integrals,
    \begin{equation}
        \begin{aligned}
            I_1 = \int d\omega G_a(\omega) = f_a
        \end{aligned}
    \end{equation}
    \begin{equation}
        \begin{aligned}
            I_2(\omega_1) = \int d\omega G_a(\omega)G_b(\omega + \omega_1) = f_{ab} d^{\omega_1}_{ab}
        \end{aligned}
    \end{equation}
    \begin{equation}
        \begin{aligned}
            I_3(\omega_1, \omega_2) = \int d\omega G_a(\omega)G_b(\omega + \omega_1)G_c(\omega + \omega_1+\omega_2) = f_a d^{\omega_1}_{ab}d^{\omega_{12}}_{ac} - f_b d^{\omega_1}_{ab}d^{\omega_2}_{bc} + f_c d^{\omega_{12}}_{ac}d^{\omega_2}_{bc} 
        \end{aligned}
    \end{equation}
    \begin{equation}
        \begin{aligned}
            I_4(\omega_1, \omega_2, \omega_3) =& \int d\omega G_a(\omega)G_b(\omega + \omega_1)G_c(\omega + \omega_1+\omega_2)G_c(\omega + \omega_1+\omega_2 + \omega_3) \\
            =& f_a d^{\omega_{1}}_{ab} d^{\omega_{12}}_{ac} d^{\omega_{123}}_{ad} - f_b d^{\omega_1}_{ab}d^{\omega_{2}}_{bc}d^{\omega_{23}}_{bd} + f_c d^{\omega_{12}}_{ac} d^{\omega_{2}}_{bc} d^{\omega_3}_{cd} - f_d d^{\omega_{123}}_{ad}d^{\omega_{23}}_{bd}d^{\omega_{3}}_{cd},
        \end{aligned}
    \end{equation}
    where $d^{\omega}_{ab} = (\varepsilon_{ab} + \omega + i\gamma)^{-1}$, $f_{a} = (e^{\varepsilon_a /kT} + 1)^{-1}$ is the Fermi-Dirac distribution function. As discussed in Ref. \cite{Diagrammatic_Parker_Moore_2019}, the factor before $\gamma$ in $d^{\omega}_{ab}$ is important for low-frequency and intraband contribution, requiring careful consideration. In this study, we can generally omit this factor except when analyzing injection currents, as detailed in Appendix \ref{appendix: Derivation of Injection Current}. The term $d^{\omega}_{ab}$ can be further decomposed into real and imaginary components, $d^{\omega}_{ab} = \text{Re}d^{\omega}_{ab} + i\text{Im}d^{\omega}_{ab} \equiv \tilde{d}^{\omega}_{ab} - i\pi \delta^{\omega}_{ab}$. In the clean limit, the real part is the principle value of $d^{\omega}_{ab}$, and the imaginary part is proportional to Dirac delta function 
    $\lim_{\gamma \rightarrow 0^{+}}\delta^{\omega}_{ab} = \delta(\hbar\omega + \varepsilon_{ab})$. Here, $\delta^{\omega}_{ab}$ denotes a regularized Dirac delta function with a smearing factor $\gamma$. 
    \par
    When considering bicircular light as the input, the possible frequencies in the integral with multiple Green's functions can be
    \begin{equation}
        \begin{aligned}
            &I_2: \omega_1 \in \{\Omega, -2\Omega, \Omega, 2\Omega, 0\} \\
            &I_3: \{\omega_1, \omega_2\} \in \left\{\{\Omega, -2\Omega\}, \{\Omega, \Omega\}, \{\Omega, -\Omega\},  \{2\Omega, -2\Omega\}\right\} \\
            &I_4: \{\omega_1, \omega_2, \omega_3\} \in \left\{\{\Omega, \Omega, -2\Omega\}\right\}.
        \end{aligned}
    \end{equation}
    \par
    For the injection current conductivity, two criteria must be met: (a) proportional to $\gamma^{-1}$, and (b) the presence of at least one Dirac delta functions. To obtain $\gamma^{-1}$, the only way is to extract terms of the form $d^{0}_{aa}$. Consequently, $I_1$ and $I_2$ can be excluded, as they fail to satisfy both conditions. For $I_3$ and $I_4$, these two conditions are fulfilled only when the band indices connected the output vertices coincide and the sum of the frequency integral's function arguments is zero. Thus, the frequency integrals that contribute to the injection current are:
    \begin{equation}
        \begin{aligned}
            &I_3: a=c \text{  and  } \{\omega_1, \omega_2\} \in \left\{\{\Omega, -\Omega\},  \{2\Omega, -2\Omega\}\right\} \\
            &I_4: a=d \text{  and  } \{\omega_1, \omega_2, \omega_3\} \in \left\{\{\Omega, \Omega, -2\Omega\}\right\}
        \end{aligned}
    \end{equation}
    
    \par
    For the shift current conductivity, the two required conditions are: (a) independent from the relaxation time, and (b) the presence of at least one Dirac delta function.  Initially, $I_1$ can be eliminated, as it cannot produce a Dirac delta function. For $I_2$, it is imperative that $\omega_1$ be non-zero to generate a Dirac delta function. For $I_3$ and $I_4$, it is observed that all terms, excluding those contributing to the injection current, contribute to the shift current. Thus, the frequency integrals of diagrams contributing to the shift current must satisfy:
    \begin{equation}
        \begin{aligned}
            &I_2: \omega_1 \in \left\{\Omega, -2\Omega, \Omega, 2\Omega\right\} \\
            &I_3: a \neq c \text{  and  } \{\omega_1, \omega_2\} \in \left\{\{\Omega, -\Omega\},  \{2\Omega, -2\Omega\}\right\} \text{  ;  }\{\omega_1, \omega_2\} \in \left\{\{\Omega, -2\Omega\}, \{\Omega, \Omega\}\right\}\\
            &I_4: a \neq d \text{  and  } \{\omega_1, \omega_2, \omega_3\} \in \left\{\{\Omega, \Omega, -2\Omega\}\right\}
        \end{aligned}
    \end{equation}

\section{Properties of Regularized Dirac delta function} \label{appendix: Properties of Regularized Dirac delta function}
    In this section, we demonstrate that properties inherent to the Dirac delta function are also valid approximations for the regularized Dirac delta function. Specifically, it will be shown that, for sufficiently small  broadening factor $\gamma$, substituting the Dirac delta function with its regularized form yields equivalent results. Two useful approximation relations, which will be employed in subsequent sections, are presented.
    \par
    The first approximation relation is given by $\tilde{d}^{\omega_2}_{bc}\delta^{\omega_1}_{ab} = \tilde{d}^{\omega_{12}}_{ac}\delta^{\omega_1}_{ab}$. For sufficiently small $\gamma$, this product dominates at resonant point $\hbar\omega_{1} + \varepsilon_{ab} = 0$, rapidly diminishing elsewhere. Therefore, we focus on the domain around the resonant point where this approximation holds with high accuracy,
        \begin{equation}
		  \begin{aligned}
			\lim\limits_{\hbar\omega_{1} + \varepsilon_{ab} \rightarrow 0}[\tilde{d}^{\omega_{2}}_{bc}- \tilde{d}^{\omega_2}_{bc}]\delta^{\omega_1}_{ab} =&\lim\limits_{\hbar\omega_{1} + \varepsilon_{ab} \rightarrow 0} -\left[\frac{\hbar\omega_2 + \varepsilon_{bc}}{(\hbar\omega_2 + \varepsilon_{bc})^2 + \gamma^2} - \frac{\hbar\omega_{12} + \varepsilon_{ac}}{(\hbar\omega_{12} + \varepsilon_{ac})^2 + \gamma^2}\right]\frac{1}{\pi}\frac{\gamma}{(\hbar\omega_{1} + \varepsilon_{ab})^2 + \gamma^2} \\
            =& \lim\limits_{\hbar\omega_{1} + \varepsilon_{ab} \rightarrow 0}-\left[\frac{(\hbar\omega_2 + \varepsilon_{bc})((\hbar\omega_2 + \varepsilon_{bc})^2 + \gamma^2) - (\hbar\omega_2 + \varepsilon_{bc})((\hbar\omega_2 + \varepsilon_{bc})^2 + \gamma^2)}{(\hbar\omega_2 + \varepsilon_{bc})^2 + \gamma^2)^2} \right]\frac{1}{\pi}\frac{1}{\gamma}  \\
		  =& 0
		\end{aligned}
	\end{equation}
    The second approximation relation we prove is $\delta^{\omega_2}_{bc}\delta^{\omega_1}_{ab} = \delta^{\omega_{12}}_{ac}\delta^{\omega_1}_{ab}$. This approximation becomes significant only when the resonant points of both regularized Dirac delta functions coincide, namely when $\hbar\omega_{1} + \varepsilon_{ab} = \hbar\omega_{2} + \varepsilon_{bc} = 0$. In the vicinity of this point, the approximation exhibits high precision for numerical calculations,
        \begin{equation}
		  \begin{aligned}
                \lim\limits_{\hbar\omega_{1} + \varepsilon_{ab} \rightarrow 0}\lim\limits_{\hbar\omega_{2} + \varepsilon_{bc}}[\delta^{\omega_{2}}_{bc}- \delta^{\omega_2}_{bc}]\delta^{\omega_1}_{ab} =&\lim\limits_{\hbar\omega_{1} + \varepsilon_{ab} \rightarrow 0}\lim\limits_{\hbar\omega_{2} + \varepsilon_{bc}} \frac{1}{\pi^2} \left[\frac{\gamma}{(\hbar\omega_2 + \varepsilon_{bc})^2 + \gamma^2} - \frac{\gamma}{(\hbar\omega_{12} + \varepsilon_{ac})^2 + \gamma^2}\right]\frac{\gamma}{(\hbar\omega_{1} + \varepsilon_{ab})^2 + \gamma^2} \\
                =&\frac{1}{\pi^2}\left[\frac{1}{\gamma} - \frac{1}{\gamma}\right]\frac{1}{\gamma}  \\
		  =& 0
		  \end{aligned}
	\end{equation}
\section{Derivation of Injection Current} \label{appendix: Derivation of Injection Current}
	\subsection{$I_3$ Contribution} \label{appendix: Derivation of Injection Current & $I_3$ Contribution}
		We first compute the contribution related to the frequency integral $I_3$. Using the Feynman rules, the expression for this contribution is
		\begin{equation}
			\sigma^{\mu\alpha\beta\gamma}_{I_3} = -\frac{1}{3!}\mathcal{S}_3\frac{e(ie)^3}{2\hbar^4\omega_1\omega_2\omega_3}\sum\limits_{abc, a=c} \int [d\boldsymbol{k}] h^{\mu}_{ac} \left[h^{\alpha\beta}_{ba}h^{\gamma}_{cb}I_3(\omega_{12}, \omega_3) + h^{\alpha}_{ba}h^{\beta\gamma}_{cb}I_3(\omega_{1}, \omega_{23})\right].
		\end{equation}
		To account for all possible combinations of incoming photons, we perform a symmetrization over the six permutations of $(\alpha, \omega_1)$, $(\beta, \omega_2)$, and $(\gamma, \omega_3)$. This symmetrization operation is denoted as $\frac{1}{3!} S_3$.
		When BCL is used as the input, four distinct frequency integrals arise: $I_3(\Omega, -\Omega)$, $I_3(-\Omega, \Omega)$, $I_3(2\Omega, -2\Omega)$, and $I_3(-2\Omega, 2\Omega)$. As detailed in Appendix \ref{appendix: Frequency Integral}, terms proportional to the relaxation time can be extracted. We use $I_3(\Omega, -\Omega)$ as an illustrative example:
		\begin{equation}
			\begin{aligned}
				I_3(\Omega, -\Omega)
                &= f_{a} d^{\Omega}_{ab} d^{0}_{ac} - f_b d^{\Omega}_{ab} d^{-\Omega}_{bc} + f_c d^{0}_{ac} d^{-\Omega}_{bc} \\
				&= f_{ab} d^{\Omega}_{ab} d^{0}_{ac} + f_{cb} d^{0}_{ac} d^{-\Omega}_{bc}.
			\end{aligned}
		\end{equation}
        \par
        By setting the band indices $a = c$, in the clean limit $\gamma \rightarrow 0^+$, $I_3(\Omega, -\Omega)$ transforms into a Dirac delta function peaked at $-\varepsilon_{ab}$. Additionally, in $d^{0}_{aa}$, it is necessary to reinstate the factor preceding $\gamma$  due to the consideration of both intraband contributions (where $a = c$) and the low-frequency condition ($\omega_{123} = 0$). This factor is 3, reflecting the fact that $\omega_{123}$ corresponds to a three-photon pole. The expression $I_3$ in this scenario can be written as
		\begin{equation}
			\begin{aligned}
				\lim\limits_{\gamma \rightarrow 0}I_{3, a=c}(\Omega, -\Omega) = \frac{f_{ab}}{3i\gamma}\left(d^{\Omega}_{ab} - \left(d^{\Omega}_{ab}\right)^*\right) = -\frac{2\pi f_{ab}}{3\gamma}\delta^{\Omega}_{ab}.
			\end{aligned}
		\end{equation}
		The other frequency integrals can be derived analogously:
		\begin{equation}
			\begin{aligned}
				\lim\limits_{\gamma \rightarrow 0}I_{3, a=c}(-\Omega, \Omega)			&= -\frac{2\pi f_{ab}}{3\gamma}\delta^{-\Omega}_{ab} \\
				\lim\limits_{\gamma \rightarrow 0}I_{3, a=c}(2\Omega, -2\Omega)			&= -\frac{2\pi f_{ab}}{3\gamma}\delta^{2\Omega}_{ab} \\
				\lim\limits_{\gamma \rightarrow 0}I_{3, a=c}(-2\Omega, 2\Omega)			&= -\frac{2\pi f_{ab}}{3\gamma}\delta^{-2\Omega}_{ab}.
			\end{aligned}
		\end{equation}
		\par
        For clarity, the symmetrization operation $\mathcal{S}_3$ is substituted with an explicit summation over all permutations of the electric field components $\{(\alpha, \omega_1), (\beta, \omega_2), (\gamma, \omega_3)\}$. The conductivity is expressed by
		\begin{equation}
			\begin{aligned}
				\sigma^{\mu\alpha\beta\gamma}_{I_3}
				=& \frac{1}{3!}\frac{i\pi e^4}{6\hbar^4\Omega^3\gamma}\sum\limits_{ab} \int [d\boldsymbol{k}] f_{ab} \left(h^{\mu}_{aa} - h^{\mu}_{bb}\right) \left[ h^{\alpha\beta}_{ba}h^{\gamma}_{ab} \delta^{2\Omega}_{ab} + h^{\beta\alpha}_{ba}h^{\gamma}_{ab} \delta^{2\Omega}_{ab}\right.\\
				& \left. + h^{\alpha}_{ba}h^{\beta\gamma}_{ab} \delta^{\Omega}_{ab} + h^{\alpha}_{ba}h^{\gamma\beta}_{ab} \delta^{\Omega}_{ab}+ h^{\beta}_{ba}h^{\alpha\gamma}_{ab} \delta^{\Omega}_{ab}
				+ h^{\beta}_{ba}h^{\gamma\alpha}_{ab} \delta^{\Omega}_{ab}\right].\\
			\end{aligned}
		\end{equation}
        As previously stated in Section \ref{BCL Induced Current}, we adopt a symmetric conductivity and sum over all permutations of the input Cartesian indices. Following this summation, the resulting conductivity is
		\begin{equation}
			\begin{aligned}
				\sigma^{\mu\alpha\beta\gamma}_{I_3,\text{ BCL, 1S}} =& \frac{1}{3!}\frac{i\pi e^4}{3\hbar^4\Omega^3\gamma}\sum\limits_{\{\alpha\beta\gamma\}} \sum\limits_{ab} \int [d\boldsymbol{k}] f_{ab} \left(h^{\mu}_{aa} - h^{\mu}_{bb}\right) \left[  2h^{\alpha}_{ba}h^{\beta\gamma}_{ab} \delta(\varepsilon_{ab} + \Omega) + 
				h^{\alpha\beta}_{ba}h^{\gamma}_{ab} \delta(\varepsilon_{ab} + 2\Omega)\right].\\
			\end{aligned}
		\end{equation}
        This contribution is designated with the subscript `1S' due to its inclusion of one singularity.
    
	\subsection{$I_4$ Contribution}
        The contribution associated with the frequency integral $I_4$ can be written as
		\begin{equation}
			\sigma^{\mu\alpha\beta\gamma}_{I_4} = -\frac{1}{3!}\mathcal{S}_3\frac{e(ie)^3}{\hbar^4 \omega_1\omega_2\omega_3} \sum\limits_{abcd, a=d} \int [d\boldsymbol{k}] h^{\alpha}_{ba}h^{\beta}_{cb}h^{\gamma}_{dc}h^{\mu}_{ad}I_4(\omega_1, \omega_2, \omega_3).
		\end{equation}
		When the BCL is utilized as the input, three frequency integrals are encountered: $I_4(\Omega, \Omega, -2\Omega)$, $I_4(\Omega, -2\Omega, \Omega)$, and $I_4(-2\Omega, \Omega, \Omega)$. The relaxation time can be extracted analogously to the procedure detailed in Appendix \ref{appendix: Derivation of Injection Current & $I_3$ Contribution}. Taking $I_4(\Omega, \Omega, -2\Omega)$ as an illustrative example,
		\begin{equation}
			\begin{aligned}
				I_4(\Omega, \Omega, -2\Omega) 
                =& f_a d^{\Omega}_{ab} d^{2\Omega}_{ac} d^{0}_{ad} - f_b d^{\Omega}_{ab}d^{\Omega}_{bc}d^{-\Omega}_{bd} + f_c d^{2\Omega}_{ac} d^{\Omega}_{bc} d^{-2\Omega}_{cd} - f_d d^{0}_{ad}d^{-\Omega}_{bd}d^{-2\Omega}_{cd} \\
                =& d^{0}_{ad}\left[f_a d^{\Omega}_{ab} d^{2\Omega}_{ac} - f_b d^{\Omega}_{ab}d^{\Omega}_{bc} - f_b d^{\Omega}_{bc}d^{-\Omega}_{bd} + f_c d^{2\Omega}_{ac} d^{\Omega}_{bc} + f_c d^{\Omega}_{bc}d^{-2\Omega}_{cd} - f_d d^{-\Omega}_{bd}d^{-2\Omega}_{cd}\right]. \\
			\end{aligned}
		\end{equation}
		By setting the band index $a = d$ and reinstating the factor 3 preceding $\gamma$. Here, we focus on the resonant part of $I_4$ with Dirac delta function(s), and this resonant part is denoted as $I^{R}_{4}$. In the limit $\gamma \rightarrow 0$, $I_4(\Omega, \Omega, -2\Omega)$ can be written as
        \begin{equation}
			\begin{aligned}
				\lim\limits_{\gamma \rightarrow 0^+} I^{R}_{4, a=d}(\Omega, \Omega, -2\Omega)
				=& -\frac{2\pi}{3\gamma}\left(f_{a}\tilde{d}^{\Omega}_{ab}\delta^{2\Omega}_{ac} + f_{a}\tilde{d}^{2\Omega}_{ac}\delta^{\Omega}_{ab}- f_{b}\tilde{d}^{\Omega}_{bc}\delta^{\Omega}_{ab} - f_{c}\tilde{d}^{-\Omega}_{cb}\delta^{\Omega}_{ac}+i \pi f_b \delta^{\Omega}_{ab}\delta^{\Omega}_{bc}- i \pi f_c \delta^{\Omega}_{bc}\delta^{2\Omega}_{ac}\right),
			\end{aligned}
		\end{equation}
		where $\tilde{d}^{\Omega}_{ab} = \text{Re} d^{\Omega}_{ab}$, as defined in Appendix \ref{appendix: Frequency Integral}. The term $I_4$ can be divided into two contributions: one associated with a single singularity and the other with double singularities.
		The other frequency integrals can be computed similarly:
        \begin{equation}
			\begin{aligned}
				\lim\limits_{\gamma \rightarrow 0^+} I^{R}_{4, a=d}(\Omega, -2\Omega, \Omega)
                =& -\frac{2\pi}{3\gamma}\left(f_{a}\tilde{d}^{\Omega}_{ab}\delta^{\Omega}_{ca} + f_{a}\tilde{d}^{-2\Omega}_{ac}\delta^{\Omega}_{ab}- f_{b}\tilde{d}^{-2\Omega}_{bc}\delta^{\Omega}_{ab} - f_{c}\tilde{d}^{2\Omega}_{cb}\delta^{\Omega}_{ca}+i \pi f_b \delta^{\Omega}_{ab}\delta^{2\Omega}_{cb}- i \pi f_c \delta^{2\Omega}_{cb}\delta^{\Omega}_{ca}\right),
			\end{aligned}
		\end{equation}
        \begin{equation}
		\begin{aligned}
				\lim\limits_{\gamma \rightarrow 0^+} I^{R}_{4, a=d}(-2\Omega, \Omega, \Omega)
                =& -\frac{2\pi}{3\gamma}\left(f_{a}\tilde{d}^{-2\Omega}_{ab}\delta^{\Omega}_{ca} + f_{a}\tilde{d}^{-2\Omega}_{ac}\delta^{2\Omega}_{ba}- f_{b}\tilde{d}^{\Omega}_{bc}\delta^{2\Omega}_{ba} - f_{c}\tilde{d}^{-\Omega}_{cb}\delta^{\Omega}_{ca}+i \pi f_b \delta^{2\Omega}_{ba}\delta^{\Omega}_{bc}- i \pi f_c \delta^{\Omega}_{bc}\delta^{\Omega}_{ca}\right).
			\end{aligned}
		\end{equation}
		\par
        The symmetrization $\mathcal{S}_3$ is expressed as a sum over all permutations of the electric field components $\{(\alpha, \omega_1), (\beta, \omega_2), (\gamma, \omega_3)\}$. The symmetrized conductivity is expressed as
		\begin{equation}
			\begin{aligned}
				\sigma^{\mu\alpha\beta\gamma}_{I_4} =& \frac{1}{3!}\frac{e(ie)^3}{2\hbar^4 \Omega^3}  \sum\limits_{abcd} \int [d\boldsymbol{k}] \left[( h^{\alpha}_{ba}h^{\beta}_{cb}h^{\gamma}_{dc}h^{\mu}_{aa} + h^{\beta}_{ba}h^{\alpha}_{cb}h^{\gamma}_{dc}h^{\mu}_{aa})I_4(\Omega, \Omega, -2\Omega)  \right. \\
				& \left. + (h^{\beta}_{ba}h^{\gamma}_{cb}h^{\alpha}_{dc}h^{\mu}_{aa} + h^{\alpha}_{ba}h^{\gamma}_{cb}h^{\beta}_{dc}h^{\mu}_{aa})I_4(\Omega, -2\Omega, \Omega) + \left( h^{\gamma}_{ba}h^{\alpha}_{cb}h^{\beta}_{dc}h^{\mu}_{aa} + h^{\gamma}_{ba}h^{\beta}_{cb}h^{\alpha}_{dc}h^{\mu}_{aa}\right)I_4(-2\Omega, \Omega, \Omega) \right].
			\end{aligned}
		\end{equation}
		\par
		To obtain a symmetric conductivity, we sum over all permutations of the input Cartesian indices. The conductivity $\sigma^{\mu\alpha\beta\gamma}_{I_4}$ is decomposed into two contributions: one corresponding to a single singularity and the other to double singularities. The one singularity contribution is
		\begin{equation}
			\begin{aligned}
				\sigma^{\mu\alpha\beta\gamma}_{I_4, \text{ BCL, 1S}}
    			=&   \frac{1}{3!}\frac{2i\pi e^4}{3\hbar^4 \Omega^3\gamma} \sum\limits_{\{\alpha, \beta, \gamma\}} \sum\limits_{abc} \int [d\boldsymbol{k}]\left(f_ah^{\mu}_{aa}+f_bh^{\mu}_{bb} - f_ah^{\mu}_{bb} - f_bh^{\mu}_{aa}\right) \\
                &\times\left[h^{\alpha}_{ba}h^{\beta}_{cb}h^{\gamma}_{ac} \delta^{\Omega}_{ab}\left(\tilde{d}^{2\Omega}_{ac} + \tilde{d}^{-\Omega}_{ac} \right) + h^{\alpha}_{ab}h^{\beta}_{bc}h^{\gamma}_{ca}\delta^{2\Omega}_{ab}\tilde{d}^{\Omega}_{ac} \right] \\
				=&  \frac{1}{3!}\frac{2i\pi e^4}{3\hbar^4 \Omega^3} \sum\limits_{\{\alpha, \beta, \gamma\}} \sum\limits_{abc} \int [d\boldsymbol{k}] f_{ab}(h^{\mu}_{aa} - h^{\mu}_{bb})\left[h^{\alpha}_{ba}h^{\beta}_{cb}h^{\gamma}_{ac} \delta^{\Omega}_{ab}\left(\tilde{d}^{2\Omega}_{ac} + \tilde{d}^{-\Omega}_{ac} \right) + h^{\alpha}_{ab}h^{\beta}_{bc}h^{\gamma}_{ca}\delta^{2\Omega}_{ab}\tilde{d}^{\Omega}_{ac} \right].
			\end{aligned}
		\end{equation}
		Here, we apply the properties established in Appendix \ref{appendix: Properties of Regularized Dirac delta function} to simplify our result, e.g., $\tilde{d}^{\omega_2}_{bc} \delta^{\omega_1}_{ab} = \tilde{d}^{\omega_{12}}_{ac} \delta^{\omega_1}_{ab}$. The final expression exhibits two resonant peaks at $\varepsilon_{ab} = -\Omega$ and $\varepsilon_{ab} = -2\Omega$.
        The contribution featuring double singularities is expressed as
        \begin{equation}
            \begin{aligned}
                \sigma^{\mu\alpha\beta\gamma}_{I_4, \text{ BCL, 2S}}
                =&  -\frac{1}{3!}\frac{2\pi^2 e^4}{3\hbar^4 \Omega^3\gamma} \sum\limits_{\{\alpha, \beta, \gamma\}} \sum\limits_{a\neq b \neq c} \int [d\boldsymbol{k}] h^{\mu}_{aa}h^{\alpha}_{ba}h^{\beta}_{cb}h^{\gamma}_{ac}\left\{f_b\left[\delta^{\Omega}_{ab}\delta^{2\Omega}_{ac}+ \delta^{\Omega}_{ab}\delta^{-\Omega}_{ac} + \delta^{\Omega}_{bc}\delta^{-2\Omega}_{ab}\right]\right. \\
                &\left.- f_c\left[\delta^{-\Omega}_{cb}\delta^{2\Omega}_{ac}+ \delta^{2\Omega}_{cb}\delta^{-\Omega}_{ac} + \delta^{-\Omega}_{cb}\delta^{2\Omega}_{ca}\right]\right\}\\
                =& -\frac{1}{3!}\frac{2\pi^2 e^4}{3\hbar^4 \Omega^3\gamma} \sum\limits_{\{\alpha, \beta, \gamma\}} \sum\limits_{a\neq b \neq c} \int [d\boldsymbol{k}] h^{\alpha}_{ba}h^{\beta}_{cb}h^{\gamma}_{ac}\left[f_{a}\left(h^{\mu}_{cc} - h^{\mu}_{bb}\right) + f_{b}\left(h^{\mu}_{aa} - h^{\mu}_{cc}\right)  + f_{c}\left(h^{\mu}_{bb} - h^{\mu}_{aa}\right)\right]\delta^{\Omega}_{ab}\delta^{\Omega}_{bc}.
            \end{aligned}
        \end{equation}
        Here, we employ the second property established in Appendix \ref{appendix: Properties of Regularized Dirac delta function}, $\delta^{\omega_1}_{ab} \delta^{\omega_2}_{bc} = \delta^{\omega_1}_{ab} \delta^{\omega_{12}}_{ac}$. When the input light possesses a non-zero frequency, only terms with distinct band indices, $a \neq b \neq c$, yield a nonvanishing contribution.

\section{Derivation of Shift Current} \label{appendix: Derivation of Shift Current}
    As discussed in Section \ref{Resonant Effect II: Shift Current}, the shift current contribution can be divided into two distinct diagram sets, and we demonstrate that cancellation occurs within each set.
	\subsection{Diagram Set I}
       Diagram Set I, depicted in Fig. \ref{feyn diagram for BCL current}(a, b, d, e), is our starting point. We demonstrate that a subset of the two-vertex diagram contributions precisely negates the contributions from the remaining two three-vertex diagrams. The mathematical expression for the two-vertex diagram contributions is:
		\begin{equation}
			\sigma^{\mu\alpha\beta\gamma}_{\text{I, 2v}} = -\frac{1}{3!}\mathcal{S}_3\frac{e(ie)^3}{2\hbar^4\omega_1\omega_2\omega_3} \sum\limits_{ab} \int [d\boldsymbol{k}]\left[h^{\alpha}_{ba}h^{\mu\beta\gamma}_{ab}I_2(\omega_{1}) + h^{\alpha\beta}_{ba}h^{\mu\gamma}_{ab}I_2(\omega_{12})\right].
		\end{equation}
		The contribution of three-vertices is
		\begin{equation}
			\sigma^{\mu\alpha\beta\gamma}_{\text{I, 3v}} = -\frac{1}{3!}\mathcal{S}_3\frac{e(ie)^3}{2\hbar^4\omega_1\omega_2\omega_3}\sum\limits_{abc, a\neq c} \int [d\boldsymbol{k}] h^{\mu}_{ac} \left[h^{\alpha\beta}_{ba}h^{\gamma}_{cb}I_3(\omega_{12}, \omega_2) + h^{\alpha}_{ba}h^{\beta\gamma}_{cb}I_3(\omega_{1}, \omega_{23})\right].
		\end{equation}
        \par
        In the clean limit and under the non-degenerate condition, the term $(\varepsilon_{ac} + i\gamma)^{-1}$ can be approximated by $\varepsilon_{ac}^{-1}$ in $I_3$. We have
		\begin{equation}
			I_{3, a\neq c}(\omega_1, \omega_2) = \frac{1}{\varepsilon_{ac}} \left(f_{ab}d^{\omega_1}_{ab} + f_{cb} d^{\omega_2}_{bc}\right).
		\end{equation}
            \par
     To facilitate subsequent calculations, the second- and third-order covariant derivatives are expanded according to their respective definitions
		\begin{equation}
			\begin{aligned}
				h^{\mu\beta\gamma}_{ab} =& [D^{\mu}, h^{\beta\gamma}]_{ab} = \partial_{\mu}h^{\beta\gamma}_{ab} - i[r^{\mu}, h^{\beta\gamma}]_{ab} \\
				=& \partial_{\mu}h^{\beta\gamma}_{ab} - i \left(r^{\mu}_{aa} - r^{\mu}_{bb}\right)h^{\beta\gamma}_{ab} -i \sum_{c\neq a} r^{\mu}_{ac}h^{\beta\gamma}_{cb} + i\sum_{c\neq b}h^{\beta\gamma}_{ac}r^{\mu}_{cb} \\
				\equiv & h^{\beta\gamma}_{ab; \mu} + \mathcal{M}^{\mu\beta\gamma}_{ab},
			\end{aligned}
		\end{equation}
		\begin{equation}
			\begin{aligned}
                \label{second order derivative expansion}
				h^{\mu\gamma}_{ab} =& [D^{\mu}, h^{\gamma}]_{ab} = \partial_{\mu}h^{\gamma}_{ab} - i[r^{\mu}, h^{\gamma}]_{ab} \\
                =& \partial_{\mu}h^{\gamma}_{ab} - i \left(r^{\mu}_{aa} - r^{\mu}_{bb}\right)h^{\gamma}_{ab} -i \sum_{c\neq a} r^{\mu}_{ac}h^{\gamma}_{cb} + i\sum_{c\neq b}h^{\gamma}_{ac}r^{\mu}_{cb} \\
                \equiv & h^{\gamma}_{ab; \mu} + \mathcal{M}^{\mu\gamma}_{ab},
			\end{aligned}
		\end{equation}
        where $r^{\mu}_{aa} = \braket{a|i\partial_{\mu}|a}$ and $r^{\mu}_{ab} = \braket{a|i\partial_{\mu}|b}$ represent the intraband and interband Berry connections, respectively. The interband Berry connection can further be related to the off-diagonal velocity matrix component as $h^{\mu}_{ab} = i\varepsilon_{ab}r^{\mu}_{ab}$. We adopt the shorthand notation $\mathcal{O}_{ab;\mu} = \partial_{\mu}\mathcal{O}_{ab} - \left(r^{\mu}_{aa} - r^{\mu}_{bb}\right)\mathcal{O}_{ab}$ for the generalized derivative introduced by Sipe et al. \cite{Generalized_derivative_Sipe_1995}. The $\mathcal{M}$-terms denote the remaining contributions involving an additional band summation.
        \par
        For $\sigma^{\mu\alpha\beta\gamma}_{\text{I, 2v}}$, we define the component with $\mathcal{M}$-terms as $\sigma^{\mu\alpha\beta\gamma}_{\text{I, 2v''}}$, while the remaining part is denoted as $\sigma^{\mu\alpha\beta\gamma}_{\text{I, 2v'}}$. The expression for $\sigma^{\mu\alpha\beta\gamma}_{\text{I, 2v''}}$ can be simplified to
		\begin{equation}
			\begin{aligned}
				\sigma^{\mu\alpha\beta\gamma}_{\text{I, 2v''}}
				=& \frac{1}{3!}\mathcal{S}_3\frac{e(ie)^3}{2\hbar^4\omega_1\omega_2\omega_3} \sum\limits_{abc, a\neq c} \int [d\boldsymbol{k}]\left[\left( \frac{h^{\alpha}_{ba}h^{\mu}_{ac}h^{\beta\gamma}_{cb}}{\varepsilon_{ac}}f_{ab}d^{\omega_1}_{ab} + \frac{h^{\alpha}_{cb}h^{\beta\gamma}_{ba}h^{\mu}_{ac}}{\varepsilon_{ac}}f_{cb}d^{\omega_1}_{bc}\right) \right.\\
				&\left. + \left( \frac{h^{\alpha\beta}_{ba}h^{\mu}_{ac}h^{\gamma}_{cb}}{\varepsilon_{ac}}f_{ab}d^{\omega_{12}}_{ab} +\frac{h^{\alpha\beta}_{cb}h^{\gamma}_{ba}h^{\mu}_{cb}}{\varepsilon_{ac}}f_{cb}d^{\omega_{12}}_{bc}\right)\right]. \\
			\end{aligned}
		\end{equation}
	By employing a symmetric conductivity, the indices of the input electric field become interchangeable. Consequently, the preceding expression can be reformulated as:
		\begin{equation}
			\begin{aligned}
				\sigma^{\mu\alpha\beta\gamma}_{\text{I, 2v''}} =& \frac{1}{3!}\mathcal{S}_3\frac{\hbar^4 e(ie)^3}{2\hbar^4\omega_1\omega_2\omega_3}\sum\limits_{\{\alpha, \beta, \gamma\}} \sum\limits_{abc, a\neq c} \int [d\boldsymbol{k}] h^{\mu}_{ac}\left[\frac{h^{\alpha\beta}_{ba}h^{\gamma}_{cb}}{\varepsilon_{ac}}\left(f_{ab}d^{\omega_{12}}_{ab} + f_{cb} d^{\omega_{1}}_{bc}\right)+ \frac{h^{\alpha}_{ba}h^{\beta\gamma}_{cb}}{\varepsilon_{ac}}\left(f_{ab}d^{\omega_1}_{ab} + f_{cb} d^{\omega_{12}}_{bc}\right)\right].
			\end{aligned}
		\end{equation}
		It can be readily verified that this term precisely cancels all contributions from the three-vertex diagrams upon appropriate permutations of the input frequencies. Following this cancellation, the remaining term, designated as $\sigma^{\mu\alpha\beta\gamma}_{\text{BCL, 2v'}}$, is the only contribution from Diagram Set I.
		\begin{equation}
			\begin{aligned}
				\sigma^{\mu\alpha\beta\gamma}_{\text{I}}
                =& -\frac{1}{3!}\mathcal{S}_3\frac{e(ie)^3}{2\hbar^4\omega_1\omega_2\omega_3}\sum\limits_{\{\alpha, \beta, \gamma\}} \sum\limits_{ab} \int [d\boldsymbol{k}]\left[f_{ab}h^{\alpha}_{ba}h^{\beta\gamma}_{ab;\mu}d^{\omega_1}_{ab} + f_{ab}h^{\alpha\beta}_{ba}h^{\gamma}_{ab;\mu}d^{\omega_{12}}_{ab}\right] ,
			\end{aligned}
		\end{equation}
        \par
        For a BCL-induced photocurrent, the input frequencies $\{\omega_1, \omega_2, \omega_3\}$ are set to $\{\Omega, \Omega, -2\Omega\}$, and the symmetrization $\mathcal{S}_3$ is replaced by a summation over all permutations of the input electric field. We further extract the contribution containing Dirac delta functions, as we focus on the resonant term, e.g., $d^{\omega_1}_{ab} \rightarrow -i\pi \delta^{\omega_1}_{ab}$. The final expression is
            \begin{equation}
                \begin{aligned}
                    \sigma^{\mu\alpha\beta\gamma}_{\text{I, BCL, 1S}} &= -\frac{1}{3!}\frac{\pi e^4}{2\hbar^4\Omega^3}\sum\limits_{\{\alpha, \beta, \gamma\}} \sum\limits_{ab} \int [d\boldsymbol{k}] f_{ab}\left[h^{\alpha}_{ba}h^{\beta\gamma}_{ab;\mu}\left(2\delta^{\Omega}_{ab} + \delta^{-2\Omega}_{ab}\right) + h^{\alpha\beta}_{ba}h^{\gamma}_{ab;\mu}\left(2\delta^{-\Omega}_{ab} + \delta^{2\Omega}_{ab}\right)\right] \\
                    &= -\frac{1}{3!}\frac{\pi e^4}{2\hbar^4\Omega^3}\sum\limits_{\{\alpha, \beta, \gamma\}} \sum\limits_{ab} \int [d\boldsymbol{k}] f_{ab}\left[\left(h^{\alpha}_{ba}h^{\beta\gamma}_{ab;\mu} - h^{\alpha}_{ba; \mu}h^{\beta\gamma}_{ab}\right)2\delta^{\Omega}_{ab} -\left(h^{\alpha}_{ab}h^{\beta\gamma}_{ba;\mu} - h^{\alpha}_{ab; \mu}h^{\beta\gamma}_{ba}\right)\delta^{2\Omega}_{ab}\right].
                \end{aligned}
            \end{equation}

	\subsection{Diagram Set II}
	    The Diagram Set II corresponds to Fig. \ref{feyn diagram for BCL current}(c, f). We will show that a portion of the three-vertex diagrams in this set cancels with the remaining four-vertex diagram. The three-vertex diagram can be expressed as
		\begin{equation}
			\sigma^{\mu\alpha\beta\gamma}_{\text{II, 3v}} = -\frac{1}{3!}\mathcal{S}_3\frac{e(ie)^3}{\hbar^4\omega_1\omega_2\omega_3}\sum\limits_{abc} \int [d\boldsymbol{k}] h^{\mu\gamma}_{ac}h^{\alpha}_{ba}h^{\beta}_{cb}I_3(\omega_{1}, \omega_2).
		\end{equation}
		The four-vertex diagram contribution is expressed as
		\begin{equation}
			\sigma^{\mu\alpha\beta\gamma}_{\text{II, 4v}} = -\frac{1}{3!}\mathcal{S}_3\frac{e(ie)^3}{\hbar^4\omega_1\omega_2\omega_3} \sum\limits_{abcd, a\neq d} \int [d\boldsymbol{k}] h^{\alpha}_{ba}h^{\beta}_{cb}h^{\gamma}_{dc}h^{\mu}_{ad}I_4(\omega_1, \omega_2, \omega_3).
		\end{equation}
		In the clean limit and under the non-degenerated condition, we replace $(\varepsilon_{ad} + i\gamma)^{-1} $ with $ \varepsilon_{ad}^{-1}$ and obtain
		\begin{equation}
			I_{4, a\neq d}(\omega_1, \omega_2, \omega_3) = \frac{1}{\varepsilon_{ad}}\left[f_a d^{\omega_1}_{ab}d^{\omega_{12}}_{ac} - f_{b}\left(d^{\omega_1}_{ab}d^{\omega_2}_{bc} + d^{\omega_{2}}_{bc}d^{\omega_{23}}_{bd} \right) + f_{c}\left(d^{\omega_{12}}_{ac}d^{\omega_2}_{bc} + d^{\omega_{2}}_{bc}d^{\omega_{3}}_{cd} \right) -
			f_{d}d^{\omega_{23}}_{bd}d^{\omega_{3}}_{cd}\right].
		\end{equation}
        Then, we expand the second derivative using Equation \ref{second order derivative expansion}. We define the term containing $\mathcal{M}$-terms as $\sigma^{\mu\alpha\beta\gamma}_{\text{II, 3v''}}$, which can be simplified to
		\begin{equation}
			\begin{aligned}
				\sigma^{\mu\alpha\beta\gamma}_{\text{II, 3v''}}
				=&\frac{1}{3!}\mathcal{S}_3\frac{e(ie)^3}{\hbar^4\omega_1\omega_2\omega_3} \sum\limits_{abcd, a\neq d} \int [d\boldsymbol{k}] \frac{h^{\mu}_{ad}h^{\gamma}_{dc}h^{\alpha}_{ba}h^{\beta}_{cb}}{\varepsilon_{ad}}\left(f_a d^{\omega_1}_{ab} d^{\omega_{12}}_{ac} - f_b d^{\omega_1}_{ab}d^{\omega_{2}}_{bc} + f_c d^{\omega_{12}}_{ac}d^{\omega_2}_{bc}\right) \\
				&- \frac{h^{\gamma}_{ba}h^{\mu}_{ad}h^{\alpha}_{cb}h^{\beta}_{dc}}{\varepsilon_{ad}}\left(f_b d^{\omega_1}_{bc} d^{\omega_{12}}_{bd} - f_c d^{\omega_1}_{bc}d^{\omega_{2}}_{cd} + f_d d^{\omega_{12}}_{bd}d^{\omega_2}_{cd}\right) .
			\end{aligned}
		\end{equation}
		Utilizing the permutation symmetry, which allows for arbitrary permutation of the input frequency indices, we demonstrate that $\sigma^{\mu\alpha\beta\gamma}_{\text{II, 3v''}}$ precisely cancels $\sigma^{\mu\alpha\beta\gamma}_{\text{II, 4v}}$:
		\begin{equation}
			\begin{aligned}
				\sigma^{\mu\alpha\beta\gamma}_{\text{II, 3v''}}
				=&\frac{1}{3!}\mathcal{S}_3\frac{e(ie)^3}{\hbar^4\omega_1\omega_2\omega_3}\sum\limits_{\{\alpha, \beta, \gamma\}}\sum\limits_{abcd, a\neq d} \int [d\boldsymbol{k}] \frac{h^{\mu}_{ad}h^{\gamma}_{dc}h^{\alpha}_{ba}h^{\beta}_{cb}}{\varepsilon_{ad}}\\
				& \left[f_a d^{\omega_1}_{ab} d^{\omega_{12}}_{ac} - f_b \left(d^{\omega_1}_{ab}d^{\omega_{2}}_{bc} + d^{\omega_1}_{bc} d^{\omega_{12}}_{bd}\right) + f_c \left(d^{\omega_{12}}_{ac}d^{\omega_2}_{bc}  + d^{\omega_1}_{bc}d^{\omega_{2}}_{cd}\right) - f_d d^{\omega_{12}}_{bd}d^{\omega_2}_{cd}\right] \\
                =&-\sigma^{\mu\alpha\beta\gamma}_{\text{II, 4v}}.
			\end{aligned}
		\end{equation}
		The remaining term, denoted as $\sigma^{\mu\alpha\beta\gamma}_{\text{II, 3v'}}$, represents the exclusive contribution to to Diagram Set II:
		\begin{equation}
			\begin{aligned}
				\sigma^{\mu\alpha\beta\gamma}_{\text{II}} =-&\frac{1}{3!}\mathcal{S}_3\frac{e(ie)^3}{\hbar^4\omega_1\omega_2\omega_3}\sum\limits_{\{\alpha, \beta, \gamma\}}\sum\limits_{abc} \int [d\boldsymbol{k}]h^{\gamma}_{ac; \mu}h^{\alpha}_{ba}h^{\beta}_{cb}I_3(\omega_1, \omega_2).
			\end{aligned}
		\end{equation}
        \par
        For a BCL-induced photocurrent, the input frequencies $\{\omega_1, \omega_2, \omega_3\}$ are set to $\{\Omega, \Omega, -2\Omega\}$. There are three types of frequency integrals for $I_3$: $I_3(\Omega, \Omega)$, $I_3(\Omega, -2\Omega)$, and $I_3(-2\Omega, \Omega)$. We further extract the contribution containing Dirac delta function(s), as we focus on the resonant term. The resonant part of $I_3$, denoted as $I^R_3$, is given by
            \begin{equation}
                \begin{aligned}
                    I^{R}_{3}(\Omega, \Omega) =& -i\pi \left[f_{a}\left(\tilde{d}^{\Omega}_{ab}\delta^{2\Omega}_{ac} + \tilde{d}^{2\Omega}_{ac}\delta^{\Omega}_{ab} - i\pi\delta^{\Omega}_{ab}\delta^{2\Omega}_{ac}\right) - f_{b}\left(\tilde{d}^{\Omega}_{ab}\delta^{\Omega}_{bc} + \tilde{d}^{\Omega}_{bc}\delta^{\Omega}_{ab} - i\pi\delta^{\Omega}_{ab}\delta^{\Omega}_{bc}\right) \right.\\
                    &\left.+ f_{c}\left(\tilde{d}^{2\Omega}_{ac}\delta^{\Omega}_{bc} + \tilde{d}^{\Omega}_{bc}\delta^{2\Omega}_{ac} - i\pi\delta^{2\Omega}_{ac}\delta^{\Omega}_{bc}\right)\right],
                \end{aligned}
            \end{equation}
            \begin{equation}
                \begin{aligned}
                    I^{R}_{3}(\Omega, -2\Omega) =& -i\pi \left[f_{a}\left(\tilde{d}^{\Omega}_{ab}\delta^{-\Omega}_{ac} + \tilde{d}^{-\Omega}_{ac}\delta^{\Omega}_{ab} - i\pi\delta^{\Omega}_{ab}\delta^{-\Omega}_{ac}\right) - f_{b}\left(\tilde{d}^{\Omega}_{ab}\delta^{-2\Omega}_{bc} + \tilde{d}^{-2\Omega}_{bc}\delta^{\Omega}_{ab} - i\pi\delta^{\Omega}_{ab}\delta^{-2\Omega}_{bc}\right) \right.\\
                    &\left.+ f_{c}\left(\tilde{d}^{-\Omega}_{ac}\delta^{-2\Omega}_{bc} + \tilde{d}^{-2\Omega}_{bc}\delta^{-\Omega}_{ac} - i\pi\delta^{-\Omega}_{ac}\delta^{-2\Omega}_{bc}\right)\right],
                \end{aligned}
            \end{equation}
            \begin{equation}
                \begin{aligned}
                    I^{R}_{3}(\Omega, -2\Omega) =& -i\pi \left[f_{a}\left(\tilde{d}^{-2\Omega}_{ab}\delta^{-\Omega}_{ac} + \tilde{d}^{-\Omega}_{ac}\delta^{-2\Omega}_{ab} - i\pi\delta^{-2\Omega}_{ab}\delta^{-\Omega}_{ac}\right) - f_{b}\left(\tilde{d}^{-2\Omega}_{ab}\delta^{\Omega}_{bc} + \tilde{d}^{\Omega}_{bc}\delta^{-2\Omega}_{ab} - i\pi\delta^{-2\Omega}_{ab}\delta^{\Omega}_{bc}\right)  \right.\\
                    &\left.+ f_{c}\left(\tilde{d}^{-\Omega}_{ac}\delta^{\Omega}_{bc} + \tilde{d}^{\Omega}_{bc}\delta^{-\Omega}_{ac} - i\pi\delta^{-\Omega}_{ac}\delta^{\Omega}_{bc}\right)\right].
                \end{aligned}
            \end{equation}
        By replacing the symmetrization operation $\mathcal{S}_3$, with an explicit summation over all permutations of the input electric field components, we obtain the resonant part of Diagram Set II
        \begin{equation}
			\begin{aligned}
				\sigma^{\mu\alpha\beta\gamma}_{\text{II, BCL}}
                =-&\frac{1}{3!}\frac{ie^4}{\hbar^4\Omega^3}\sum\limits_{\{\alpha, \beta, \gamma\}}\sum\limits_{abc} \int [d\boldsymbol{k}]h^{\gamma}_{ac; \mu}h^{\alpha}_{ba}h^{\beta}_{cb}\left[I^R_3(\Omega, \Omega)+I^R_3(\Omega, -2\Omega)+I^R_3(-2\Omega, \Omega)\right].
			\end{aligned}
		\end{equation}
        \par
        Similar to the injection current, we observe the presence of both single-singularity and double-singularity terms. The term containing one singularity is
            \begin{equation}
                \begin{aligned}
                    \sigma^{\mu\alpha\beta\gamma}_{\text{II, BCL, 1S}} &= -\frac{1}{3!}\frac{\pi e^4}{\hbar^4\Omega^3}\sum\limits_{\{\alpha, \beta, \gamma\}} \sum\limits_{abc} \int [d\boldsymbol{k}] f_{ab}\left[ h^{\alpha}_{ba}h^{\beta}_{cb}h^{\gamma}_{ac}\left(\frac{h^{\gamma}_{ac; \mu}}{h^{\gamma}_{ac}} + \frac{h^{\beta}_{cb; \mu}}{h^{\beta}_{cb}} - \frac{h^{\alpha}_{ba; \mu}}{h^{\alpha}_{ba}}\right)\left(\tilde{d}^{2\Omega}_{ac} +  \tilde{d}^{-\Omega}_{ac}\right)\delta^{\Omega}_{ab}\right.\\
                    &\left. - h^{\alpha}_{ab}h^{\beta}_{bc}h^{\gamma}_{ca}\left(\frac{h^{\gamma}_{ca; \mu}}{h^{\gamma}_{ca}} + \frac{h^{\beta}_{bc; \mu}}{h^{\beta}_{bc}} - \frac{h^{\alpha}_{ab; \mu}}{h^{\alpha}_{ab}}\right)\tilde{d}^{\Omega}_{ac}\delta^{2\Omega}_{ab}\right].
                \end{aligned}
            \end{equation}
            The term with double singularities is
            \begin{equation}
                \begin{aligned}
                    \sigma^{\mu\alpha\beta\gamma}_{\text{II, BCL, 2S}} 				=& \frac{1}{3!}\frac{i \pi^2 e^4}{\hbar^4\Omega^3}\sum\limits_{\{\alpha\beta\gamma\}}\sum\limits_{a \neq b \neq c} \int [d\boldsymbol{k}] h^{\alpha}_{ba}h^{\beta}_{cb}h^{\gamma}_{ac} \\
                &\left[f_a\left( \frac{h^{\alpha}_{ba; \mu}}{h^{\alpha}_{ba}}+ \frac{h^{\gamma}_{ac; \mu}}{h^{\gamma}_{ac}} - \frac{h^{\beta}_{cb; \mu}}{h^{\beta}_{cb}}\right) + f_b \left(\frac{h^{\beta}_{cb; \mu}}{h^{\beta}_{cb}} + \frac{h^{\alpha}_{ba; \mu}}{h^{\alpha}_{ba}} - \frac{h^{\gamma}_{ac; \mu}}{h^{\gamma}_{ac}}\right)+ f_c \left(\frac{h^{\gamma}_{ac; \mu}}{h^{\gamma}_{ac}} + \frac{h^{\beta}_{cb; \mu}}{h^{\beta}_{cb}} - \frac{h^{\alpha}_{ba; \mu}}{h^{\alpha}_{ba}}\right)\right] \delta^{\Omega}_{ab}\delta^{\Omega}_{bc}.
                \end{aligned}
            \end{equation}
\section{Identification of Geometric Quantities for Injection Current} \label{Identification of Geometry Quantities for Injection Current}
        The injection current arises from composite geometric and physical quantities, which are initially obscured within the covariant derivative. To elucidate these contributions, we categorize the injection current conductivity into three distinct groups, each directly associated with a specific composite geometric structure: (group velocity) $\times$ (shift vector) $\times$ (quantum geometric tensor), (group velocity) $\times$ (group velocity) $\times$ (quantum geometric tensor), and (group velocity) $\times$ (triple phase product).
        \par
        Prior to detailing the derivation, we outline the methodology for decomposing the covariant derivatives. Equations \ref{Injection current expression 1S} and \ref{Injection current expression 2S} involve both first- and second-order covariant derivatives. The first-order covariant derivative directly corresponds to a component of the velocity matrix. When the band indices satisfy $a \neq b$, the second-order covariant derivative can be decomposed into three parts:
        \begin{equation}
			h^{\alpha\beta}_{ab} = \left\{-iR^{\alpha\beta}_{ab}h^{\beta}_{ab}\right\} + \left\{ \frac{h^{\beta}_{ab}}{\varepsilon_{ab}}\Delta^{\alpha}_{ab}+  \frac{h^{\alpha}_{ab}}{\varepsilon_{ab}}\Delta^{\beta}_{ab}\right\} -  \left\{ \sum\limits_{c, a\neq b\neq c}\left(\frac{h^{\alpha}_{ac}h^{\beta}_{cb}}{\varepsilon_{ac}} - \frac{h^{\beta}_{ac}h^{\alpha}_{cb}}{\varepsilon_{cb}}\right)\right\},
		\end{equation}
        where $R^{\alpha, \beta}_{ab} = i\partial_{\alpha} \ln r^{\beta}_{ab} + \left(r^{\alpha}_{aa} - r^{\alpha}_{bb}\right)$ is the shift vector, and $\Delta^{\alpha}_{ab} = h^{\alpha}_{aa} - h^{\alpha}_{bb}$ represents the group velocity difference between bands $a$ and $b$.
        
		\paragraph{(group velocity)$\times$(shift vector)$\times$(quantum geometric tensor)}
		It is evident that the shift vector arises exclusively from the decomposition of the second-order covariant derivative, specifically within the $I_3$ contribution. The corresponding expression for this category is:
		\begin{equation}
			\begin{aligned}
				\sigma^{\mu\alpha\beta\gamma}_{\text{BCL, } \Delta\times R \times Q}
				=& \frac{1}{3!}\frac{\pi e^4}{3\hbar^4\Omega^3\gamma}\sum\limits_{\{\alpha\beta\gamma\}} \sum\limits_{ab} \int [d\boldsymbol{k}] f_{ab} \Delta^{\mu}_{ab}h^{\alpha}_{ba}h^{\gamma}_{ab} \left[  2R^{\beta\gamma}_{ab}\delta^{\Omega}_{ab} + 
				R^{\beta\alpha}_{ba} \delta^{2\Omega}_{ab}\right]\\
                =& \frac{1}{3!}\frac{\pi e^4}{3\hbar^4\Omega^3\gamma}\sum\limits_{\{\alpha\beta\gamma\}} \sum\limits_{ab} \int [d\boldsymbol{k}] f_{ab}\varepsilon^2_{ab} \Delta^{\mu}_{ab}Q^{\alpha\gamma}_{ab} \left[  2R^{\beta\gamma}_{ab}\delta^{\Omega}_{ab} + 
				R^{\beta\alpha}_{ba} \delta^{2\Omega}_{ab}\right].\\
			\end{aligned}
		\end{equation}
        
		\paragraph{(group velocity)$\times$(group velocity)$\times$(quantum geometric tensor)}
		This category is identifiable within both $I_3$ and $I_4$ contributions. In the $I_3$ contribution, this term originates from the second term in the second-order covariant expansion:
        \begin{equation}
			\begin{aligned}
				\sigma^{\mu\alpha\beta\gamma}_{I_3 , \text{ } \Delta\times\Delta\times Q}
				=& \frac{1}{3!}\frac{i\pi e^4}{3\hbar^4\Omega^3\gamma}\sum\limits_{\{\alpha\beta\gamma\}} \sum\limits_{ab} \int [d\boldsymbol{k}] f_{ab} \Delta^{\mu}_{ab}\Delta^{\gamma}_{ab}\left[4\frac{h^{\alpha}_{ba}h^{\beta}_{ab}}{\varepsilon_{ab}} \delta^{\Omega}_{ab} -
				2\frac{h^{\alpha}_{ba}h^{\beta}_{ab}}{\varepsilon_{ba}} \delta^{2\Omega}_{ab}\right]\\
				=& \frac{1}{3!}\frac{i\pi e^4}{3\hbar^4\Omega^3\gamma}\sum\limits_{\{\alpha\beta\gamma\}} \sum\limits_{ab} \int [d\boldsymbol{k}] f_{ab} \varepsilon_{ab} \Delta^{\mu}_{ab}\Delta^{\gamma}_{ab}Q^{\alpha\beta}_{ab}\left[4\delta^{\Omega}_{ab} + 
				2\delta^{2\Omega}_{ab}\right].\\
			\end{aligned}
		\end{equation}
		\par
		In the $I_4$ contribution, this term can be extracted when the band indices satisfy $b = c$ and $a = c$:
        \begin{equation}
			\begin{aligned}
				\sigma^{\mu\alpha\beta\gamma}_{I_4,  \text{ }\Delta^2\times Q} 
				=&  \frac{1}{3!}\frac{2i\pi e^4}{3\hbar^4\Omega^3\gamma} \sum\limits_{\{\alpha, \beta, \gamma\}} \sum\limits_{ab} \int [d\boldsymbol{k}] f_{ab}\Delta^{\mu}_{ab}\\ &\times\left\{\left[\frac{h^{\alpha}_{ba}h^{\beta}_{ab}h^{\gamma}_{aa}}{2\Omega} + \frac{h^{\alpha}_{ba}h^{\beta}_{ab}h^{\gamma}_{aa}}{ -\Omega} + \frac{h^{\alpha}_{ba}h^{\beta}_{bb}h^{\gamma}_{ab}}{\varepsilon_{ab}+2\Omega} + \frac{h^{\alpha}_{ba}h^{\beta}_{bb}h^{\gamma}_{ab}}{\varepsilon_{ab}- \Omega}\right]\delta^{\Omega}_{ab} + \left[\frac{h^{\alpha}_{aa}h^{\beta}_{ba}h^{\gamma}_{ab}}{\Omega} + \frac{h^{\alpha}_{ba}h^{\beta}_{bb}h^{\gamma}_{ab}}{\varepsilon_{ab} + \Omega}\right]\delta^{2\Omega}_{ab}\right\} \\
				=&  \frac{1}{3!}\frac{2i\pi e^4}{3\hbar^4\Omega^3\gamma} \sum\limits_{\{\alpha, \beta, \gamma\}} \sum\limits_{ab} \int [d\boldsymbol{k}] f_{ab}\Delta^{\mu}_{ab}\Delta^{\gamma}_{ab}h^{\alpha}_{ba}h^{\beta}_{ab}\left[-\frac{1}{2\Omega}\delta^{\Omega}_{ab}+ \frac{1}{\Omega}\delta^{2\Omega}_{ab}\right] \\
				=& \frac{1}{3!}\frac{2i\pi e^4}{3\hbar^4\Omega^3\gamma} \sum\limits_{\{\alpha, \beta, \gamma\}} \sum\limits_{ab} \int [d\boldsymbol{k}] f_{ab}\varepsilon_{ab}\Delta^{\mu}_{ab}\Delta^{\gamma}_{ab}Q^{\alpha\beta}_{ab}\left[\frac{1}{2}\delta^{\Omega}_{ab}-2\delta^{2\Omega}_{ab}\right].
			\end{aligned}
		\end{equation}
		In the first line, we neglect the relaxation time in the denominator of the principal value $\tilde{d}$, as the frequencies consistently exceed $\gamma$ under the constraint imposed by the Dirac delta function in the case of a cold semiconductor.
        Combining these components, the complete expression for this category is:
        \begin{equation}
			\begin{aligned}
				\sigma^{\mu\alpha\beta\gamma}_{\text{BCL, } \Delta^2\times Q }
				=& \frac{1}{3!}\frac{i\pi e^4}{3\hbar^4\Omega^3\gamma}\sum\limits_{\{\alpha\beta\gamma\}} \sum\limits_{ab} \int [d\boldsymbol{k}] f_{ab} \varepsilon_{ab}\Delta^{\mu}_{ab}\Delta^{\gamma}_{ab}Q^{\alpha\beta}_{ab}\left[5\delta^{\Omega}_{ab} - 
				2\delta^{2\Omega}_{ab}\right].\\
			\end{aligned}
		\end{equation}

		\paragraph{(group velocity)$\times$(triple phase product)}
		This term involves contributions with both one singularity and double singularities. The contribution with one singularity originates from  two sources: one from the $I_3$ contribution,
		\begin{equation}
			\begin{aligned}
				\sigma^{\mu\alpha\beta\gamma}_{I_3, \text{ }\Delta\times T, \text{1S}}
				=& -\frac{1}{3!}\frac{\pi e^4}{3\hbar^4\Omega^3\gamma}\sum\limits_{\{\alpha\beta\gamma\}} \sum\limits_{a \neq b \neq c} \int [d\boldsymbol{k}] f_{ab} \Delta^{\mu}_{ab}\varepsilon_{ba}\varepsilon_{cb}\varepsilon_{ac} \\
                &\times\left[2\left(\frac{T^{\alpha\beta\gamma}_{abc}}{\varepsilon_{ac}} - \frac{T^{\alpha\beta\gamma}_{abc}}{\varepsilon_{cb}}\right) \delta^{\Omega}_{ab} +
				\left(\frac{\left(T^{\alpha\beta\gamma}_{abc}\right)^*}{\varepsilon_{bc}} - \frac{\left(T^{\alpha\beta\gamma}_{abc}\right)^*}{\varepsilon_{ca}}\right)\delta^{2\Omega}_{ab}\right]\\
				=& \frac{1}{3!}\frac{\pi e^4}{3\hbar^4\Omega^3\gamma}\sum\limits_{\{\alpha\beta\gamma\}} \sum\limits_{a \neq b \neq c} \int [d\boldsymbol{k}] \varepsilon_{ba}\varepsilon_{cb}\varepsilon_{ac}f_{ab} \Delta^{\mu}_{ab}\left(\frac{1}{\varepsilon_{cb}} - \frac{1}{\varepsilon_{ac}}\right)\left[2T^{\alpha\beta\gamma}_{abc} \delta^{\Omega}_{ab} - 
				\left(T^{\alpha\beta\gamma}_{abc}\right)^*\delta^{2\Omega}_{ab}\right],\\
			\end{aligned}
		\end{equation}
        and the other from the $I_4$ contribution,
		\begin{equation}
			\begin{aligned}
				\sigma^{\mu\alpha\beta\gamma}_{I_4, \text{ } \Delta\times T, \text{1S}}
				=&  \frac{1}{3!}\frac{2\pi e^4}{3\hbar^4\Omega^3\gamma} \sum\limits_{\{\alpha, \beta, \gamma\}} \sum\limits_{a \neq b \neq c} \int [d\boldsymbol{k}] \varepsilon_{ba}\varepsilon_{cb}\varepsilon_{ac}f_{ab}\Delta^{\mu}_{ab} \left[T^{\alpha\beta\gamma}_{abc}\left(\tilde{d}^{2\Omega}_{ac} + \tilde{d}^{-\Omega}_{ac}\right)\delta^{\Omega}_{ab} -\left(T^{\alpha\beta\gamma}_{abc}\right)^*\tilde{d}^{\Omega}_{ac}\delta^{2\Omega}_{ab}\right].
			\end{aligned}
		\end{equation}
        The overall expression for the component with  single singularity is
        \begin{equation}
			\begin{aligned}
				\sigma^{\mu\alpha\beta\gamma}_{\text{BCL, } \Delta\times T, \text{1S}}
				=&\frac{1}{3!}\frac{\pi e^4}{3\hbar^4\Omega^3\gamma}\sum\limits_{\{\alpha\beta\gamma\}} \sum\limits_{a \neq b \neq c} \int [d\boldsymbol{k}] \varepsilon_{ba}\varepsilon_{cb}\varepsilon_{ac}f_{ab} \Delta^{\mu}_{ab}\left\{\left(\frac{1}{\varepsilon_{cb}} - \frac{1}{\varepsilon_{ac}}\right)\left[2T^{\alpha\beta\gamma}_{abc} \delta^{\Omega}_{ab} -
				\left(T^{\alpha\beta\gamma}_{abc}\right)^*\delta^{2\Omega}_{ab}\right]\right.
                \\
                &\left.+ 2 \left[T^{\alpha\beta\gamma}_{abc}\left(\tilde{d}^{2\Omega}_{ac} + \tilde{d}^{-\Omega}_{ac}\right)\delta^{\Omega}_{ab} -\left(T^{\alpha\beta\gamma}_{abc}\right)^*\tilde{d}^{\Omega}_{ac}\delta^{2\Omega}_{ab}\right]\right\}.
			\end{aligned}
		\end{equation}
        The component containing double singularities corresponds exactly to Equation \ref{Injection current expression 2S}:
        \begin{equation}
			\begin{aligned}
				\sigma^{\mu\alpha\beta\gamma}_{\text{BCL, } \Delta\times T \text{, 2S}}
				=& \frac{1}{3!}\frac{2i\pi e^4}{3\hbar^4\Omega^3\gamma}\sum\limits_{\{\alpha\beta\gamma\}}\int [d\boldsymbol{k}] \sum\limits_{a \neq b \neq c} \varepsilon_{ba}\varepsilon_{cb}\varepsilon_{ac}T^{\alpha\beta\gamma}_{abc}\left(f_a \Delta^{\mu}_{cb} + f_b \Delta^{\mu}_{ac} +  f_c \Delta^{\mu}_{ba}\right) \delta^{\Omega}_{ab}\delta^{\Omega}_{bc}.
			\end{aligned}
		\end{equation}
        \par

\section{Identification of Geometric Quantities for Shift Current} \label{Identification of Geometry Quantities for Shift Current}
        Similar to the injection current, the shift current can also be categorized based on its composite geometry. We classify the shift current conductivity into seven categories: (shift vector dipole) $\times$ (quantum geometric tensor), (shift vector) $\times$ (shift vector) $\times$ (quantum geometric tensor), (shift vector) $\times$ (group velocity) $\times$ (quantum geometric tensor), (group velocity) $\times$ (group velocity) $\times$ (quantum geometric tensor), (inverse mass difference) $\times$ (quantum geometric tensor), (group velocity) $\times$ (triple phase product), and (shift vector) $\times$ (triple phase product).
        \par
        The shift current conductivity is characterized by two types of generalized derivatives, $h^{\gamma}_{ab; \mu}$ and $h^{\beta\gamma}_{ab; \mu}$. Diagram Set I contains both types, whereas Diagram Set II includes only the first type. To identify geometric quantities, we introduce a decomposition strategy for generalized derivatives. When a generalized derivative acts on a gauge-covariant quantity $\mathcal{O}_{ab}$, we can always distinguish between gauge-covariant and gauge-invariant terms within $\mathcal{O}$. After differentiation, we combine the derivative of the gauge-covariant term with $(r^{\mu}_{aa} - r^{\mu}_{bb})$ to construct a shift vector.  
        \par
        We first examine the first generalized derivative, $h^{\gamma}_{ab; \mu}$. This derivative contains both diagonal ($a = b$) and off-diagonal ($a \neq b$) components. The diagonal components of the generalized derivative are
        \begin{equation}
            \label{expansion of first order generalized derivative: D}
            h^{\gamma}_{aa; \mu} = \partial_{\mu}\partial_{\gamma}\varepsilon_{a}.
        \end{equation}
        For off-diagonal components, the corresponding gauge-covariant quantity is given by $h^{\gamma}_{ab} = i \varepsilon_{ab} r^{\gamma}_{ab}$. Here, the gauge-invariant term is $\varepsilon_{ab}$, while the gauge-covariant term is $r^{\gamma}_{ab}$. Applying the aforementioned separation strategy, we can decompose this term into  
        \begin{equation}
            \label{expansion of first order generalized derivative}
            h^{\gamma}_{ab; \mu} = \left\{\Delta^{\mu}_{ab}\frac{h^{\gamma}_{ab}}{\varepsilon_{ab}}\right\} + \left\{-iR^{\mu, \gamma}_{ab}h^{\gamma}_{ab}\right\}.
        \end{equation}
		Next, we consider the second generalized derivative, $h^{\beta\gamma}_{ab; \mu}$. Only off-diagonal components exist in this case. The corresponding gauge-covariant quantity is given by the off-diagonal components of the second-order covariant derivative:  
		\begin{equation}
        \label{expansion of second order covariant derivative}
			\begin{aligned}
				h^{\beta\gamma}_{ab} = \left\{\varepsilon_{ab}R^{\beta, \gamma}_{ab} r^{\gamma}_{ab}\right\} + \left\{i\Delta^{\beta}_{ab}r^{\gamma}_{ab} + i\Delta^{\gamma}_{ab}r^{\beta}_{ab} \right\} + \left\{\sum\limits_{c, a\neq c \neq b}\left(\varepsilon_{cb}r^{\beta}_{ac} r^{\gamma}_{cb} - \varepsilon_{ac}r^{\gamma}_{ac}r^{\beta}_{cb}\right)\right\}.
			\end{aligned}
		\end{equation}
		The derivative can act on gauge-invariant terms, such as $\varepsilon$, $R$, and $\Delta$, or on gauge-covariant terms, such as $r$. The second generalized derivative, $h^{\beta\gamma}_{ab; \mu}$, is then decomposed into  
        \begin{equation}
            \label{expansion of second order generalized derivative}
            \begin{aligned}
                h^{\beta\gamma}_{ab; \mu}
                =& \left\{-ih^{\gamma}_{ab}\partial_{\mu}R^{\beta, \gamma}_{ab}\right\}  + \left\{-i\frac{\Delta^{\mu}_{ab}}{\varepsilon_{ab}}R^{\beta, \gamma}_{ab} h^{\gamma}_{ab} - i\Delta^{\beta}_{ab}R^{\mu, \gamma}_{ab}\frac{h^{\gamma}_{ab}}{\varepsilon_{ab}} - i\Delta^{\gamma}_{ab}R^{\mu, \beta}_{ab}\frac{h^{\beta}_{ab}}{\varepsilon_{ab}} \right\} \\
                &+ \left\{-R^{\beta, \gamma}_{ab} R^{\mu,\gamma}_{ab}h^{\gamma}_{ab}\right\} + \left\{ w^{\mu\beta}_{ab}\frac{h^{\gamma}_{ab}}{\varepsilon_{ab}}  +  w^{\mu\gamma}_{ab}\frac{h^{\beta}_{ab}}{\varepsilon_{ab}}\right\} \\
                &+ \sum\limits_{c, a\neq c \neq b} \left[\left\{-\Delta^{\mu}_{cb}\frac{h^{\beta}_{ac} h^{\gamma}_{cb}}{\varepsilon_{ac}\varepsilon_{cb}} + \Delta^{\mu}_{ac}\frac{h^{\gamma}_{ac} h^{\beta}_{cb}}{\varepsilon_{ac}\varepsilon_{cb}}\right\}  + \left\{i\left(R^{\mu, \beta}_{ac}+ R^{\mu, \gamma}_{cb}\right) \frac{h^{\beta}_{ac}h^{\gamma}_{cb}}{\varepsilon_{ac}} - i\left(R^{\mu, \gamma}_{ac}+ R^{\mu, \beta}_{cb}\right)\frac{h^{\gamma}_{ac}h^{\beta}_{cb}}{\varepsilon_{cb}}\right\}\right].
            \end{aligned}
        \end{equation}
        By employing this strategy, the underlying quantum geometric contributions become apparent. The detailed categorization is presented below.
        
		\paragraph{(quantum geometric tensor)$\times$(shift vector dipole)}
		This term is obtained by differentiating the gauge-invariant term $R^{\beta, \gamma}_{ab}$, which corresponds to the first term in Equation \ref{expansion of second order generalized derivative}. 
		\begin{equation}
			\begin{aligned}
				\sigma^{\mu\alpha\beta\gamma}_{\text{BCL, } \partial R \times Q}
				=& \frac{1}{3!}\frac{i\pi e^4}{2\hbar^4\Omega^3}\sum\limits_{\{\alpha, \beta, \gamma\}} \sum\limits_{ab} \int [d\boldsymbol{k}]f_{ab}\varepsilon^2_{ab}Q^{\alpha\gamma}_{ab} \left[2\partial_{\mu}R^{\beta,\gamma}_{ab} \delta^{\Omega}_{ab} -  \partial_{\mu}R^{\beta,\alpha}_{ba} \delta^{2\Omega}_{ab}\right].
			\end{aligned}
		\end{equation}

        \paragraph{(shift vector)$\times$(shift vector)$\times$(quantum geometric tensor)}
        This term originates from two distinct sources within Diagram Set I: (i) the extraction of the third term from Equation \ref{expansion of second order generalized derivative}, (ii) the extraction of the second term from Equation \ref{expansion of first order generalized derivative} combined with the first term from Equation \ref{expansion of second order covariant derivative}.  The overall expression is:
		\begin{equation}
			\begin{aligned}
				\sigma^{\mu\alpha\beta\gamma}_{\text{BCL, } R\times R\times Q }
				=& -\frac{1}{3!}\frac{\pi e^4}{2\hbar^4\Omega^3}\sum\limits_{\{\alpha, \beta, \gamma\}} \sum\limits_{ab} \int [d\boldsymbol{k}]f_{ab}\varepsilon^2_{ab}
                \left[Q^{\alpha\gamma}_{ab}R^{\beta,\gamma}_{ab}\left(R^{\mu,\gamma}_{ab} - R^{\mu,\alpha}_{ba}\right)2\delta^{\Omega}_{ab}  -Q^{\gamma\alpha}_{ab}R^{\beta,\gamma}_{ba}\left(R^{\mu,\gamma}_{ba} - R^{\mu,\alpha}_{ab}\right)\delta^{2\Omega}_{ab} \right] \\
				=& -\frac{1}{3!}\frac{\pi e^4}{2\hbar^4\Omega^3}\sum\limits_{\{\alpha, \beta, \gamma\}} \sum\limits_{ab} \int [d\boldsymbol{k}]f_{ab}\varepsilon^2_{ab}Q^{\alpha\gamma}_{ab}\left(R^{\mu,\gamma}_{ab} - R^{\mu,\alpha}_{ba}\right)
				\left[ 2R^{\beta,\gamma}_{ab}\delta(\varepsilon_{ab} + \Omega) + R^{\beta,\alpha}_{ba}\delta(\varepsilon_{ab} + 2\Omega) \right].			
			\end{aligned}
		\end{equation}

		\paragraph{(group velocity)$\times$(shift vector)$\times$(quantum geometric tensor)} Both Diagram Set contribute to this term. For Diagram Set I, it originates from two sources: (i) extracting the second term in Equation \ref{expansion of second order generalized derivative}, (ii) extracting the first term in Equation \ref{expansion of first order generalized derivative} and combining it with the first term in Equation \ref{expansion of second order covariant derivative}, then extracting the second part of Equation \ref{expansion of first order generalized derivative} and combining it with the second term in Equation \ref{expansion of second order covariant derivative}. Summing up both contributions, the overall result from Diagram Set I is:
		\begin{equation}
			\begin{aligned}
				\sigma^{\mu\alpha\beta\gamma}_{\text{BCL, } \Delta\times R\times Q\text{, I}} =& -\frac{1}{3!}\frac{\pi e^4}{2\hbar^4\Omega^3}\sum\limits_{\{\alpha, \beta, \gamma\}} \sum\limits_{ab} \int [d\boldsymbol{k}]f_{ab} \\
                &\left[\left(i\varepsilon_{ba} Q^{\alpha\gamma}_{ab}\Delta^{\mu}_{ab}R^{\beta,\gamma}_{ab} + 2i\varepsilon_{ba}Q^{\alpha\gamma}_{ab}\Delta^{\beta}_{ab}R^{\mu,\gamma}_{ab} - i\varepsilon_{ab}Q^{\alpha\gamma}_{ab}\Delta^{\mu}_{ba}R^{\beta,\gamma}_{ab} - 2i\varepsilon_{ba}Q^{\alpha\gamma}_{ab}\Delta^{\beta}_{ab}R^{\mu,\alpha}_{ba} \right)2\delta^{\Omega}_{ab} \right.\\
                & \left.- \left(i\varepsilon_{ab} Q^{\alpha\gamma}_{ba}\Delta^{\mu}_{ba}R^{\beta,\gamma}_{ba} + 2i\varepsilon_{ab}Q^{\alpha\gamma}_{ba}\Delta^{\beta}_{ba}R^{\mu,\gamma}_{ba} - i\varepsilon_{ba}Q^{\alpha\gamma}_{ba}\Delta^{\mu}_{ab}R^{\beta,\gamma}_{ba} - 2i\varepsilon_{ab}Q^{\alpha\gamma}_{ba}\Delta^{\beta}_{ba}R^{\mu,\alpha}_{ab} \right)2\delta^{\Omega}_{ab}\right] \\
                =& -\frac{1}{3!}\frac{i\pi e^4}{2\hbar^4\Omega^3}\sum\limits_{\{\alpha, \beta, \gamma\}} \sum\limits_{ab} \int [d\boldsymbol{k}]f_{ab} \varepsilon_{ab}\Delta^{\beta}_{ab}Q^{\alpha\gamma}_{ab}\left(R^{\mu,\alpha}_{ba} - R^{\mu,\gamma}_{ab}\right)\left(2\delta^{\Omega}_{ab} + \delta^{2\Omega}_{ab}\right).
			\end{aligned}
		\end{equation}
        \par
        For Diagram Set II, this term exists when the band indices satisfy $(a = c, a \neq b)$ or $(b \neq c, a = b)$. We then extract the second term in Equation \ref{expansion of first order generalized derivative}. Summing over both cases, the overall expression is:
		\begin{equation}
			\begin{aligned}
				\sigma^{\mu\alpha\beta\gamma}_{\text{BCL, } \Delta\times R\times Q\text{, II}} =&-\frac{1}{3!}\frac{\pi e^4}{\hbar^4\Omega^3}\sum\limits_{\{\alpha, \beta, \gamma\}} \sum\limits_{ab} \int [d\boldsymbol{k}]f_{ab} \\
                &\left[h^{\alpha}_{ba}h^{\beta}_{ab}h^{\gamma}_{aa}\left(-iR^{\mu,\beta}_{ab} + i R^{\mu,\alpha}_{ba}\right)\frac{1}{-2\Omega}\delta^{\Omega}_{ab} -h^{\alpha}_{ab}h^{\beta}_{ba}h^{\gamma}_{aa}\left(-iR^{\mu,\beta}_{ba} + i R^{\mu,\alpha}_{ab}\right)\frac{1}{\Omega}\delta^{2\Omega}_{ab} \right. \\
                &\left. + h^{\alpha}_{ba}h^{\beta}_{bb}h^{\gamma}_{ab}\left(-iR^{\mu,\gamma}_{ab} + i R^{\mu,\alpha}_{ba}\right)\frac{1}{2\Omega}\delta^{\Omega}_{ab} -h^{\alpha}_{ab}h^{\beta}_{bb}h^{\gamma}_{ba}\left(-iR^{\mu,\gamma}_{ba} + i R^{\mu,\alpha}_{ab}\right)\frac{1}{-\Omega}\delta^{2\Omega}_{ab} \right] \\
                =&\frac{1}{3!}\frac{i\pi e^4}{\hbar^4\Omega^3}\sum\limits_{\{\alpha, \beta, \gamma\}} \sum\limits_{ab} \int [d\boldsymbol{k}]f_{ab} \varepsilon_{ab}Q^{\alpha\gamma}_{ab}\Delta^{\beta}_{ab}\left(R^{\mu,\gamma}_{ab} - R^{\mu,\alpha}_{ba}\right)\left(\frac{1}{2}\delta^{\Omega}_{ab} - 2\delta^{2\Omega}_{ab}\right).\\
			\end{aligned}
		\end{equation}
        Here, we neglect the relaxation time term in the denominator of the principal value $\tilde{d}^{\omega}_{aa}$ in the case of a cold semiconductor.
        \par
        Now, we have found all the related terms and the overall expression is:
        \begin{equation}
			\begin{aligned}
				\sigma^{\mu\alpha\beta\gamma}_{\text{BCL, } R\times\Delta\times Q}
				=&\frac{1}{3!}\frac{i\pi e^4}{2\hbar^4\Omega^3}\sum\limits_{\{\alpha, \beta, \gamma\}} \sum\limits_{ab} \int [d\boldsymbol{k}]f_{ab} \varepsilon_{ab}Q^{\alpha\gamma}_{ab}\Delta^{\beta}_{ab}\left(R^{\mu,\gamma}_{ab} - R^{\mu,\alpha}_{ba}\right)\left(5\delta^{\Omega}_{ab} - 2\delta^{2\Omega}_{ab}\right).\\
			\end{aligned}
		\end{equation}
        
		\paragraph{(group velocity)$\times$(group velocity)$\times$(quantum geometric tensor)}Both Diagram Sets contribute to this term. For Diagram Set I, we obtain this term by extracting the first term in Equation \ref{expansion of first order generalized derivative} and combining it with the second term in Equation \ref{expansion of second order covariant derivative}. The expression is:
		\begin{equation}
			\begin{aligned}
				\sigma^{\mu\alpha\beta\gamma}_{\text{BCL, } \Delta\times \Delta\times\text{Q, I}} =& -\frac{1}{3!}\frac{\pi e^4}{2\hbar^4\Omega^3}\sum\limits_{\{\alpha, \beta, \gamma\}} \sum\limits_{ab} \int [d\boldsymbol{k}]f_{ab} \\
                &\left[\left( \Delta^{\mu}_{ba}\Delta^{\beta}_{ab}Q^{\alpha\gamma}_{ab} + \Delta^{\mu}_{ba}\Delta^{\gamma}_{ab}Q^{\alpha\beta}_{ab}\right)2\delta^{\Omega}_{ab}- \left(\Delta^{\mu}_{ab}\Delta^{\beta}_{ba}Q^{\alpha\gamma}_{ba} + \Delta^{\mu}_{ab}\Delta^{\gamma}_{ba}Q^{\alpha\beta}_{ba}\right)\delta^{2\Omega}_{ab}\right] \\
                =& \frac{1}{3!}\frac{\pi e^4}{\hbar^4\Omega^3}\sum\limits_{\{\alpha, \beta, \gamma\}} \sum\limits_{ab} \int [d\boldsymbol{k}]f_{ab}\Delta^{\mu}_{ab}\Delta^{\alpha}_{ab}Q^{\beta\gamma}_{ab}\left(2\delta^{\Omega}_{ab}- \delta^{2\Omega}_{ab}\right). \\
			\end{aligned}
		\end{equation}
		\par
        For Diagram Set II, we obtain this term by requiring the band indices to satisfy either $(a = c, a \neq b)$ or $(b \neq c, a = b)$. We then extract the first term in Equation \ref{expansion of first order generalized derivative}. Summing over both cases, we find that the contribution from Diagram Set II is zero:
        \begin{equation}
			\begin{aligned}
				\sigma^{\mu\alpha\beta\gamma}_{\text{BCL, } \Delta\times \Delta\times\text{Q, II}} =&-\frac{1}{3!}\frac{\pi e^4}{\hbar^4\Omega^3}\sum\limits_{\{\alpha, \beta, \gamma\}} \sum\limits_{ab} \int [d\boldsymbol{k}]f_{ab} \\
                &\left[h^{\alpha}_{ba}h^{\beta}_{ab}h^{\gamma}_{aa}\left( \frac{\Delta^{\mu}_{ab}}{\varepsilon_{ab}} - \frac{\Delta^{\mu}_{ba}}{\varepsilon_{ba}}\right)\frac{1}{-2\Omega}\delta^{\Omega}_{ab} -h^{\alpha}_{ab}h^{\beta}_{ba}h^{\gamma}_{aa}\left(\frac{\Delta^{\mu}_{ba}}{\varepsilon_{ba}} - \frac{\Delta^{\mu}_{ab}}{\varepsilon_{ab}}\right)\frac{1}{\Omega}\delta^{2\Omega}_{ab} \right. \\
                &\left. + h^{\alpha}_{ba}h^{\beta}_{bb}h^{\gamma}_{ab}\left(\frac{\Delta^{\mu}_{ab}}{\varepsilon_{ab}} - \frac{\Delta^{\mu}_{ba}}{\varepsilon_{ba}}\right)\frac{1}{2\Omega}\delta^{\Omega}_{ab} -h^{\alpha}_{ab}h^{\beta}_{bb}h^{\gamma}_{ba}\left(\frac{\Delta^{\mu}_{ba}}{\varepsilon_{ba}} - \frac{\Delta^{\mu}_{ab}}{\varepsilon_{ab}}\right)\frac{1}{-\Omega}\delta^{2\Omega}_{ab} \right]
                =0.\\
			\end{aligned}
		\end{equation}
        Thus the overall contribution is solely from Diagram Set I.
        
		\paragraph{(inverse mass difference)$\times$(quantum geometric tensor)} This contribution appears in both Diagram Sets. In Diagram Set I, we can obtain this term by extracting the fourth term in Equation \ref{expansion of second order generalized derivative}.
		\begin{equation}
			\begin{aligned}
				\sigma^{\mu\alpha\beta\gamma}_{\text{BCL, $\text{m}\times$Q, I}}    =& -\frac{1}{3!}\frac{\pi e^4}{2\hbar^4\Omega^3}\sum\limits_{\{\alpha, \beta, \gamma\}} \sum\limits_{ab} \int [d\boldsymbol{k}]f_{ab} \\
                &\left[\left(\varepsilon_{ab}w^{\mu\beta}_{ab}Q^{\alpha\gamma}_{ab} + \varepsilon_{ab}w^{\mu\gamma}_{ab}Q^{\alpha\beta}_{ab}\right)2\delta^{\Omega}_{ab} - \left(\varepsilon_{ba}w^{\mu\beta}_{ba}Q^{\alpha\gamma}_{ba} + \varepsilon_{ba}w^{\mu\gamma}_{ba}Q^{\alpha\beta}_{ba}\right)\delta^{2\Omega}_{ab}\right] \\
				=& -\frac{1}{3!}\frac{\pi e^4}{\hbar^4\Omega^3}\sum\limits_{\{\alpha, \beta, \gamma\}} \sum\limits_{ab} \int [d\boldsymbol{k}]f_{ab}\varepsilon_{ab}Q^{\alpha\gamma}_{ab}w^{\mu\beta}_{ab}\left[2\delta^{\Omega}_{ab} - \delta^{2\Omega}_{ab}\right].\\
			\end{aligned}
		\end{equation} 
        \par
		For Diagram Set II,  we can obtain such term by requiring band indices $(a = c, a \neq b)$ or $(b \neq c, a=b)$. Then we extract diagonal components of generalized derivative which equals to a second order derivative of band energy, e.g. $h^{\gamma}_{aa; \mu} = \partial_{\mu}\partial_{\gamma}\varepsilon_{a}$. Sum up both case, the overall expression from Diagram Set II is
		\begin{equation}
			\begin{aligned}
				\sigma^{\mu\alpha\beta\gamma}_{\text{BCL, $\text{m}\times$Q, II}} 
				=&-\frac{1}{3!}\frac{\pi e^4}{\hbar^4\Omega^3}\sum\limits_{\{\alpha, \beta, \gamma\}} \sum\limits_{ab} \int [d\boldsymbol{k}]f_{ab} \\
                &\left[h^{\alpha}_{ba}h^{\beta}_{ab}h^{\gamma}_{aa}\frac{\partial_{\mu}\partial_{\gamma}\varepsilon_{a}}{h^{\gamma}_{aa}}\frac{1}{-2\Omega}2\delta^{\Omega}_{ab} - h^{\alpha}_{ab}h^{\beta}_{ba}h^{\gamma}_{aa}\frac{\partial_{\mu}\partial_{\gamma}\varepsilon_{a}}{h^{\gamma}_{aa}}\frac{1}{\Omega}\delta^{2\Omega}_{ab} \right.\\
                &\left.+ h^{\alpha}_{ba}h^{\beta}_{bb}h^{\gamma}_{ab}\frac{\partial_{\mu}\partial_{\beta}\varepsilon_{b}}{h^{\beta}_{bb}}\frac{1}{2\Omega}2\delta^{\Omega}_{ab} - h^{\alpha}_{ab}h^{\beta}_{bb}h^{\gamma}_{ba}\frac{\partial_{\mu}\partial_{\beta}\varepsilon_{b}}{h^{\beta}_{bb}}\frac{1}{-\Omega}\delta^{2\Omega}_{ab}\right] \\
                =&-\frac{1}{3!}\frac{\pi e^4}{\hbar^4\Omega^3}\sum\limits_{\{\alpha, \beta, \gamma\}} \sum\limits_{ab} \int [d\boldsymbol{k}]f_{ab}\varepsilon_{ab}w^{\mu\beta}_{ab}Q^{\alpha\mu}_{ab}\left(\frac{1}{2}\delta^{\Omega}_{ab} + 2\delta^{2\Omega}_{ab} \right).
			\end{aligned}
		\end{equation}
        \par
        The overall expression is
        \begin{equation}
			\begin{aligned}
				\sigma^{\mu\alpha\beta\gamma}_{\text{BCL, $\text{m}\times$Q}}
                =&-\frac{1}{3!}\frac{\pi e^4}{2\hbar^4\Omega^3}\sum\limits_{\{\alpha, \beta, \gamma\}} \sum\limits_{ab} \int [d\boldsymbol{k}]f_{ab}\varepsilon_{ab}w^{\mu\beta}_{ab}Q^{\alpha\mu}_{ab}\left(5\delta^{\Omega}_{ab} + 2\delta^{2\Omega}_{ab} \right).
			\end{aligned}
		\end{equation}

		\paragraph{(shift vector)$\times$(triple phase product)}
        Both Diagram Sets contribute to this term. For Diagram Set I, there are two distinct sources for this term. (i) Extract the last term of Equation \ref{expansion of second order generalized derivative}. (ii) Extract the second term of Equation \ref{expansion of first order generalized derivative} and combine it with the last term in Equation \ref{expansion of second order covariant derivative}. It's obvious that overall contribution from Diagram Set I all has one singularity. The overall expression for Diagram Set I is
        \begin{equation}
		\begin{aligned}
			\sigma^{\mu\alpha\beta\gamma}_{\text{BCL, R}\times\text{T, I, 1S}}
                =&-\frac{1}{3!}\frac{\pi e^4}{2\hbar^4\Omega^3}\sum\limits_{\{\alpha, \beta, \gamma\}} \sum\limits_{a \neq b \neq c} \int [d\boldsymbol{k}]f_{ab} \\
                &\left\{\varepsilon_{ba}\varepsilon_{cb}\varepsilon_{ac}r^{\alpha}_{ba}\left[\left(R^{\mu,\beta}_{ac}+ R^{\mu,\gamma}_{cb}\right)\frac{r^{\beta}_{ac}r^{\gamma}_{cb}}{\varepsilon_{ac}} - \left(R^{\mu,\gamma}_{ac}+ R^{\mu,\beta}_{cb}\right)\frac{r^{\gamma}_{ac}r^{\beta}_{cb}}{\varepsilon_{cb}} \right]2\delta^{\Omega}_{ab}\right. \\
                & -\varepsilon_{ab}\varepsilon_{bc}\varepsilon_{ca}r^{\alpha}_{ab}\left[\left(R^{\mu,\beta}_{bc}+ R^{\mu,\gamma}_{ca}\right)\frac{r^{\beta}_{bc}r^{\gamma}_{ca}}{\varepsilon_{bc}} - \left(R^{\mu,\gamma}_{bc}+ R^{\mu,\beta}_{ca}\right)\frac{r^{\gamma}_{bc}r^{\beta}_{ca}}{\varepsilon_{ca}} \right]\delta^{2\Omega}_{ab} \\
                &\left.-\varepsilon_{ba}\varepsilon_{cb}\varepsilon_{ac}r^{\alpha}_{ba}R^{\mu,\alpha}_{ba}\left[\frac{r^{\beta}_{ac}r^{\gamma}_{cb}}{\varepsilon_{ac}} - \frac{r^{\gamma}_{ac}r^{\beta}_{cb}}{\varepsilon_{cb}} \right]2\delta^{\Omega}_{ab} +\varepsilon_{ab}\varepsilon_{bc}\varepsilon_{ca}r^{\alpha}_{ab}R^{\mu,\alpha}_{ab}\left[\frac{r^{\beta}_{bc}r^{\gamma}_{ca}}{\varepsilon_{bc}} - \frac{r^{\gamma}_{bc}r^{\beta}_{ca}}{\varepsilon_{ca}} \right]\delta^{2\Omega}_{ab}\right\} \\
                =&\frac{1}{3!}\frac{\pi e^4}{2\hbar^4\Omega^3}\sum\limits_{\{\alpha, \beta, \gamma\}}  \sum\limits_{a \neq b \neq c}  \int [d\boldsymbol{k}]f_{ab} \varepsilon_{ba}\varepsilon_{cb}\varepsilon_{ac}\left(\frac{1}{\varepsilon_{cb}} - \frac{1}{\varepsilon_{ac}} \right)\\
                &T^{\alpha\beta\gamma}_{abc}\left(R^{\mu,\beta}_{ac} + R^{\mu,\gamma}_{cb} - R^{\mu,\alpha}_{ba}\right)2\delta^{\Omega}_{ab} + \left(T^{\alpha\beta\gamma}_{abc}\right)^* \left(R^{\mu,\beta}_{ca} + R^{\mu,\gamma}_{bc} - R^{\mu,\alpha}_{ab}\right)\delta^{2\Omega}_{ab}.\\
		\end{aligned}
	\end{equation}

        \par
        For Diagram Set II, there exist both parts with one singularity and part with double singularities. Only $a \neq b \neq c$ contributed to such term. Then extract the second term in Equation \ref{expansion of first order generalized derivative}. The part containing one singularity is
        \begin{equation}
            \label{shift current RT, II, 1S}
		  \begin{aligned}
				\sigma^{\mu\alpha\beta\gamma}_{\text{BCL, R}\times\text{T, II, 1S}}
                =&\frac{1}{3!}\frac{i\pi e^4}{\hbar^4\Omega^3} \sum\limits_{\{\alpha, \beta, \gamma\}} \sum\limits_{a\neq b \neq c} \int [d\boldsymbol{k}] f_{ab} \\             &\left[h^{\alpha}_{ba}h^{\beta}_{cb}h^{\gamma}_{ac}\left(R^{\mu,\gamma}_{ac} + R^{\mu,\beta}_{cb} - R^{\mu,\alpha}_{ba}\right)\left(\tilde{d}^{-\Omega}_{ac} + \tilde{d}^{2\Omega}_{ac}\right)\delta^{\Omega}_{ab} - h^{\alpha}_{ab}h^{\beta}_{bc}h^{\gamma}_{ca}\left(R^{\mu,\gamma}_{ca} + R^{\mu,\beta}_{bc} - R^{\mu,\alpha}_{ab}\right)\tilde{d}^{\Omega}_{ac}\delta^{2\Omega}_{ab}\right] \\
                =&\frac{1}{3!}\frac{\pi e^4}{\hbar^4\Omega^3} \sum\limits_{\{\alpha, \beta, \gamma\}} \sum\limits_{a\neq b \neq c} \int [d\boldsymbol{k}] f_{ab} \varepsilon_{ba}\varepsilon_{cb}\varepsilon_{ac}\\             &\left[T^{\alpha\beta\gamma}_{abc}\left(R^{\mu,\gamma}_{ac} + R^{\mu,\beta}_{cb} - R^{\mu,\alpha}_{ba}\right)\left(\tilde{d}^{-\Omega}_{ac} + \tilde{d}^{2\Omega}_{ac}\right)\delta^{\Omega}_{ab} + \left(T^{\alpha\beta\gamma}_{abc}\right)^*\left(R^{\mu,\gamma}_{ca} + R^{\mu,\beta}_{bc} - R^{\mu,\alpha}_{ab}\right)\tilde{d}^{\Omega}_{ac}\delta^{2\Omega}_{ab}\right].
	   	\end{aligned}
		\end{equation}
        \par
        The overall expression for terms with one singularity is:
        \begin{equation}
		  \begin{aligned}
				\sigma^{\mu\alpha\beta\gamma}_{\text{BCL, R}\times\text{T, 1S}}
                =&\frac{1}{3!}\frac{\pi e^4}{\hbar^4\Omega^3}\sum\limits_{\{\alpha, \beta, \gamma\}}  \sum\limits_{a \neq b \neq c}  \int [d\boldsymbol{k}]f_{ab} \varepsilon_{ba}\varepsilon_{cb}\varepsilon_{ac}\\
                &\left\{\frac{1}{2}\left(\frac{1}{\varepsilon_{cb}} - \frac{1}{\varepsilon_{ac}} \right)\left[T^{\alpha\beta\gamma}_{abc}\left(R^{\mu,\beta}_{ac} + R^{\mu,\gamma}_{cb} - R^{\mu,\alpha}_{ba}\right)2\delta^{\Omega}_{ab} + \left(T^{\alpha\beta\gamma}_{abc}\right)^* \left(R^{\mu,\beta}_{ca} + R^{\mu,\gamma}_{bc} - R^{\mu,\alpha}_{ab}\right)\delta^{2\Omega}_{ab}\right]\right.\\            &\left.+\left[T^{\alpha\beta\gamma}_{abc}\left(R^{\mu,\gamma}_{ac} + R^{\mu,\beta}_{cb} - R^{\mu,\alpha}_{ba}\right)\left(\tilde{d}^{-\Omega}_{ac} + \tilde{d}^{2\Omega}_{ac}\right)\delta^{\Omega}_{ab} + \left(T^{\alpha\beta\gamma}_{abc}\right)^*\left(R^{\mu,\gamma}_{ca} + R^{\mu,\beta}_{bc} - R^{\mu,\alpha}_{ab}\right)\tilde{d}^{\Omega}_{ac}\delta^{2\Omega}_{ab}\right]\right\}.
	   	\end{aligned}
		\end{equation} 
        \par
        And the overall expression for terms with double singularity is solely from Diagram Set II, 
        \begin{equation}
		  \begin{aligned}
				\sigma^{\mu\alpha\beta\gamma}_{\text{BCL, R}\times\text{T, 2S}}
                =&-\frac{1}{3!}\frac{i\pi^2 e^4}{\hbar^4\Omega^3} \sum\limits_{\{\alpha, \beta, \gamma\}} \sum\limits_{a\neq b \neq c} \int [d\boldsymbol{k}] \varepsilon_{ba}\varepsilon_{cb}\varepsilon_{ac}T^{\alpha\beta\gamma}_{abc}\\
                &\left[f_{a}\left(R^{\mu, \alpha}_{ba} + R^{\mu, \gamma}_{ac} - R^{\mu, \beta}_{cb}\right)+f_{b}\left(R^{\mu, \beta}_{cb} + R^{\mu, \alpha}_{ba} - R^{\mu, \gamma}_{ac}\right)+f_{c}\left(R^{\mu,\gamma}_{ac} + R^{\mu, \beta}_{cb} - R^{\mu, \alpha}_{ba}\right)\right]\delta^{\Omega}_{ab}\delta^{\Omega}_{bc}.
	   	\end{aligned}
        \end{equation}

		\paragraph{(group velocity)$\times$(triple phase product)} Both Diagram Set I and Diagram Set II contribute to this term. In Diagram Set I, there are two sources for this contribution: (i) extracting the fifth term of Equation \ref{expansion of second order generalized derivative}, (ii) extracting the first term of Equation \ref{expansion of first order generalized derivative} and combining it with the last term of Equation \ref{expansion of second order covariant derivative}. This contribution only contains one singularity. The expression is:
		\begin{equation}
			\begin{aligned}
				\sigma^{\mu\alpha\beta\gamma}_{\text{BCL, }\Delta\times\text{T, I, 1S}} 
                =&\frac{1}{3!}\frac{\pi e^4}{2\hbar^4\Omega^3}\sum\limits_{\{\alpha, \beta, \gamma\}}\sum\limits_{a\neq b\neq c} \int [d\boldsymbol{k}] f_{ab}\\
                &\left[i\varepsilon_{ba}r^{\alpha}_{ba}\left(\Delta^{\mu}_{cb}r^{\beta}_{ac}r^{\gamma}_{cb} - \Delta^{\mu}_{ac}r^{\gamma}_{ac}r^{\beta}_{cb}\right)2\delta^{\Omega}_{ab} - i\varepsilon_{ab}r^{\alpha}_{ab}\left(\Delta^{\mu}_{ca}r^{\beta}_{bc}r^{\gamma}_{ca} - \Delta^{\mu}_{bc}r^{\gamma}_{bc}r^{\beta}_{ca}\right)\delta^{2\Omega}_{ab}\right.\\
                &\left.-i\Delta^{\mu}_{ba}r^{\alpha}_{ba}\left(\varepsilon_{cb}r^{\beta}_{ac}r^{\gamma}_{cb} - \varepsilon_{ac}r^{\gamma}_{ac}r^{\beta}_{cb}\right)2\delta^{\Omega}_{ab} +i\Delta^{\mu}_{ab}r^{\alpha}_{ab}\left(\varepsilon_{ca}r^{\beta}_{bc}r^{\gamma}_{ca} - \varepsilon_{bc}r^{\gamma}_{bc}r^{\beta}_{ca}\right)\delta^{2\Omega}_{ab}\right]\\
                =&\frac{1}{3!}\frac{i\pi e^4}{2\hbar^4\Omega^3}\sum\limits_{\{\alpha, \beta, \gamma\}}\sum\limits_{a\neq b\neq c} \int [d\boldsymbol{k}] f_{ab}\varepsilon_{ba}\varepsilon_{cb}\varepsilon_{ac}\left(\frac{1}{\varepsilon_{cb}} - \frac{1}{\varepsilon_{ac}}\right)\left(\frac{\Delta^{\mu}_{cb} + \Delta^{\mu}_{ac}}{\varepsilon_{cb} - \varepsilon_{ac}} - \frac{\Delta^{\mu}_{ba}}{\varepsilon_{ba}}\right) \\
                &\left[2T^{\alpha\beta\gamma}_{abc} \delta^{\Omega}_{ab} + \left(T^{\alpha\beta\gamma}_{abc}\right)^* \delta^{\Omega}_{ab}\delta^{2\Omega}_{ab}\right]. \\
			\end{aligned}
		\end{equation}
        \par
        For Diagram Set II, both contributions with one singularity and those with double singularities exist. Only cases where $a \neq b \neq c$ contribute to this term. We extract the first term from Equation \ref{expansion of first order generalized derivative}. The expression for the contribution with one singularity is given by:
        \begin{equation}
            \label{shift current DeltaT, II, 1S}
		  \begin{aligned}
				\sigma^{\mu\alpha\beta\gamma}_{\text{BCL,  $\Delta\times$T, II, 1S}}
                =&-\frac{1}{3!}\frac{\pi e^4}{\hbar^4\Omega^3}  \sum\limits_{\{\alpha, \beta, \gamma\}} \sum\limits_{a\neq b \neq c} \int [d\boldsymbol{k}] f_{ab}\\
                &\left[h^{\alpha}_{ba}h^{\beta}_{cb}h^{\gamma}_{ac}\left(\frac{\Delta^{\mu}_{ac}}{\varepsilon_{ac}} + \frac{\Delta^{\mu}_{cb}}{\varepsilon_{cb}} - \frac{\Delta^{\mu}_{ba}}{\varepsilon_{ba}}\right)\left(\tilde{d}^{-\Omega}_{ac} + \tilde{d}^{2\Omega}_{ac} \right)\delta^{\Omega}_{ab} \right.\\
                &\left.- h^{\alpha}_{ab}h^{\beta}_{bc}h^{\gamma}_{ca}\left(\frac{\Delta^{\mu}_{ca}}{\varepsilon_{ca}} + \frac{\Delta^{\mu}_{bc}}{\varepsilon_{bc}} - \frac{\Delta^{\mu}_{ab}}{\varepsilon_{ab}}\right)\tilde{d}^{\Omega}_{ac}\delta^{2\Omega}_{ab}\right]\\
                =&\frac{1}{3!}\frac{i\pi e^4}{\hbar^4\Omega^3}  \sum\limits_{\{\alpha, \beta, \gamma\}} \sum\limits_{a\neq b \neq c} \int [d\boldsymbol{k}] f_{ab} \varepsilon_{ba}\varepsilon_{cb}\varepsilon_{ac}\left(\frac{\Delta^{\mu}_{ac}}{\varepsilon_{ac}} + \frac{\Delta^{\mu}_{cb}}{\varepsilon_{cb}} - \frac{\Delta^{\mu}_{ba}}{\varepsilon_{ba}}\right)\\
                &\left[ T^{\alpha\beta\gamma}_{abc}\left(\tilde{d}^{-\Omega}_{ac} + \tilde{d}^{2\Omega}_{ac} \right)\delta^{\Omega}_{ab} +\left(T^{\alpha\beta\gamma}_{abc}\right)^* \tilde{d}^{\Omega}_{ac}\delta^{2\Omega}_{ab}\right].
	   	\end{aligned}
		\end{equation}
        The overall expression for contribution with one singularity is
        \begin{equation}
		  \begin{aligned}
				\sigma^{\mu\alpha\beta\gamma}_{\text{BCL,  $\Delta\times$T, 1S}}
                =&\frac{1}{3!}\frac{i\pi e^4}{\hbar^4\Omega^3}\sum\limits_{\{\alpha, \beta, \gamma\}}\sum\limits_{a\neq b\neq c} \int [d\boldsymbol{k}] f_{ab}\varepsilon_{ba}\varepsilon_{cb}\varepsilon_{ac}\\
                &\left\{\frac{1}{2}\left(\frac{1}{\varepsilon_{cb}} - \frac{1}{\varepsilon_{ac}}\right)\left(\frac{\Delta^{\mu}_{cb} + \Delta^{\mu}_{ac}}{\varepsilon_{cb} - \varepsilon_{ac}} - \frac{\Delta^{\mu}_{ba}}{\varepsilon_{ba}}\right)\left[2T^{\alpha\beta\gamma}_{abc} \delta^{\Omega}_{ab} + \left(T^{\alpha\beta\gamma}_{abc}\right)^* \delta^{\Omega}_{ab}\delta^{2\Omega}_{ab}\right]\right.\\
                & \left.+\left(\frac{\Delta^{\mu}_{ac}}{\varepsilon_{ac}} + \frac{\Delta^{\mu}_{cb}}{\varepsilon_{cb}} - \frac{\Delta^{\mu}_{ba}}{\varepsilon_{ba}}\right)\left[ T^{\alpha\beta\gamma}_{abc}\left(\tilde{d}^{-\Omega}_{ac} + \tilde{d}^{2\Omega}_{ac} \right)\delta^{\Omega}_{ab} +\left(T^{\alpha\beta\gamma}_{abc}\right)^* \tilde{d}^{\Omega}_{ac}\delta^{2\Omega}_{ab}\right]\right\}.
	   	\end{aligned}
		\end{equation}                 
        The overall expression for contribution with double singularities is solely from Diagram Set II,
        \begin{equation}
		  \begin{aligned}
				\sigma^{\mu\alpha\beta\gamma}_{\text{BCL,  $\Delta\times$T, II}}                                 =&\frac{1}{3!}\frac{\pi^2 e^4}{\hbar^4\Omega^3} \sum\limits_{\{\alpha, \beta, \gamma\}} \sum\limits_{a\neq b \neq c} \int [d\boldsymbol{k}] \varepsilon_{ba}\varepsilon_{cb}\varepsilon_{ac}T^{\alpha\beta\gamma}_{abc}\\
                &\left[f_{a}\left(\frac{\Delta^{\mu}_{ba}}{\varepsilon_{ba}} + \frac{\Delta^{\mu}_{ac}}{\varepsilon_{ac}} - \frac{\Delta^{\mu}_{cb}}{\varepsilon_{cb}}\right)+f_{b}\left(\frac{\Delta^{\mu}_{cb}}{\varepsilon_{cb}} + \frac{\Delta^{\mu}_{ba}}{\varepsilon_{ba}} - \frac{\Delta^{\mu}_{ac}}{\varepsilon_{ac}}\right)+f_{c}\left(\frac{\Delta^{\mu}_{ac}}{\varepsilon_{ac}} + \frac{\Delta^{\mu}_{cb}}{\varepsilon_{cb}} - \frac{\Delta^{\mu}_{ba}}{\varepsilon_{ba}}\right)\right]\delta^{\Omega}_{ab}\delta^{\Omega}_{bc}.
	   	\end{aligned}
        \end{equation}

\section{Symmetry Consideration of Geometry Quantities}
	\label{Symmetry Consideration of Geometry Quantities}
		Here, we analyze the properties of various geometric quantities in a spinless system with time reversal symmetry ($\mathcal{T}$) or combined time reversal symmetry and centrosymmetry ($\mathcal{T} + \mathcal{P}$).
        We begin by examining the constraints imposed by $\mathcal{T}$ and $\mathcal{P}$. The eigenvectors of the Hamiltonian satisfy
		\begin{equation}
		    \begin{aligned}
		        &\mathcal{T}: && u_a (\boldsymbol{k}) = u^*_a (-\boldsymbol{k}), \\
		        &\mathcal{P}: && u_a (\boldsymbol{k}) = u_a (-\boldsymbol{k}), \\
		    \end{aligned}
		\end{equation}
        The geometric quantities relevant to our derivation are also constrained by $\mathcal{T}$. For the Berry connection, the following property holds:
		\begin{equation}
		    \begin{aligned}
		        &\mathcal{T}: && r^{\alpha}_{ab}(\boldsymbol{k}) = \int[d\boldsymbol{k}]u^*_{a}(\boldsymbol{k}) i\partial_{k_{\alpha}}u_{b}(\boldsymbol{k}) = \int[d\boldsymbol{k}]u_{a}(-\boldsymbol{k}) i\partial_{k_{\alpha}}u^*_{b}(-\boldsymbol{k}) = \int[d\boldsymbol{k}]u^*_{b}(-\boldsymbol{k}) i\partial_{-k_{\alpha}}u_{a}(-\boldsymbol{k}) = r^{\alpha}_{ba}(-\boldsymbol{k}), \\
		        &\mathcal{P}: && r^{\alpha}_{ab}(\boldsymbol{k}) = \int[d\boldsymbol{k}]u^*_{a}(\boldsymbol{k}) i\partial_{k_{\alpha}}u_{b}(\boldsymbol{k}) = \int[d\boldsymbol{k}]u^{*}_{a}(-\boldsymbol{k}) i\partial_{k_{\alpha}}u_{b}(-\boldsymbol{k}) = -\int[d\boldsymbol{k}]u^*_{a}(-\boldsymbol{k}) i\partial_{-k_{\alpha}}u_{a}(-\boldsymbol{k}) = -r^{\alpha}_{ab}(-\boldsymbol{k}).
		    \end{aligned}
		\end{equation}
		Here, we define $\partial_{k_{\alpha}} \equiv \partial_{\alpha} = \partial/\partial_{k_{\alpha}} $, and note that $\partial_{-k_{\alpha}} = -\partial/\partial_{k_{\alpha}} $.
        For the group velocity difference $\Delta^{\mu}_{ab}$ and the inverse mass dipole $w^{\mu\alpha}_{ab}$, both are purely real and obey the same symmetry constraints in materials with either $\mathcal{T}$ or $\mathcal{P}$ symmetry,
        \begin{equation}
            \begin{aligned}
                &\mathcal{T} \text{ or } \mathcal{P}: && \Delta^{\mu}_{ab}(\boldsymbol{k}) = \partial_{k_{\mu}}\varepsilon_{ab}(\boldsymbol{k}) = -\partial_{-k_{\mu}}\varepsilon_{ab}(-\boldsymbol{k}) = -\Delta^{\mu}_{ab}(-\boldsymbol{k}), \\
            \end{aligned}
        \end{equation}
        \begin{equation}
            \begin{aligned}
                &\mathcal{T} \text{ or } \mathcal{P}: && w^{\mu\alpha}_{ab}(\boldsymbol{k}) = \partial_{k_{\mu}}\partial_{k_{\alpha}}\varepsilon_{ab}(\boldsymbol{k}) = \partial_{-k_{\mu}}\partial_{-k_{\alpha}}\varepsilon_{ab}(-\boldsymbol{k}) = w^{\mu}_{ab}(-\boldsymbol{k}).
            \end{aligned}
        \end{equation}
        \par
        For the shift vector $R^{\alpha, \beta}_{ab} = i\partial_{\alpha} \ln r^{\beta}_{ab} + r^{\alpha}_{aa} - r^{\alpha}_{bb}$, it can be decomposed into real and imaginary parts as $R^{\alpha, \beta}_{ab} = \tilde{R}^{\alpha, \beta}_{ab} + i\tilde{I}^{\alpha, \beta}_{ab}$. First, we express the Berry connection in the form of a complex exponential, $r^{\beta}_{ab}= |r^{\beta}_{ab}| e^{-i\phi_{ab}}$. The explicit expressions for the real and imaginary parts are given by $\tilde{I}^{\alpha, \beta}_{ab} = \partial_{\alpha} \ln |r^{\beta}_{ab}|$ and $\tilde{R}^{\alpha, \beta}_{ab} = \partial_{\alpha} \phi_{ab} + r^{\alpha}_{aa} - r^{\alpha}_{bb}$. These quantities satisfy the following constraints:
        \begin{equation}
		    \begin{aligned}
		        &\mathcal{T}: &&\tilde{R}^{\alpha, \beta}_{ab}(\boldsymbol{k}) = \partial_{k_{\alpha}}\phi_{ab}(\boldsymbol{k}) + r^{\alpha}_{aa}(\boldsymbol{k}) - r^{\alpha}_{bb}(\boldsymbol{k}) = \partial_{k_{\alpha}}\phi_{ab}(-\boldsymbol{k}) + r^{\alpha}_{aa}(-\boldsymbol{k}) - r^{\alpha}_{bb}(-\boldsymbol{k}) = \tilde{R}^{\alpha, \beta}_{ab}(-\boldsymbol{k}), \\
		        &\mathcal{P}: &&\tilde{R}^{\alpha, \beta}_{ab}(\boldsymbol{k}) = \partial_{k_{\alpha}}\phi_{ab}(\boldsymbol{k}) + r^{\alpha}_{aa}(\boldsymbol{k}) - r^{\alpha}_{bb}(\boldsymbol{k}) = -\partial_{-k_{\alpha}}\phi_{ab}(-\boldsymbol{k}) - r^{\alpha}_{aa}(-\boldsymbol{k}) + r^{\alpha}_{bb}(-\boldsymbol{k}) = -\tilde{R}^{\alpha, \beta}_{ab}(-\boldsymbol{k}), \\
		    \end{aligned}
		\end{equation}
		\begin{equation}
		    \begin{aligned}
		        &\mathcal{T}: &&\tilde{I}^{\alpha, \beta}_{ab}(\boldsymbol{k}) = \partial_{k_{\alpha}} \ln(|r^{\beta}_{ab}(\boldsymbol{k})|) = -\partial_{-k_{\alpha}} \ln(|r^{\beta}_{ba}(-\boldsymbol{k})|) = -\tilde{I}^{\alpha, \beta}_{ab}(-\boldsymbol{k}), \\
		        &\mathcal{P}: &&\tilde{I}^{\alpha, \beta}_{ab}(\boldsymbol{k}) = \partial_{k_{\alpha}} \ln(|r^{\beta}_{ab}(\boldsymbol{k})|) = -\partial_{-k_{\alpha}} \ln(|r^{\beta}_{ab}(-\boldsymbol{k})|) = -\tilde{I}^{\alpha, \beta}_{ab}(-\boldsymbol{k}). 
		    \end{aligned}
		\end{equation}
        \par
        The properties of the shift vector dipole are similar. It can be decomposed into real and imaginary parts, i.e., $\partial_{\mu} R^{\alpha, \beta}_{ab} = \partial_{\mu} \tilde{R}^{\alpha, \beta}_{ab} + i\partial_{\mu} \tilde{I}^{\alpha, \beta}_{ab}$. These two components satisfy
        \begin{equation}
            \begin{aligned}
                &\mathcal{T}: && \partial_{\mu}\tilde{R}^{\alpha, \beta}_{ab}(\boldsymbol{k}) = \partial_{k_{\mu}}\tilde{R}^{\alpha, \beta}_{ab}(\boldsymbol{k}) =
                -\partial_{-k_{\mu}}\tilde{R}^{\alpha, \beta}_{ab}(-\boldsymbol{k}) = -\partial_{\mu}\tilde{R}^{\alpha, \beta}_{ab}(-\boldsymbol{k}), \\
                &\mathcal{P}: && \partial_{\mu}\tilde{R}^{\alpha, \beta}_{ab}(\boldsymbol{k}) = \partial_{k_{\mu}}\tilde{R}^{\alpha, \beta}_{ab}(\boldsymbol{k}) =
                \partial_{-k_{\mu}}\tilde{R}^{\alpha, \beta}_{ab}(-\boldsymbol{k}) = \partial_{\mu}\tilde{R}^{\alpha, \beta}_{ab}(-\boldsymbol{k}), \\
            \end{aligned}
        \end{equation}
        \begin{equation}
            \begin{aligned}
                &\mathcal{T}: && \partial_{\mu}\tilde{I}^{\alpha, \beta}_{ab}(\boldsymbol{k}) = \partial_{k_{\mu}}\tilde{I}^{\alpha, \beta}_{ab}(\boldsymbol{k}) =
                \partial_{-k_{\mu}}\tilde{I}^{\alpha, \beta}_{ab}(-\boldsymbol{k}) = \partial_{\mu}\tilde{I}^{\alpha, \beta}_{ab}(-\boldsymbol{k}), \\
                &\mathcal{P}: && \partial_{\mu}\tilde{I}^{\alpha, \beta}_{ab}(\boldsymbol{k}) = \partial_{k_{\mu}}\tilde{I}^{\alpha, \beta}_{ab}(\boldsymbol{k}) =
                \partial_{-k_{\mu}}\tilde{I}^{\alpha, \beta}_{ab}(-\boldsymbol{k}) = \partial_{\mu}\tilde{I}^{\alpha, \beta}_{ab}(-\boldsymbol{k}). \\
            \end{aligned}
        \end{equation}
        \par
        For the quantum geometric tensor $Q^{\alpha\beta}_{ba} = r^{\alpha}_{ba} r^{\beta}_{ab}$, we decompose it into two components:  $Q^{\alpha\beta}_{ba} = \text{Re} Q^{\alpha\beta}_{ba} + i \text{Im} Q^{\alpha\beta}_{ba} \equiv g^{\alpha\beta}_{ba} - \frac{i}{2} \Omega^{\alpha\beta}_{ba}$.  The real part corresponds to the quantum metric, while the imaginary part represents the Berry curvature. They satisfy the following constraints:
		\begin{equation}
			\begin{aligned}
			    &\mathcal{T}: && g^{\alpha\beta}_{ba}(\boldsymbol{k}) = \frac{r^{\alpha}_{ba}(\boldsymbol{k})r^{\beta}_{ab}(\boldsymbol{k}) + r^{\beta}_{ba}(\boldsymbol{k})r^{\alpha}_{ab}(\boldsymbol{k})}{2} = \frac{r^{\alpha}_{ab}(-\boldsymbol{k})r^{\beta}_{ba}(-\boldsymbol{k}) + r^{\beta}_{ab}(-\boldsymbol{k})r^{\alpha}_{ba}(-\boldsymbol{k})}{2} = g^{\alpha\beta}_{ba}(-\boldsymbol{k}), \\
			    &\mathcal{P}: && g^{\alpha\beta}_{ba}(\boldsymbol{k}) = \frac{r^{\alpha}_{ba}(\boldsymbol{k})r^{\beta}_{ab}(\boldsymbol{k}) + r^{\beta}_{ba}(\boldsymbol{k})r^{\alpha}_{ab}(\boldsymbol{k})}{2} = \frac{r^{\alpha}_{ba}(-\boldsymbol{k})r^{\beta}_{ab}(-\boldsymbol{k}) + r^{\beta}_{ba}(-\boldsymbol{k})r^{\alpha}_{ab}(-\boldsymbol{k})}{2} = g^{\alpha\beta}_{ba}(-\boldsymbol{k}), \\
			\end{aligned}
		  \end{equation}
		  \begin{equation}
			\begin{aligned}
				&\mathcal{T}: && \Omega^{\alpha\beta}_{ba}(\boldsymbol{k}) = i\left(r^{\alpha}_{ba}(\boldsymbol{k})r^{\beta}_{ab}(\boldsymbol{k}) - r^{\beta}_{ba}(\boldsymbol{k})r^{\alpha}_{ab}(\boldsymbol{k})\right) = i\left(r^{\alpha}_{ab}(-\boldsymbol{k})r^{\beta}_{ba}(-\boldsymbol{k}) - r^{\beta}_{ab}(-\boldsymbol{k})r^{\alpha}_{ba}(-\boldsymbol{k})\right) = -\Omega^{\alpha\beta}_{ba}(-\boldsymbol{k}), \\
				&\mathcal{P}: && \Omega^{\alpha\beta}_{ba}(\boldsymbol{k}) = i\left(r^{\alpha}_{ba}(\boldsymbol{k})r^{\beta}_{ab}(\boldsymbol{k}) - r^{\beta}_{ba}(\boldsymbol{k})r^{\alpha}_{ab}(\boldsymbol{k})\right) = i\left(r^{\alpha}_{ba}(-\boldsymbol{k})r^{\beta}_{ab}(-\boldsymbol{k}) - r^{\beta}_{ba}(-\boldsymbol{k})r^{\alpha}_{ab}(-\boldsymbol{k})\right) = \Omega^{\alpha\beta}_{ba}(-\boldsymbol{k}). \\
			\end{aligned}
		  \end{equation}
		\par
        For the triple phase product $T^{\alpha\beta\gamma} = r^{\alpha}_{ba} r^{\beta}_{cb} r^{\gamma}_{ac}$, we similarly decompose it into real and imaginary components:  $T^{\alpha\beta\gamma}_{abc} = \text{Re} T^{\alpha\beta\gamma}_{abc} + i \text{Im} T^{\alpha\beta\gamma}_{abc}$.  These components satisfy the following constraints:
		\begin{equation}
			\begin{aligned}
				&\mathcal{T}: & \text{Re}T^{\alpha\beta\gamma}_{abc}(\boldsymbol{k}) &= \frac{r^{\alpha}_{ba}(\boldsymbol{k})r^{\beta}_{cb}(\boldsymbol{k})r^{\gamma}_{ac}(\boldsymbol{k}) + r^{\alpha}_{ab}(\boldsymbol{k})r^{\beta}_{bc}(\boldsymbol{k})r^{\gamma}_{ca}(\boldsymbol{k})}{2} \\
				& & &=  \frac{r^{\alpha}_{ab}(-\boldsymbol{k})r^{\beta}_{bc}(-\boldsymbol{k})r^{\gamma}_{ca}(-\boldsymbol{k}) + r^{\alpha}_{ba}(-\boldsymbol{k})r^{\beta}_{cb}(-\boldsymbol{k})r^{\gamma}_{ac}(-\boldsymbol{k})}{2} = \text{Re}T^{\alpha\beta\gamma}_{abc}(-\boldsymbol{k}), \\
				&\mathcal{P}: & \text{Re}T^{\alpha\beta\gamma}_{abc}(\boldsymbol{k}) &= \frac{r^{\alpha}_{ba}(\boldsymbol{k})r^{\beta}_{cb}(\boldsymbol{k})r^{\gamma}_{ac}(\boldsymbol{k}) + r^{\alpha}_{ab}(\boldsymbol{k})r^{\beta}_{bc}(\boldsymbol{k})r^{\gamma}_{ca}(\boldsymbol{k})}{2} \\
				& & &=  -\frac{r^{\alpha}_{ba}(-\boldsymbol{k})r^{\beta}_{cb}(-\boldsymbol{k})r^{\gamma}_{ac}(-\boldsymbol{k}) + r^{\alpha}_{ab}(-\boldsymbol{k})r^{\beta}_{bc}(-\boldsymbol{k})r^{\gamma}_{ca}(-\boldsymbol{k})}{2} = -\text{Re}T^{\alpha\beta\gamma}_{abc}(-\boldsymbol{k}),
			\end{aligned}
		\end{equation}
		\begin{equation}
			\begin{aligned}
				&\mathcal{T}: & \text{Im}T^{\alpha\beta\gamma}_{abc}(\boldsymbol{k}) &= \frac{r^{\alpha}_{ba}(\boldsymbol{k})r^{\beta}_{cb}(\boldsymbol{k})r^{\gamma}_{ac}(\boldsymbol{k}) - r^{\alpha}_{ab}(\boldsymbol{k})r^{\beta}_{bc}(\boldsymbol{k})r^{\gamma}_{ca}(\boldsymbol{k})}{2i} \\
				& & &=  \frac{r^{\alpha}_{ab}(-\boldsymbol{k})r^{\beta}_{bc}(-\boldsymbol{k})r^{\gamma}_{ca}(-\boldsymbol{k}) - r^{\alpha}_{ba}(-\boldsymbol{k})r^{\beta}_{cb}(-\boldsymbol{k})r^{\gamma}_{ac}(-\boldsymbol{k})}{2i} = -\text{Im}T^{\alpha\beta\gamma}_{abc}(-\boldsymbol{k}), \\
				&\mathcal{P}: & \text{Im}T^{\alpha\beta\gamma}_{abc}(\boldsymbol{k}) &= \frac{r^{\alpha}_{ba}(\boldsymbol{k})r^{\beta}_{cb}(\boldsymbol{k})r^{\gamma}_{ac}(\boldsymbol{k}) - r^{\alpha}_{ab}(\boldsymbol{k})r^{\beta}_{bc}(\boldsymbol{k})r^{\gamma}_{ca}(\boldsymbol{k})}{2i} \\
				& & &=  -\frac{r^{\alpha}_{ba}(-\boldsymbol{k})r^{\beta}_{cb}(-\boldsymbol{k})r^{\gamma}_{ac}(-\boldsymbol{k}) - r^{\alpha}_{ab}(-\boldsymbol{k})r^{\beta}_{bc}(-\boldsymbol{k})r^{\gamma}_{ca}(-\boldsymbol{k})}{2i} = -\text{Im}T^{\alpha\beta\gamma}_{abc}(-\boldsymbol{k}).
			\end{aligned}
		\end{equation}
        \par

\begin{table*}[ht]
    \renewcommand{\arraystretch}{1.3}
    \centering
    \begin{tabular}{|m{2cm}<{\centering} |m{2cm}<{\centering} | m{2cm}<{\centering} |m{2cm}<{\centering} |m{5.5cm}<{\centering}|m{3.5cm}<{\centering} |}
        \hline
		~ & \multirow{2}*{\textbf{Band Num.}} & \multirow{2}*{\textbf{Singularity}} &\textbf{Composite} &\multirow{2}*{$\boldsymbol{\mathcal{T}}$}& \multirow{2}*{$\boldsymbol{\mathcal{T}} + \boldsymbol{\mathcal{P}}$}\\
		~ & ~ & ~ & \textbf{Geometry} & ~ & ~ \\
        \hline
        \multirow{4}*{\textbf{Injection}} & \multirow{2}*{Pairwise} & \multirow{3}*{1S} & $\Delta\times R \times Q$ & $\Delta \times \tilde{I} \times g + \Delta\times\tilde{R}\times\Omega$  & $\Delta \times \tilde{I} \times g$ \\
		~ & ~ & ~ & $\Delta\times \Delta \times Q$ & $\Delta \times \Delta \times g$ & $\Delta \times \Delta \times g$ \\
        \cline{2-2}
        ~& \multirow{2}*{Multi}  & ~ & $\Delta\times T$ & $\Delta\times\text{Im}T$  & $\Delta \times \text{Im} T$ \\
        \cline{3-6}
        ~ &~& 2S & $\Delta\times T$ & $\Delta\times\text{Im}T$  & $\Delta \times \text{Im} T$\\
        \hline
        \multirow{7}*{\textbf{Shift}} & \multirow{5}*{Pairwise}& \multirow{7}*{1S}&$\partial R \times Q$ & $\partial \tilde{I}\times g + \partial\tilde{R}\times\Omega$ & $\partial \tilde{I} \times g$ \\
        ~ & ~ & ~ & $ R \times R \times Q$ & $\tilde{R}\times\tilde{R}\times g + \tilde{R}\times\tilde{I}\times \Omega + \tilde{I}\times\tilde{I}\times g $& $\tilde{I}\times \tilde{I} \times g$ \\
        ~ & ~ & ~ & $\Delta \times R \times Q$ & $\Delta\times\tilde{I}\times g + \Delta\times\tilde{R}\times\Omega$ & $\Delta \times \tilde{I} \times g$ \\
        ~ & ~ & ~ & $\Delta \times \Delta \times Q$ & $\Delta \times \Delta \times g$& $\Delta \times \Delta \times g$ \\
        ~ & ~ & ~ & $w \times Q$ & $ w \times g$& $w \times g$ \\
        \cline{2-2}
        ~ & \multirow{4}*{Multi} & ~ & $ R \times T$ & $ \tilde{R} \times \text{Re}T + \tilde{I} \times \text{Im}T$& $\tilde{I} \times \text{Im} T$ \\
        ~ & ~ & ~ &  $ \Delta \times T$ & $ \Delta \times \text{Im} T $& $\Delta \times \text{Im} T$ \\
        \cline{3-6}
        ~ &~ & \multirow{2}*{2S}& $ R \times T$ & $ \tilde{R} \times \text{Re}T + \tilde{I} \times \text{Im}T$ & $\tilde{I} \times \text{Im} T$ \\
        ~ & ~ & ~ & $ \Delta \times T$ & $ \Delta \times \text{Im}T$& $\Delta \times \text{Im} T$ \\
		\hline
    \end{tabular}
    \setlength{\abovecaptionskip}{0.4cm}
    \caption{The non-vanishing conductivity components for each category induced by BCL in materials with time reversal symmetry ($\mathcal{T}$) and those with both time reversal symmetry and centrosymmetry ($\mathcal{T} + \mathcal{P}$) are identified based on the surviving composite geometric quantities.}
    \label{surviving term in time reversal symetry}
\end{table*}

        For a material with $\mathcal{T}$ or $\mathcal{P} + \mathcal{T}$ symmetry, the above properties allow us to identify the non-vanishing conductivity components in each category of Table \ref{injection currenty contribution} and Table \ref{shift currenty contribution}, as each category is characterized by a distinct composite geometry. The overall results are summarized in Table \ref{surviving term in time reversal symetry}.  
        As a consequence of symmetry constraints, the composite geometry is constrained to be either purely real or purely imaginary. The properties of conductivity are further determined by two factors: the relaxation time dependence and the number of singularities. Specifically, both the injection current conductivity with a single singularity and the shift current conductivity with double singularities are purely imaginary, whereas both the injection current conductivity with double singularities and the shift current conductivity with a single singularity are purely real.

\end{document}